\documentclass[twocolumn, aps, prc, floatfix]{revtex4}

\usepackage{multirow}
\usepackage{epsfig}
\usepackage{amsmath}

\usepackage{subfigure}
\usepackage{latexsym}
\usepackage{feynmp}

\usepackage[normalem]{ulem}  % \sout{old text} for strikeout
\usepackage[dvips]{color} % For blue in-text comments and additions

\renewcommand\sout{\bgroup \color{red} \ULdepth=-.5ex \ULset}

\begin{document}

\title{Elliptic and triangular flow of charmonium states in heavy ion collisions}

\author{Sungtae Cho}
\affiliation{Division of Science Education, Kangwon National
University, Chuncheon 24341, Korea}

%\date{\today}
\begin{abstract}

We study the elliptic and triangular flow of charmonium states, or
$J/\psi$, $\psi(2S)$, and $\chi_c(1P)$ mesons in heavy ion
collisions. Starting from the evaluation of charmonia transverse
momentum distributions and yields, we calculate elliptic and
triangular flow of charmonium states based on the coalescence
model. We show that the internal structure, or the wave function
distribution of charmonium states plays a significant role,
especially when charmonium states are produced by charm quark
recombination, leading to the transverse momentum distribution of
the $\psi(2S)$ meson as half large as that of the $J/\psi$ meson.
We also consider the dependence of the elliptic and triangular
flow of charmonium states on internal structures of charmonium
states, and find that the wave function effects as well as
feed-down contributions are averaged out for elliptic and
triangular flow, resulting in similar elliptic and triangular flow
for all charmonium states. We investigate further the elliptic and
triangular flow of charmonium states at low transverse momentum
region, and also discuss the quark number scaling of elliptic and
triangular flow for charmonium states.

\end{abstract}

% \pacs{14.40.Pq, 25.75.Dw, 25.75.-q}

% 14.40.Pq Heavy quarkonia
% 25.75.Dw Particle and resonance production
% 25.75.-q Relativistic heavy ion collisions

\maketitle

\section{Introduction}

It has been expected that a new form of the matter composed of
deconfined partonic states, the so-called quark-gluon plasma (QGP)
is created in the environments of high temperatures enough to set
quarks and gluons free from hadrons \cite{Shuryak:1980tp}. Since
the QGP disappears very quickly, and is converted into hadrons in
a time scale of strong interactions, various ways of recognizing
the existence of the QGP, or the phase transition to the QGP in
heavy ion collisions have been sought out. As one of plausible
tools for the probe of the QGP, a heavy quarkonium, i.e., the
$J/\psi$ meson has been proposed from the expectation that the
production of the $J/\psi$ is suppressed in the QGP medium due to
the dissociation of the $J/\psi$ by the Debye color screening
between charm and anti-charm quarks \cite{Matsui:1986dk}.

With much larger achievable energy than ever before, possible in
experiments carried out at the Large Hadron Collider (LHC), other
charmonium states as well as the $J/\psi$ meson are estimated to
be more abundantly produced, thereby becoming convenient probes
and presenting us with important information on many properties of
the QGP \cite{Thews:2000rj, BraunMunzinger:2000px,
Andronic:2007bi}. It has been anticipated that the different
charmonium states dissociate in the QGP at different temperatures
depending on the strength of their binding energies, and therefore
playing significant roles as probes to the temperature variation
of the QGP during the time evolution in heavy ion collisions
\cite{Satz:2005hx, Karsch:2005nk, Mocsy:2007jz}.

The effects of the quark-gluon plasma on the $J/\psi$ meson are
mostly observed with the ratio between the yield of the $J/\psi$
in heavy ion collisions and that in $p+p$ collisions re-scaled by
the number of binary collisions, i.e., the so-called nuclear
modification factor, $R_{AA}$. In the early measurement at the
Relativistic Heavy Ion Collider (RHIC), the $R_{AA}$ of the
$J/\psi$ decreases with numbers of participating nucleons, or with
increasing centralities, implying that the $J/\psi$ production is
actually suppressed due to the presence of the QGP, and is more
suppressed with increasing size of the QGP \cite{Adare:2006ns}.

On the other hand, the same measurement on the $R_{AA}$ of the
$J/\psi$ in collision energies ten times larger at LHC by ALICE
Collaboration shows that the $R_{AA}$ of $J/\psi$ mesons is not
dependent on centralities any more, meaning the less suppression
of the $J/\psi$ production at the QGP, and supporting
possibilities of the $J/\psi$ regeneration from charm quarks in
the QGP \cite{Abelev:2012rv, Abelev:2013ila}. Especially, the
enhancement of the $R_{AA}$ of the $J/\psi$ observed at low
transverse momentum region \cite{Abelev:2013ila} strongly favors
the scenario that the significant amount of the $J/\psi$ meson is
produced from a charm and an anti-charm quarks by recombination in
the QGP.

Recent measurements by ALICE Collaboration show that the $R_{AA}$
of the $\psi(2S)$ meson is also independent of centralities as
that of the $J/\psi$, indicating that the significant amount of
$\psi(2S)$ mesons is regenerated as well from charm quarks in the
QGP \cite{ALICE:2022jeh}. Compared to the $R_{AA}$ of the
$J/\psi$, however, the $R_{AA}$ of the $\psi(2S)$ meson is as half
large as that of the $J/\psi$, raising the possibility of the
different production mechanism between the $J/\psi$ and $\psi(2S)$
when they are produced by charm quark recombination from the QGP.

% Recently, the production of the $\psi(2S)$, the radially excited
% state of the $J/\psi$ has been measured at LHC, and those
% measurements are expected to provide chances to understand in more
% detail the effects of the QGP on the charmonia production as well
% as the production mechanism of charmonium states when compared to
% the measurement of the $J/\psi$ production. The measurement of the
% $\psi(2S)$ is often compared to that of the $J/\psi$ in terms of
% the double ratio of the nuclear modification factor between the
% $\psi(2S)$ and $J/\psi$ \cite{}.

Moreover, the measurement by CMS Collaboration on the nuclear
modification factor ratio between the $\psi(2S)$ and $J/\psi$ in
the transverse momentum, $p_T$ and rapidity, $y$ ranges, 3 $<p_T<$
30 GeV and 1.6 $<|y|<$ 2.4 at $\sqrt{s_{NN}}$=2.76 TeV shows that
the ratio increases with increasing number of participants, or
with increasing centralities \cite{Khachatryan:2014bva}. The
similar measurement in different transverse momentum and rapidity
regions, 9 $<p_T<$ 40 GeV and $|y|<$ 2 by ATLAS Collaboration at
$\sqrt{s_{NN}}$=5.02 TeV also shows that the prompt $\psi(2S)$ to
$J/\psi$ double ratio increases with increasing number of
participants \cite{ATLAS:2018hqe}, implying that different amounts
of the $J/\psi$ and $\psi(2S)$ are produced when they are affected
by the QGP in heavy ion collisions.

It has been known that the yield in the coalescence production is
dependent on the overlap between the coalescence probability
function made up of the wave function of the produced hadron, or
the Wigner distribution and the phase space density functions, or
the transverse momentum distributions of constituents
\cite{Greco:2003xt, Greco:2003mm, Fries:2003vb, Fries:2003kq}.
Reminding that both the $J/\psi$ and $\psi(2S)$ share charm and
anti-charm quarks as their constituents, one may confirm different
coalescence probabilities for the cause of different amounts of
the $J/\psi$ and $\psi(2S)$ yields when the $J/\psi$ and
$\psi(2S)$ are formed from the same source in the QGP by
recombination.

The investigation on the production of not only the $J/\psi$ meson
but also $\psi(2S)$ and $\chi_{c1}(1P)$ mesons from the QGP by
recombination has already been made \cite{Cho:2014xha}, and it has
been shown that the $\psi(2S)$ meson can be produced half as much
as the $J/\psi$ meson when those charmonia are produced by charm
quark coalescence, which is a very large enhancement in the
production of the $\psi(2S)$ compared to that of the $J/\psi$ when
the mass difference between two charmonium states, about 600 MeV
is taken into account. The enhanced production of the $\psi(2S)$,
or the different amount of yields for different charmonium states
has been found to be originated from the different amounts of
overlap between the different Wigner function for each charmonium
state and the same distribution of constituents in phase space, or
the different amounts of overlap between different wave function
distributions from different internal structures, $1S, 2S$, and
$1P$ of charmonia and the same transverse momentum distribution of
charm quarks in momentum space \cite{Cho:2014xha}.

In addition to the transverse momentum distribution of charmonium
states, the elliptic flow of charmonium states is also dependent
on their wave function distributions through the coalescence
probability function; the elliptic flow of hadrons is known to
bear the elliptic flow of constituents caused by their anisotropic
configurations at the moment of heavy ion collisions when hadrons
are produced by quark coalescence \cite{Molnar:2003ff}. Since the
elliptic flow of hadrons is dependent on transverse momentum
distributions of constituents, to be exact, the transverse
momentum distribution anisotropy of quarks via the transverse
momentum distribution of hadrons, the elliptic flow of charmonium
states is also expected to be dependent on their internal
structures, thereby giving rise to different elliptic flow for
different charmonium states when charmonium states of different
internal structures are produced from charm quarks by
recombination.

The study on the elliptic flow of charmonium states has already
been carried out experimentally, e.g., \cite{ALICE:2013xna}, but
the consideration of different coalescence probability functions
for different charmonium states has been overlooked. Moreover, the
investigation on whether or not the so-called quark number scaling
of hadrons, the elliptic flow of hadrons similar in size to the
elliptic flow of a quark times the number of quarks inside hadrons
\cite{Molnar:2003ff}, is also applicable to the elliptic flow of
charmonium states has not been made when charmonium states of
different internal structures produced from the same number of
charm quarks by coalescence.

Therefore, it is necessary to re-evaluate the elliptic flow of
each charmonium state produced from charm quarks by coalescence in
heavy ion collisions in order to investigate the dependence of the
elliptic flow of charmonium states on their wave function
distribution in momentum space as well as the applicability of the
quark number scaling of elliptic flow to the case of charmonium
states. This study may help us to understand the recent
observation by CMS Collaboration on the transverse momentum
distribution of the elliptic flow of not only the $J/\psi$ but
also the $\psi(2S)$ \cite{CMS:2022gvy}, showing different
transverse momentum distributions of the elliptic flow between the
$J/\psi$ and $\psi(2S)$.

In addition, the CMS Collaboration also reported the observation
of the triangular flow of the $J/\psi$ in heavy ion collisions
\cite{CMS:2022gvy}. In line with this observation we find it is
also necessary to study the triangular flow of charmonium states
when charmonium states of different internal structures are
produced from the same charm quark triangular flow by coalescence,
and therefore we extend our discussion on elliptic flow of
charmonium states to the study on the higher harmonic flow of
charmonium states as well.

As is well known, the harmonic flow, e.g., the elliptic and
triangular flow, is originated from the initial geometry of
nucleus at the moment of heavy ion collisions; the pressure
gradient generated in the particular shape caused by the
anisotropic initial collisions creates various kinds of flows, or
harmonic flows. In addition to the anisotropy of nucleon
distributions in heavy ion collisions, event-by-event fluctuations
in heavy ion collisions are also found to be important origins,
especially giving rise to higher flow harmonics, e.g., the
triangular flow \cite{Alver:2010gr, Alver:2010dn, Petersen:2010cw,
Nahrgang:2014vza}. We discuss here mainly the contribution of the
triangular flow of charmonium states built from the triangular
flow of charm quarks due to the initial anisotropic heavy ion
collisions in the same way that the elliptic flow of charmonium
states is formed from that of charm quarks by recombination.

The paper is organized as follows. In Sec. II, we discuss
transverse momentum distributions and yields of charmonium states,
the $J/\psi$, $\psi(2S)$, and $\chi_{c1}(1P)$. We investigate the
different internal structures of charmonium states, and calculate
their yields and transverse momentum distributions which are
dependent on their internal structures, or their wave function
distributions in momentum space. In Sec. III we evaluate elliptic
and triangular flow of charmonium states when they are produced
from charm quark elliptic and triangular flow by recombination at
the quark-hadron phase boundary. We further show the quark number
scaling of elliptic and triangular flow for charmonium states, and
also present comparison between our evaluation and various
experimental measurements in Sec. III. We discuss the dependence
of transverse momentum distributions, yields and flow harmonics of
charmonium states on their internal structure, or the wave
function distribution of each charmonium state in momentum space
in Sec. IV. Finally, we present our summary and conclusion in Sec.
V.

\section{Transverse momentum distributions and yields of charmonium states}

We first consider the meson produced from the quark-gluon plasma
by quark coalescence. When a meson is formed from a quark, $q$ and
an anti-quark, $\bar{q}$, the yield of the meson in the
coalescence model is described by \cite{Greco:2003mm},
\begin{eqnarray}
&& N=g\int p_q\cdot d\sigma_{q} p_{\bar{q}}\cdot d\sigma_{\bar{q}}
\frac{d^3\vec p_q}{(2\pi)^3E_c}\frac{d^3\vec
p_{\bar{q}}}{(2\pi)^3E_{\bar{q}}} \nonumber \\
&& \qquad \times f_q(r_q, p_q)f_{\bar{q}'}(r_{\bar{q}'},
p_{\bar{q}})W(r_q, r_{\bar{q}} ; p_q, p_{\bar{q}}),
\label{CoalGen}
\end{eqnarray}
where $d\sigma$ is a hyper-surface element, and $f_{q}(r, p)$ and
$f_{\bar{q}}(r, p)$ are, respectively, covariant distribution
functions of a quark and an anti-quark satisfying the following
normalization condition,
\begin{equation}
\int p\cdot d\sigma \frac{d^3 \vec p_{q(\bar{q})}}{(2\pi)^3
E_{q(\bar{q})}} f_{q(\bar{q})} (r, p)=N_{q(\bar{q})},
\end{equation}
the number of all (anti-) quarks available in the system. The
factor $g$ takes into account the chance of generating a meson
from quarks, e.g., $g_{J/\psi}=3/(2\times 3)^2$. In Eq.
(\ref{CoalGen}), $W(r_q, r_{\bar{q}} ; p_q, p_{\bar{q}})$ is the
coalescence probability function, or the so-called Wigner function
made up of wave functions of the hadron produced by coalescence.

Under the assumption of boost-invariant longitudinal momentum
distributions of (anti-) quarks satisfying $\eta=y$, or the
Bjorken correlation between spatial and momentum rapidities, the
transverse momentum distribution of the yield for the meson can be
derived in the non-relativistic limit from Eq. (\ref{CoalGen})
into \cite{Greco:2003xt, Greco:2003mm, Greco:2003vf, Oh:2009zj},
\begin{eqnarray}
&& \frac{d^2N}{dp_T^2}=\frac{g}{V}\int d^3\vec r d^2\vec
p_{qT}d^2\vec p_{\bar{q}T}\delta^{(2)}(\vec p_T-\vec
p_{qT}-\vec p_{\bar{q}T}) \nonumber \\
&& \qquad\quad \times \frac{d^2N_q}{dp_{qT}^2}
\frac{d^2N_{\bar{q}}}{dp_{\bar{q}T}^2}W(\vec r, \vec k),
\label{CoalTrans}
\end{eqnarray}
where $d^2N_{q(\bar{q})}/d^2\vec p_{q(\bar{q})T}$ is the number of
(anti-) quarks as a function of transverse momentum, and $W(\vec
r, \vec k)$ is the Wigner function,
\begin{equation}
W(\vec r, \vec k)=\int \frac{d^3\vec q}{(2\pi)^3}
\tilde{\psi}^*\Big(\vec k+\frac{\vec q}{2}\Big)e^{-i\vec
r\cdot\vec q}\tilde{\psi}\Big(\vec k-\frac{\vec q}{2}\Big),
\label{wigner}
\end{equation}
with $\vec r$ and $\vec k$ being, respectively, the distance and
relative momentum between a quark and an anti-quark in a meson
rest frame. The Wigner function has been normalized on the
condition, $\int W(\vec r, \vec k)d^3\vec r d^3\vec k=(2\pi)^3$,
and the $\tilde{\psi}(\vec k)$ in Eq. (\ref{wigner}) is the wave
function of the meson produced by quark coalescence in momentum
representation.

In the same manner, the transverse momentum distribution of
charmonium states can be derived as represented in Eq.
(\ref{CoalTrans}),
\begin{eqnarray}
&& \frac{d^2N_M}{dp_T^2}=\frac{g_M}{V}\int d^3\vec r d^2\vec
p_{cT}d^2\vec p_{\bar{c}T}\delta^{(2)}(\vec p_T-\vec
p_{cT}-\vec p_{\bar{c}T}) \nonumber \\
&& \qquad\quad \times \frac{d^2N_q}{dp_{cT}^2}
\frac{d^2N_{\bar{c}}}{dp_{\bar{c}T}^2}W_M(\vec r, \vec k),
\label{CoalTransCharmonium}
\end{eqnarray}
where subscript $M$ in Eq. (\ref{CoalTransCharmonium}) stands for
the kind of charmonium states, e.g., $s$ for $J/\psi$, $p$ for
$\chi_{c1}(1P)$, and $10$ for $\psi(2S)$.

As shown in Eq. (\ref{CoalTransCharmonium}), yields or transverse
momentum distributions of charmonium states produced by quark
coalescence are mainly dependent on two factors, the transverse
momentum distribution of charm quarks and the Wigner function.
Reminding that all different charmonium states, the $J/\psi$,
$\chi_c(1P)$, and $\psi(2S)$ have the same charm quark components,
one can expect that the Wigner function plays a major role in Eq.
(\ref{CoalTransCharmonium}) in characterizing the production of
different charmonium states from the same charm quark constituents
by charm quark coalescence, as already pointed out in
\cite{Cho:2014xha}.

As the Wigner function is constructed from the wave function of
hadron produced by coalescence, Eq. (\ref{wigner}), the transverse
momentum distribution, and also the yield of charmonium states
should be dependent on the wave function of charmonium states
through the Wigner function, e.g., the Wigner function of the
$J/\psi$ must be different from that of the $\psi(2S)$; the
$J/\psi$ meson is an $s$-wave state whereas the $\psi(2S)$ is a
radially excited state of the $J/\psi$. It should be noted that
the internal structure of the $\chi_c(1P)$ meson is also different
from those of the $J/\psi$ and $\psi(2S)$ meson; the $\chi_c(1P)$
meson is in a $p$-wave state.

One can investigate the explicit dependence of the transverse
momentum distribution of charmonium states on the wave function by
considering various types of wave functions. In general, Gaussian
type wave functions are chosen for the description of the wave
function of hadrons. On the other hand, one can also consider
Coulomb-type wave functions, especially for charmonium states.
Since charmonium states can be considered as the charm and
anti-charm quark bound state formed by a color Coulomb interaction
between charm and anti-charm quarks \cite{Matsui:1986dk}, as an
analogue of the electromagnetic Coulomb interaction between an
electron and a proton inside a Hydrogen atom, the Coulomb wave
function can also be adopted in describing the wave function of
charmonium states.

The explicit representation of the Wigner function constructed
from Coulomb wave functions for charmnonium states as well as the
dependence of the transverse momentum distribution of charmonium
states on both Coulomb and Gaussian wave functions have already
been discussed \cite{Cho:2014xha}. It has been shown that
transverse momentum distributions of the $J/\psi$ and $\psi(2S)$
meson are clearly dependent on their wave functions, and also the
detailed comparison between transverse momentum distribution of
the $J/\psi$ and $\psi(2S)$ based on both Coulomb and Gaussian
wave functions has been made \cite{Cho:2014xha}.

With this in mind, we apply in this work different wave functions,
or different Wigner functions for different charmonium states in
evaluating both yield distributions as functions of transverse
momenta and flow harmonics such as $v_2$ and $v_3$ as an attempt
to investigate differences in the production of different
charmonium states. However, we do not adopt here both types of
wave functions, the Coulomb and Gaussian, and confine our
discussion on the results only obtained with Gaussian-type wave
functions so as to focus, without loss of generality, on
differences between charmonium states in their production by charm
quark coalescence at the quark-hadron phase boundary.

The Wigner functions based on Gaussian wave functions have already
been obtained for $s$-, $p$- \cite{Baltz:1995tv,
KanadaEn'yo:2006zk} and $d$-wave \cite{Cho:2011ew} states,
\begin{eqnarray}
&& W_s(\vec r, \vec k) = 8 e^{-\frac{r^2}{\sigma^2}-k^2
\sigma^2} \nonumber \\
&& W_p(\vec r, \vec k) = \bigg( \frac{16}{3} \frac{r^2}{\sigma^2}
-8+\frac{16}{3} \sigma^2 k^2 \bigg)
e^{-\frac{r^2}{\sigma^2}-k^2 \sigma^2} \nonumber \\
&& W_d(\vec r,\vec k) =
\frac{8}{15}\Big(4\frac{r^4}{\sigma^4}-20\frac{r^2}{\sigma^2}
+15-20\sigma^2k^2+4\sigma^4k^4 \nonumber \\
&& \qquad\quad\quad+16r^2k^2-8(\vec r\cdot\vec k)^2\Big)
e^{-\frac{r^2}{\sigma^2}-k^2\sigma^2}, \label{WigGau}
\end{eqnarray}
where an oscillator frequency $\omega$ is related to the reduced
mass $\mu$ with $\sigma^2=1/(\mu\omega)$. The $2S$ state Wigner
function based on a Gaussian wave function is also available
\cite{Cho:2014xha},
\begin{eqnarray} && W_{10}(\vec r,\vec k) =
\frac{16}{3}\Big(\frac{r^4}{\sigma^4}-2\frac{r^2}{\sigma^2}
+\frac{3}{2}-2\sigma^2k^2+\sigma^4k^4 \nonumber \\
&& \qquad\qquad\quad-2r^2k^2+4(\vec r\cdot\vec k)^2\Big)
e^{-\frac{r^2}{\sigma^2}-k^2\sigma^2}, \label{wigner10}
\end{eqnarray}
constructed from 3-dimensional harmonic oscillator wave functions.
The subscript $10$ represents the first excited state, $\psi_{10}$
from its ground state, $\psi_{00}$ with the lowest angular
momentum, describable for the wave function of the $\psi(2S)$
meson. The more Wigner function for higher excited states also
constructed from 3-dimensional harmonic oscillator wave functions
together with phase space distributions have been systematically
investigated recently \cite{Kordell:2021prk}.

The Wigner function in Eqs. (\ref{WigGau}) and (\ref{wigner10}),
can be simplified to the absolute value square of the wave
function in momentum representation when it is integrated over
coordinate space, $\vec r$ with the help of one of important
properties of the Wigner function \cite{Hillery:1983ms},
\begin{equation}
\int d^3\vec r W(\vec r, \vec k)=|\tilde{\psi}(\vec k)|^2.
\label{IntegWigner}
\end{equation}
The $\tilde{\psi}(\vec k)$ is the wave function in momentum
representation, corresponding to the coordinate space wave
function adopted in the Wigner function, $\psi(\vec r)$.

Applying Eq. (\ref{IntegWigner}) to Eq. (\ref{CoalTrans}), we
obtain the simpler yield distribution as a function of transverse
momentum,
\begin{eqnarray}
&& \frac{d^2N_M}{dp_T^2}=\frac{g_M}{V}\int d^2\vec p_{cT}d^2\vec
p_{\bar{c}T}\delta^{(2)}(\vec p_T-\vec
p_{cT}-\vec p_{\bar{c}T}) \nonumber \\
&& \qquad\quad \times \frac{d^2N_c}{dp_{cT}^2}
\frac{d^2N_{\bar{c}}}{dp_{\bar{c}T}^2}|\tilde{\psi}_M(\vec k)|^2.
\label{CoalTransSim}
\end{eqnarray}
The absolute value square of the wave function in momentum
representation for each charmonium state is given by,
\begin{eqnarray}
&& |\tilde{\psi}_M(\vec
k)|^2 \nonumber \\
&& =\left\{ \begin{array}{ll}
(2\sqrt{\pi}\sigma)^3 e^{-k^2\sigma^2} &\psi_s; J/\psi  \\
\frac{2}{3}(2\sqrt{\pi}\sigma)^3 e^{-k^2\sigma^2}\sigma^2k^2
&\psi_p; \chi_{c1}(1P)  \\
\frac{2}{3}(2\sqrt{\pi}\sigma)^3 e^{-k^2\sigma^2}\Big(
\sigma^2k^2-\frac{3}{2}\Big)^2 &\psi_{10}; \psi(2S).
\end{array} \right. \label{WigIntdr}
\end{eqnarray}

When evaluating the transverse momentum distribution of the
charmonium yield, Eq. (\ref{CoalTransSim}), we only consider the
contribution from transverse momenta by neglecting the
longitudinal momentum in the wave function at mid-rapidities,
$y=0$ \cite{Cho:2011ew, Cho:2014xha}; the relative momentum
between charm quarks becomes, $\vec k=(\vec p_{cT}^{~'}-\vec
p_{\bar{c}T}^{~'})/2$ with $\vec p_{cT}^{~'}$ and $\vec
p_{\bar{c}T}^{~'}$ being the transverse momenta in the charmonium
frame, converted from the transverse momenta of charm and
anti-charm quarks, $\vec p_{cT}$ and $\vec p_{\bar{c}T}$ in a
fire-ball frame by the Lorentz transformation
\cite{Scheibl:1998tk, Oh:2009zj}.

In order to evaluate Eq. (\ref{CoalTransSim}), it is necessary to
determine an oscillator frequency, $\omega$, which is related to
the size of charmonium states produced by charm quark coalescence
\cite{Oh:2009zj, Cho:2019lxb, Cho:2019syk}. Here, we adopt
oscillator frequencies, $\omega=0.078$ GeV at RHIC and
$\omega=0.076$ GeV at LHC \cite{Cho:2019syk} obtained on the
condition that all charm quarks at zero transverse momentum are
hadronized exclusively by quark coalescence \cite{Oh:2009zj,
Cho:2019syk}. In determining above oscillator frequencies, a total
of 14 single charmed hadrons, i,e., ten charm baryons,
$\Lambda_c$, $\Sigma_c(2455)$, $\Sigma_c(2520)$,
$\Lambda_c(2595)$, $\Lambda_c(2625)$, $\Xi_c$, $\Xi_c'$,
$\Xi_c(2645)$, $\Omega_c$, and $\Omega_c(2770)$, and four open
charm mesons, $D$, $D^*$, $D_s$, and $D_s^*$ have been taken into
account. Since transverse momentum distributions of charmonium
states, made up of charm and anti-charm quarks, are small compared
to those of single charmed hadrons mentioned above, the oscillator
frequencies are found to be almost unchanged as $\omega=0.078$ GeV
at RHIC and $\omega=0.076$ GeV at LHC even though charmonium
states are included in the above calculation of oscillator
frequencies on the same condition.

Based on the relation between the mean square radius, $\langle
r^2\rangle$ and the oscillator frequency, $\sigma^2=1/(\mu\omega)$
for charmonium states, $\langle
r^2\rangle_{J/\psi}$=3/2$\sigma_{J/\psi}^2$, $\langle
r^2\rangle_{\chi_{c1}(1P)}$ =5/2$\sigma_{\chi_{c1}(1P)}^2$, and
$\langle r^2\rangle_{\psi(2S)}$=7/2$\sigma_{\psi(2S)}^2$
\cite{Cho:2014xha} the size of charmonium states can be evaluated
when the above oscillator frequencies are adopted. Using
$\omega=0.078$ or $\omega=0.076$ GeV one obtains, respectively,
1.0, 1.3, and 1.5 fm for the root-mean-square radii $\sqrt{\langle
r^2\rangle}$ of $J/\psi$, $\chi_{c1}(1P)$, and $\psi(2S)$ states.

In addition to oscillator frequencies, it is also necessary to
have the information on the transverse momentum distribution of
charm quarks in the system, $d^2N_c/dp_{cT}^2$ in Eq.
(\ref{CoalTransSim}). Here we introduce the following transverse
momentum distributions of charm quarks at midrapidity in 0-10$\%$
centrality \cite{Plumari:2017ntm},
\begin{widetext}
\begin{eqnarray}
&& \frac{d^2N_c^R}{dp_{cT}^2}=\left\{
\begin{array}{ll}
0.69e^{(-1.22p_{cT}^{1.57})} & \quad p_{cT} \le 1.85~\textrm{GeV} \\
1.08e^{(-3.04p_{cT}^{0.71})}+3.79(1.0+p_{cT}^{2.02})^{-3.48} &
\quad p_{cT} > 1.85~\textrm{GeV} \\
\end{array} \right. \nonumber \\
&& \frac{d^2N_c^L}{dp_{cT}^2}=\left\{
\begin{array}{ll}
1.97e^{(-0.35p_{cT}^{2.47})} & ~p_{cT} \le 1.85~\textrm{GeV} \\
7.95e^{(-3.49p_{cT}^{3.59})}+87335(1.0+p_{cT}^{0.5})^{-14.31} &
~p_{cT} > 1.85~\textrm{GeV} \\
\end{array} \right., \label{dNcdpT}
\end{eqnarray}
\end{widetext}
with superscripts $R$ and $L$ being represented by a charm quark
transverse momentum distribution at RHIC and LHC, respectively.
The transverse momentum distribution of the charm quark at LHC in
Eq. (\ref{dNcdpT}) has been obtained for $\sqrt{s_{NN}}=2.76$ TeV
Pb+Pb collisions \cite{Plumari:2017ntm}. The above transverse
momentum distributions, Eq. (\ref{dNcdpT}) corresponds to
$dN_c/dy$=2.00 at RHIC and $dN_c/dy$=14.9 at LHC for the total
number of charm quarks at midrapidity in the system
\cite{Cho:2019syk}.

With the transverse momentum distribution of charm quarks, Eq.
(\ref{dNcdpT}) and also the Wigner function, Eq. (\ref{WigIntdr})
we evaluate transverse momentum distributions of charmonium
states, $J/\psi$, $\chi_{c1}(1P)$, and $\psi(2S)$ mesons produced
from charm quarks by recombination at the quark-hadron phase
transition, $d^2N_{J/\psi}/dp_T^2$, $d^2N_{\chi_{c1}(1P)}/dp_T^2$,
and $d^2N_{\psi(2S)}/dp_T^2$ at both RHIC, $\sqrt{s_{NN}}=200$ GeV
and LHC, $\sqrt{s_{NN}}=2.76$ TeV. We use the coalescence volume
1790 and 3530 $\rm{fm}^3$ for RHIC and LHC, respectively, and
assume the constituent charm quark mass as 1.5 GeV
\cite{Cho:2017dcy, Cho:2019syk}.

\begin{figure}[!t]
\begin{center}
\includegraphics[width=0.51\textwidth]{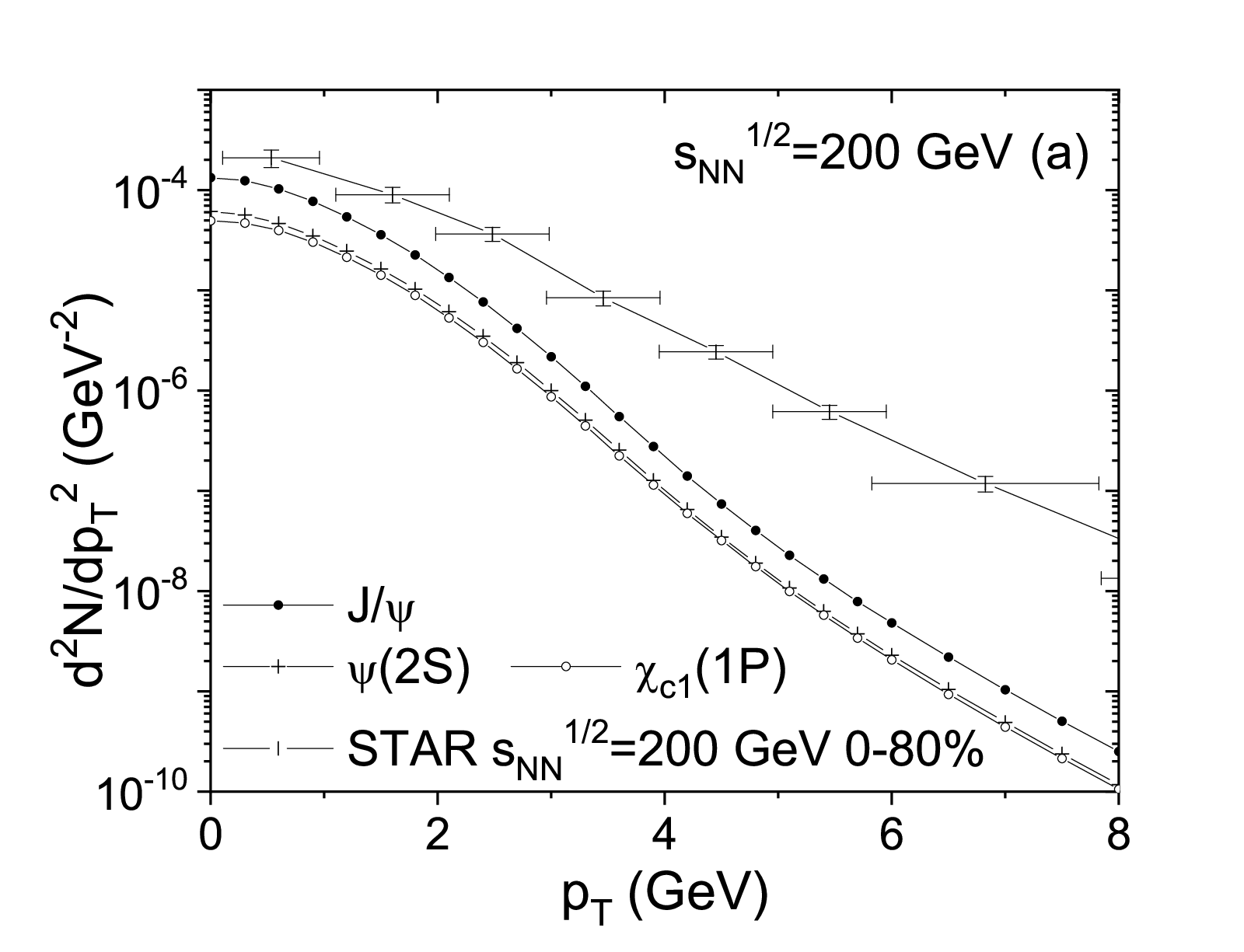}
\includegraphics[width=0.51\textwidth]{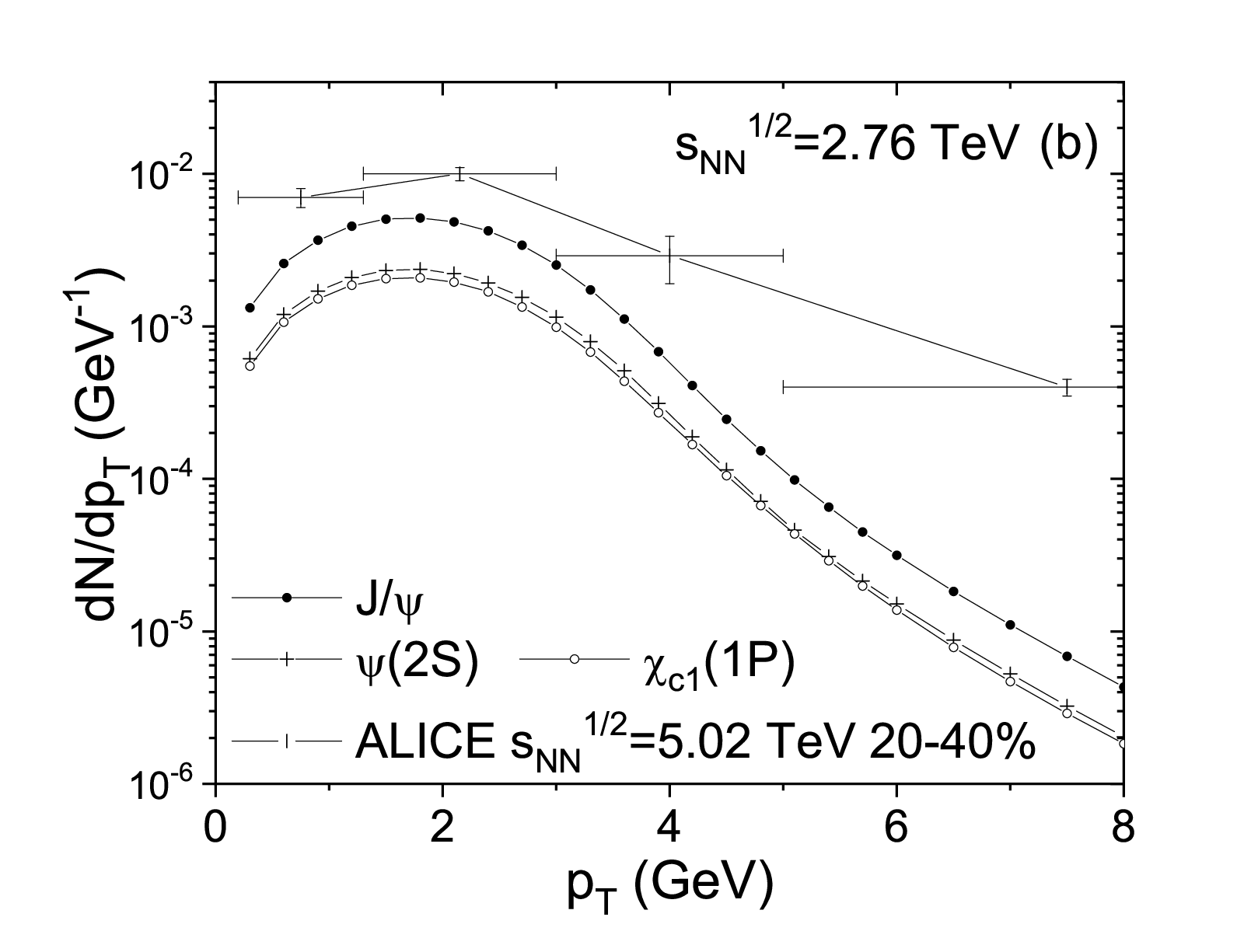}
\end{center}
\caption{Transverse momentum distributions, $d^2N/dp_T^2$ of
charmonium states at midrapidity in 0-10 $\%$ centrality at RHIC,
$\sqrt{s_{NN}}=200$ GeV (a) and those multiplied by $2\pi p_T$,
$dN/dp_T$ at LHC, $\sqrt{s_{NN}}=2.76$ TeV (b). We also show for
comparison experimental measurements of the $J/\psi$ transverse
momentum distribution at RHIC, $\sqrt{s_{NN}}=200$ GeV
\cite{Adam:2019rbk} at midrapidity in 0-80 $\%$ centrality (a) and
the transverse momentum distribution of the $J/\psi$ multiplied by
$2\pi p_T$, $dN/dp_T$ measured at LHC, $\sqrt{s_{NN}}=5.02$ TeV,
$|y|<$0.9 in 20-40 $\%$ centrality \cite{Acharya:2019lkh} (b). }
\label{pT_charmonia}
\end{figure}

\begin{figure}[!h]
\begin{center}
\includegraphics[width=0.54\textwidth]{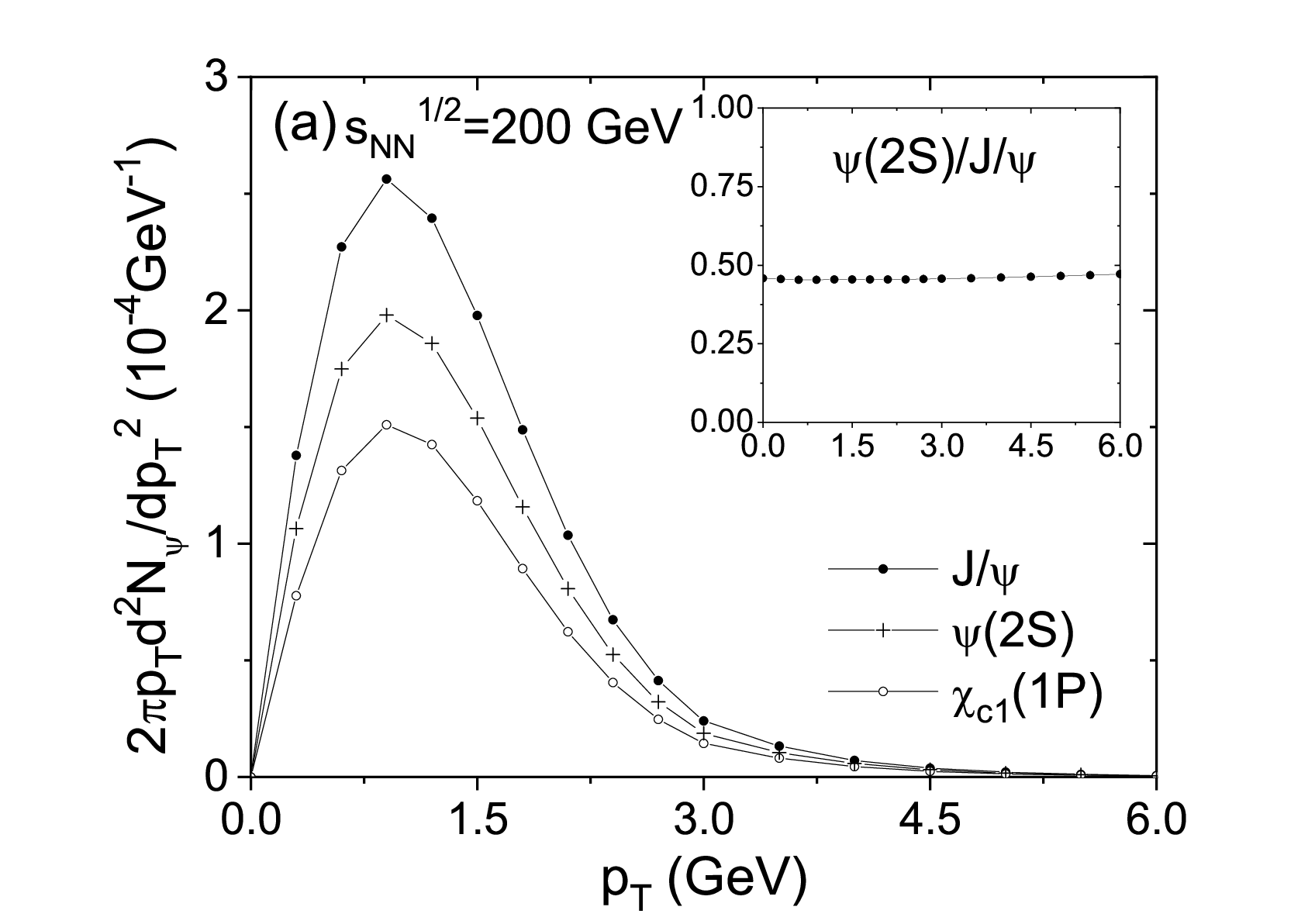}
\includegraphics[width=0.54\textwidth]{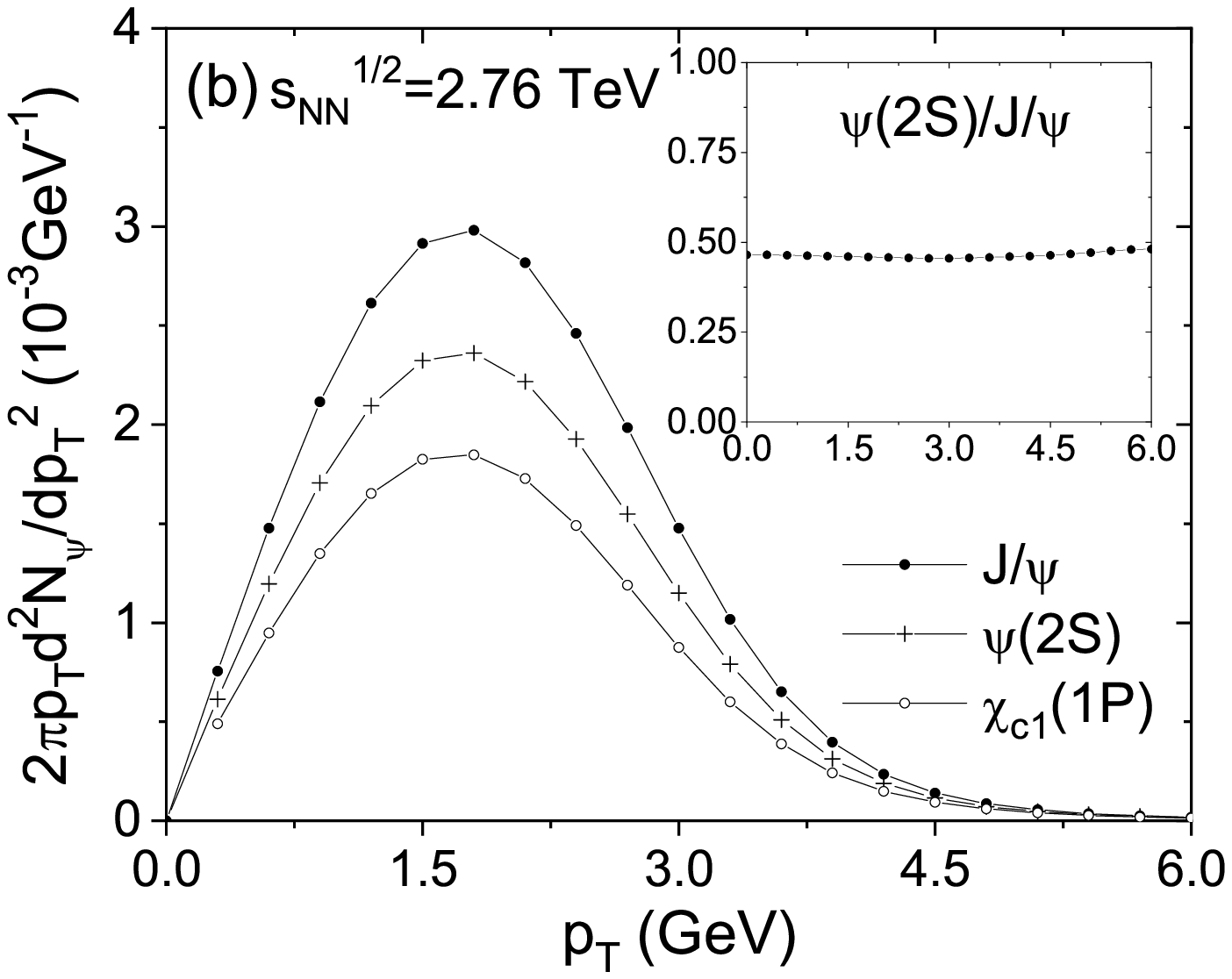}
\end{center}
\caption{Transverse momentum distributions multiplied by $2\pi
p_T$, $dN/dp_T$ of charmonium states, $J/\psi$, $\chi_{c1}(1P)$,
and $\psi(2S)$ mesons at midrapidity in 0-10 $\%$ centrality
without feed-down contributions at RHIC, $\sqrt{s_{NN}}=200$ GeV
(a) and LHC, $\sqrt{s_{NN}}=2.76$ TeV (b). In the inset of each
figure the transverse momentum distribution ratios between the
$\psi(2S)$ and $J/\psi$ obtained after including the feed-down
contributions following Eq. (\ref{feeddown}),
$d^2N_{\psi(2S)}/{dp_T^2} /d^2N_{J/\psi}^{all}/dp_T^2$, are shown.
} \label{Jpsitopsi2Sratio}
\end{figure}

We show in Fig. \ref{pT_charmonia} transverse momentum
distributions, $d^2N/dp_T^2$ of charmonium states evaluated at
midrapidity in 0-10 $\%$ centrality at RHIC, $\sqrt{s_{NN}}=200$
GeV and those multiplied by $2\pi p_T$, $dN/dp_T$ at LHC,
$\sqrt{s_{NN}}=2.76$ TeV. We also show for comparison experimental
measurements of the $J/\psi$ transverse momentum distribution at
RHIC, $\sqrt{s_{NN}}=200$ GeV, $|y|<0.5$ in 0-80 $\%$ centrality
\cite{Adam:2019rbk} and the transverse momentum distribution of
the $J/\psi$ multiplied by $2\pi p_T$, $dN/dp_T$ measured at LHC,
$\sqrt{s_{NN}}=5.02$ TeV, $|y|<$0.9 in 20-40 $\%$ centrality
\cite{Acharya:2019lkh}. In evaluating the transverse momentum
distributions of $J/\psi$ and $\chi_{c1}(1P)$ as shown in Fig.
\ref{pT_charmonia} contributions from the decay of heavier
charmonium states are also taken into account, i.e., the
significant amount of the $J/\psi$ production comes from other
heavier charmonium states, feed-downs from $\chi_{c1}(1P)$ and
$\psi(2S)$ mesons \cite{Beringer:1900zz};
\begin{eqnarray}
&& \frac{d^2N_{J/\psi}^{all}}{dp_T^2}=
\frac{d^2N_{J/\psi}}{dp_T^2}+0.614~\frac{d^2N_{\psi(2S)}}
{dp_T^2}+0.343\frac{d^2N_{\chi_{c1}}^{all}}{dp_T^2}, \nonumber \\
&& \frac{d^2N_{\chi_{c1}}^{all}}{dp_T^2}=
\frac{d^2N_{\chi_{c1}}}{dp_T^2}
+0.0975~\frac{d^2N_{\psi(2S)}}{dp_T^2}.   % 0.665=(0.614+0.0975*0.343+0.0952*0.19)
\label{feeddown}
\end{eqnarray}

We have neglected in Eq. (\ref{feeddown}) the modification of the
transverse momentum distribution of daughter particles from that
of a mother particle since decays are mostly radiative, and the
mass difference between mother and daughter particles is small
compared to the mass of daughter particles.

We also show in Fig. \ref{Jpsitopsi2Sratio} transverse momentum
distributions multiplied by $2\pi p_T$, $dN/dp_T$ of charmonium
states, $J/\psi$, $\chi_{c1}(1P)$, and $\psi(2S)$ mesons at
midrapidity in 0-10 $\%$ centrality when they are purely produced
from charm quarks by recombination at the quark-hadron phase
transition, $d^2N_{J/\psi}/dp_T^2$, $d^2N_{\chi_{c1}}/dp_T^2$, and
$d^2N_{\psi(2S)}/dp_T^2$. In addition, we show in the inset of
Fig. \ref{Jpsitopsi2Sratio} the transverse momentum distribution
ratios between the $\psi(2S)$ and $J/\psi$ obtained after
considering the feed-down contributions following Eq.
(\ref{feeddown}), $d^2N_{\psi(2S)}/{dp_T^2}
/d^2N_{J/\psi}^{all}/dp_T^2$, also at RHIC, $\sqrt{s_{NN}}=200$
GeV and LHC, $\sqrt{s_{NN}}=2.76$ TeV.

As shown in Fig. \ref{Jpsitopsi2Sratio} the transverse momentum
distribution of the $\psi(2S)$ is found to be as large as that of
the $J/\psi$ when they are initially produced from charm quarks by
recombination. Accordingly, the transverse momentum distribution
of the $\psi(2S)$ is still comparable to that of the $J/\psi$
though major feed-down contributions to the $J/\psi$ from heavier
charmonium states, the $\chi_{c1}(1P)$ and $\psi(2S)$ are
included, as shown in the inset of the Fig.
\ref{Jpsitopsi2Sratio}. It is expected, however, that the actual
transverse momentum distributions of the $J/\psi$ measured at both
RHIC and LHC would be slightly larger than those shown in Fig.
\ref{Jpsitopsi2Sratio} due to feed-down contributions from much
heavier charmed and bottomed hadrons at higher transverse momentum
regions.

It should be reminded that the yield, or the transverse momentum
distribution of the $\psi(2S)$ is expected to be smaller than that
of the $J/\psi$ by about a factor of $10^2$ in the statistical
hadronization model where the heavier mass of the $\psi(2S)$ than
that of the $J/\psi$ by about 600 MeV makes the yield of the
$\psi(2S)$ much smaller compared to that of the $J/\psi$. However,
as seen in Fig. \ref{Jpsitopsi2Sratio}, the transverse momentum
distribution of the $\psi(2S)$ based on the charm quark
coalescence is not so small compared to that of the $J/\psi$
meson, and therefore the transverse momentum distribution ratios
between the $\psi(2S)$ and $J/\psi$ meson becomes about 0.5 in the
transverse momentum range between 0 and 6 GeV at both RHIC,
$\sqrt{s_{NN}}=200$ GeV and LHC, $\sqrt{s_{NN}}=2.76$ TeV. This
becomes more evident if the yield of the $\psi(2S)$ is directly
compared to that of the $J/\psi$. Here we take the integration of
the transverse momentum distribution of charmonium states shown in
Fig. \ref{pT_charmonia} over all transverse momenta, and evaluate
yields of charmonium states. We take into account all the
feed-down contribution as well for the yield based on Eq.
(\ref{feeddown}), and summarize the result in Table
\ref{pT_yields}.

\begin{table}[!b]
\caption{Total yields of the $J/\psi$, $\psi(2S)$, and
$\chi_{c1}(1P)$ meson at midrapidity in 0-10 $\%$ centrality
obtained by integrating the transverse momentum distributions
shown in Fig. \ref{pT_charmonia} over all transverse momenta at
RHIC in $\sqrt{s_{NN}}=200$ GeV Au+Au collisions and at LHC in
$\sqrt{s_{NN}}=2.76$ TeV Pb+Pb collisions. } \label{pT_yields}
\begin{center}
\begin{tabular}{c|c|c}
\hline \hline
& RHIC & LHC   \\
\hline $J/\psi$ & $7.6\times 10^{-4}$ & $1.3\times 10^{-2}$   \\
$\psi(2S)$ & $3.5\times 10^{-4}$ & $5.8\times 10^{-3}$   \\
$\chi_{c1}(1P)$ & $3.0\times 10^{-4}$ & $5.1\times 10^{-3}$   \\
\hline \hline
\end{tabular}
\end{center}
\end{table}

We note in Table \ref{pT_yields} that the yield of the $\psi(2S)$
is smaller than that of the $J/\psi$ but is not so small as
expected. It has been found that the reason for the enhanced
production of the $\psi(2S)$ meson compared to the expectation in
the statistical hadronization model is attributable to the large
contribution at low transverse momenta from the wave function
distribution of the $\psi(2S)$ in a momentum space
\cite{Cho:2014xha}. We discuss in detail in Sec. IV what makes it
possible for the yield and transverse momentum distribution of the
$\psi(2S)$ to be as half large as those of the $J/\psi$ meson.

\section{Elliptic and triangular flow of charmonium states}

We now discuss harmonic flows of charmonium states, i.e., the
elliptic and triangular flow of the $J/\psi$, $\psi(2S)$, and
$\chi_{c1}(1P)$ meson in heavy ion collisions. The flow harmonics
are represented in general as $v_n$, the $n$-th coefficient in the
Fourier expansion of flows defined as \cite{Fries:2003kq,
Molnar:2003ff}

% The flow harmonics are known to be originated by the initial
% geometry of nucleus at the moment of heavy ion collisions; the
% pressure gradient generated in the particular shape caused by the
% anisotropic initial collisions creates various kinds of flows, and
% those flows are called harmonic flows. The first flow is called
% the direct flow, and the second flow is called the elliptic flow.
% In addition to the anisotropy of nucleon distributions in heavy
% ion collisions, event-by-event fluctuations in heavy ion
% collisions are also found to be important origins, especially
% giving rise to higher flow harmonics.

\begin{eqnarray}
&& v_n(p_T)=\langle\cos(n(\psi-\Psi_n))\rangle \nonumber \\
&& \quad\quad\quad=\frac{\int d\psi \cos(n(\psi-\Psi_n))
\frac{d^2N}{dp_T^2}}{\int d\psi \frac{d^2N}{dp_T^2}}, \label{vn}
\end{eqnarray}
where $p_T$ and $\psi$ are, respectively, the transverse momentum
and azimuthal angle of the charmonium state in the transverse
plane perpendicular to the collision axis. The $\Psi_n$ in Eq.
(\ref{vn}) is the event-plane angle, \cite{Petersen:2010cw,
Nahrgang:2014vza},

\begin{equation}
\Psi_n=\frac{1}{n}\tan^{-1}\Big(\frac{\langle p_T
\sin(n\psi)\rangle}{\langle p_T \cos(n\psi)\rangle}\Big),
\label{EPangle}
\end{equation}
defined in the region, $-\pi/n<\Psi_n<\pi/n$.
$\langle\cdots\rangle$ in Eq. (\ref{EPangle}) represents the
average over particles.

In order to evaluate the elliptic and triangular flow of
charmnoium states, Eq. (\ref{vn}) we adopt the transverse momentum
distributions of charmonium states, $d^2N/dp_T^2$, evaluated in
Eq. (\ref{CoalTransSim}). Then, the flow harmonics of charmonium
states in Eq. (\ref{vn}) are dependent on flow harmonics of charm
quarks via transverse momentum spectrum of charm quarks in Eq.
(\ref{CoalTransSim}),
\begin{equation}
\frac{d^2N_c}{dp_{cT}^2}=\frac{1}{2\pi p_{cT}}\frac{dN_c}
{dp_{cT}}\Big(1+\sum_{n=1}^{\infty}2 v_{nc}(p_{cT})
\cos(n(\phi_c-\Psi_n))\Big), \label{vn_charm}
\end{equation}
where $\phi_c$ is an azimuthal angle of a charm quark in the
transverse plane satisfying the momentum conservation condition in
the process of charmonium production, $\vec p_{cT}+\vec
p_{\bar{c}T}=\vec p_T$, in Eq. (\ref{CoalTransSim}).
$v_{nc}(p_{cT})$ in Eq. (\ref{vn_charm}) is the $n$th flow
harmonic of a charm quark.

\begin{figure}[!t]
\begin{center}
\includegraphics[width=0.480\textwidth]{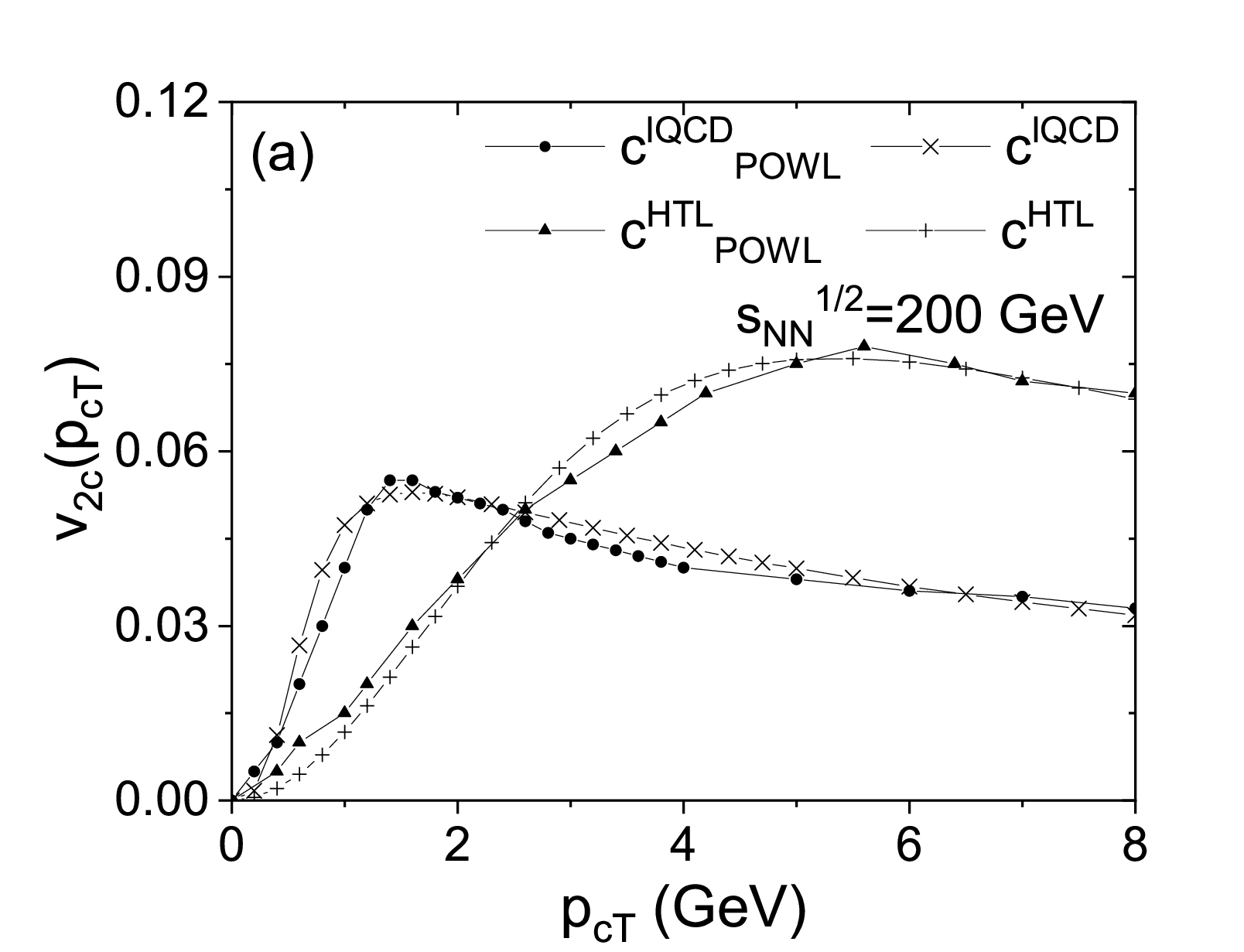}
\includegraphics[width=0.480\textwidth]{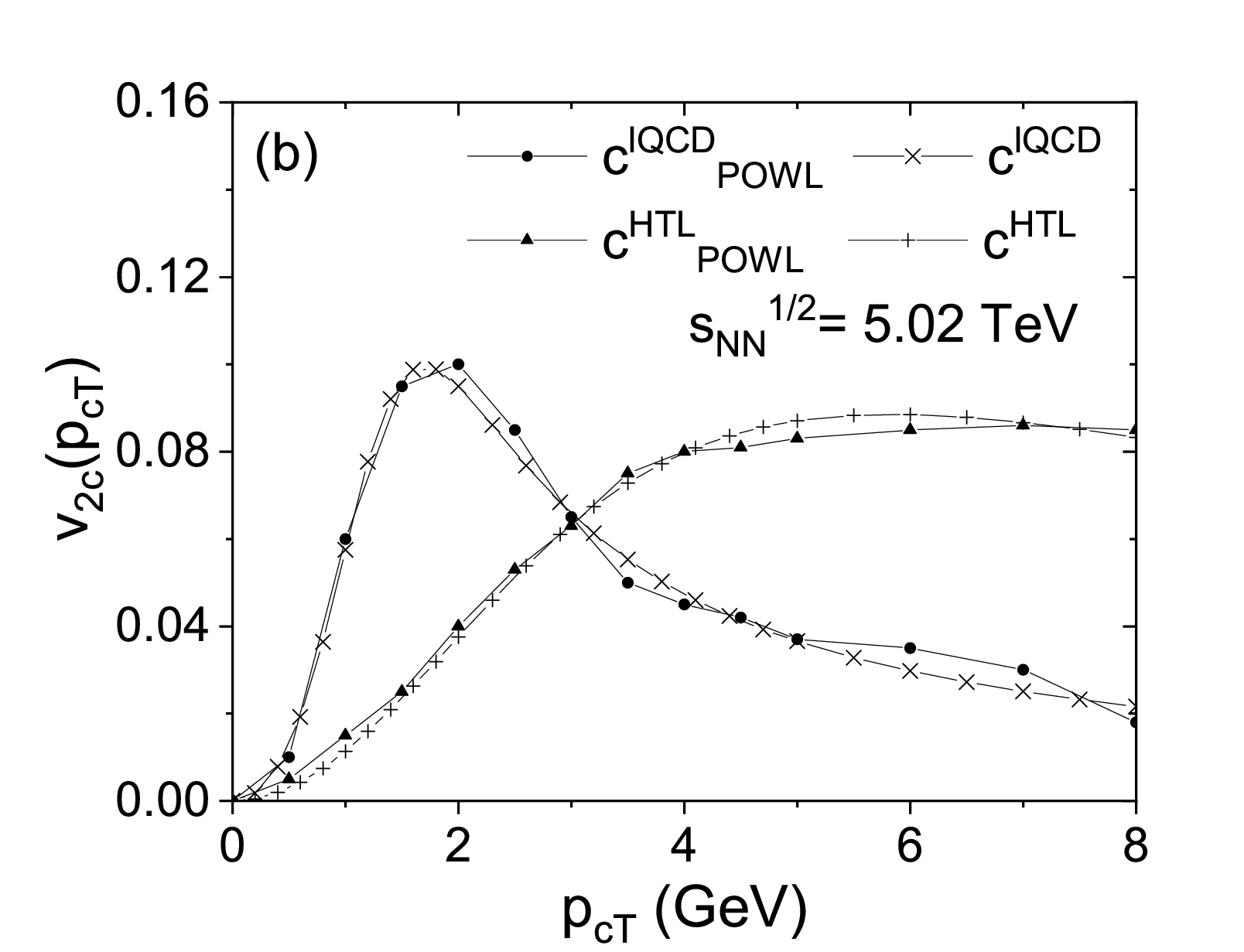}
\includegraphics[width=0.480\textwidth]{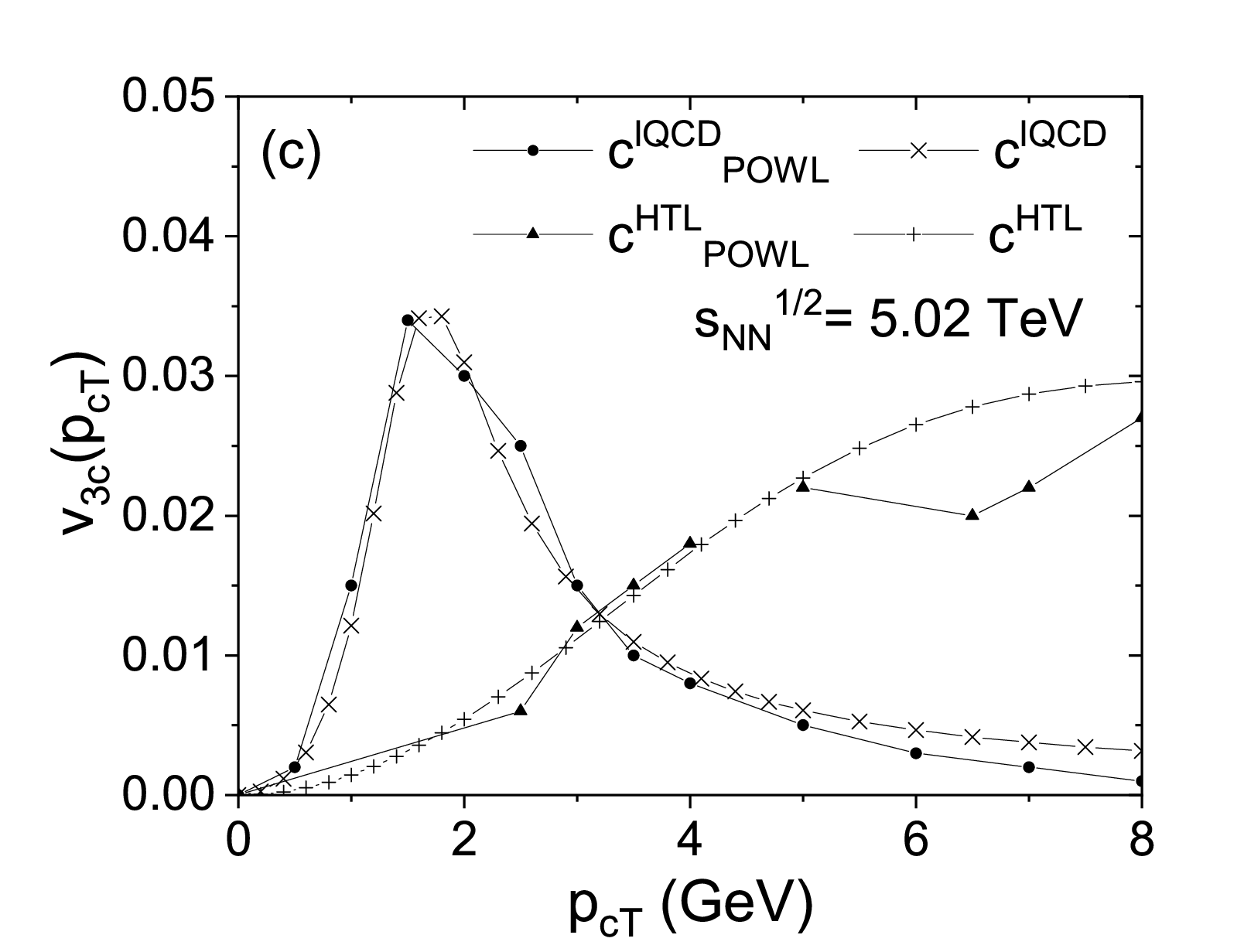}
\end{center}
\caption{Elliptic flow of charm quarks at RHIC,
$\sqrt{s_{NN}}$=200 GeV (a) and elliptic (b) and triangular (c)
flow of charm quarks at LHC, $\sqrt{s_{NN}}$=5.02 TeV from the
POWLANG transport setup based on HTL and Lattice QCD (lQCD)
transport coefficients \cite{Beraudo:2017gxw}, together with curve
fits for those flow harmonics in the Pad$\acute{\mathrm{e}}$
approximation.} \label{v23_charm}
\end{figure}

Therefore, it is also necessary to know the information on the
elliptic and triangular flow of a charm quark in order to evaluate
the elliptic and triangular flow of charmonium states. We adopt
here two kinds of flow harmonics of a charm quark obtained by the
POWLANG transport analysis \cite{Beraudo:2017gxw}, which describes
time evolutions of heavy quarks in heavy ion collisions through
the relativistic Langevin equation.

In the POWLANG transport study \cite{Beraudo:2017gxw} the time
evolution of charm-quark flow harmonics has been investigated, and
transverse momentum distributions of charm-quark elliptic and
triangular flow have been obtained for two occasions depending on
the transport coefficient, $\kappa$, governing the momentum
broadening during the propagation of heavy quarks in the medium;
one is the weak coupling transport coefficient based on Hard
Thermal Loop (HTL) re-summation analysis, and the other is the
transport coefficient based on non-perturbative Lattice Quntum
Chromodynamics (QCD) calculation.

% 1. momentum dependent $\kappa$ in a weak coupling with a hard
% thermal loop resummation of medium effects in the case of
% interactions mediated by the exchange of soft gluons 2. momentum
% independent but temperature dependent static lattice-QCD transport
% coefficient $\kappa$

In this work we apply charm quark flow harmonics obtained by both
transport coefficients in order to investigate flow harmonics of
charmonium states. We show in Fig. \ref{v23_charm} elliptic and
triangular flow of charm quarks at RHIC, $\sqrt{s_{NN}}$=200 GeV
and LHC $\sqrt{s_{NN}}$=5.02 TeV from the POWLANG analysis based
on HTL and Lattice QCD (lQCD) transport coefficients
\cite{Beraudo:2017gxw} together with our fit for those flow
harmonics in the Pad$\acute{\mathrm{e}}$ approximation,
% following the representation of the elliptic flow in terms of
% Knudsen number \cite{Nagle:2009ip},
\begin{equation}
v_{nc}(p_{cT})=\frac{a_3 p_{cT}^3+a_2 p_{cT}^2+a_1 p_{cT}}{b_4
p_{cT}^4+b_3 p_{cT}^3+b_2 p_{cT}^2+b_1 p_{cT}+1}, \label{Pade}
\end{equation}
with $a_i (i=1,2,3)$ and $b_j (j=1,2,3,4)$ being fit parameters.
In Eq. (\ref{Pade}), flow harmonics are parameterized to be zero
at both $p_{cT}=0$ and $p_{cT}\rightarrow\infty$ limits.

We observe in Fig. \ref{v23_charm} that flow harmonics based on
the Lattice QCD have peaks at low transverse momenta whereas those
based on the HTL have peaks at higher transverse momenta. It has
been found that the HTL transport coefficient gives rise to larger
flow harmonics at high transverse momentum region due to the
parton energy loss, which is different in longitudinal and
transverse directions, while the Lattice QCD transport coefficient
leads to the larger flow harmonics at low transverse momentum
region due to the larger momentum diffusion constant
\cite{Beraudo:2017gxw}. Since the coalescence production of
charmonium states are dominant at low and intermediate transverse
momentum regions, it is expected that flow harmonics of charm
quarks based on the Lattice QCD play more important roles in
understanding the flow harmonics of charmonium states.

It should be noted that the elliptic flow of charm quarks shown in
Fig. \ref{v23_charm}(a) has been evaluated in centrality 0-80 $\%$
at RHIC, $\sqrt{s_{NN}}$=200 GeV, and the elliptic and triangular
flow of charm quarks displayed in Fig. \ref{v23_charm}(b) and (c)
have been evaluated in 30-50 $\%$ centrality class at LHC,
$\sqrt{s_{NN}}$=5.02 TeV in the POWLANG analysis
\cite{Beraudo:2017gxw}. On the other hand, the transverse momentum
distribution of charm quarks adopted here, $d^2N_c^L/dp_{cT}^2$ in
Eq. (\ref{dNcdpT}) have been obtained at midrapidity in 0-10 $\%$
centrality at RHIC, $\sqrt{s_{NN}}$=200 GeV, and also at
midrapidity in 0-10 $\%$ centrality at LHC, $\sqrt{s_{NN}}$=2.76
TeV, desirable for the evaluation of both the transverse momentum
distribution, Eq. (\ref{CoalTransSim}) and flow harmonics of
charmonium states, Eq. (\ref{vn}) at midrapidity in 0-10 $\%$
centrality.

Therefore, we adjust oscillator frequencies in order to make the
transverse momentum distribution of charm quarks, Eq.
(\ref{dNcdpT}) applicable for describing the transverse momentum
distribution of charmonium states at both RHIC,
$\sqrt{s_{NN}}=200$ GeV at midrapidity in 0-80 $\%$ centrality and
LHC, $\sqrt{s_{NN}}=5.02$ TeV also at midrapidity in 20-40 $\%$
centrality.

\begin{figure}[!b]
\begin{center}
\includegraphics[width=0.51\textwidth]{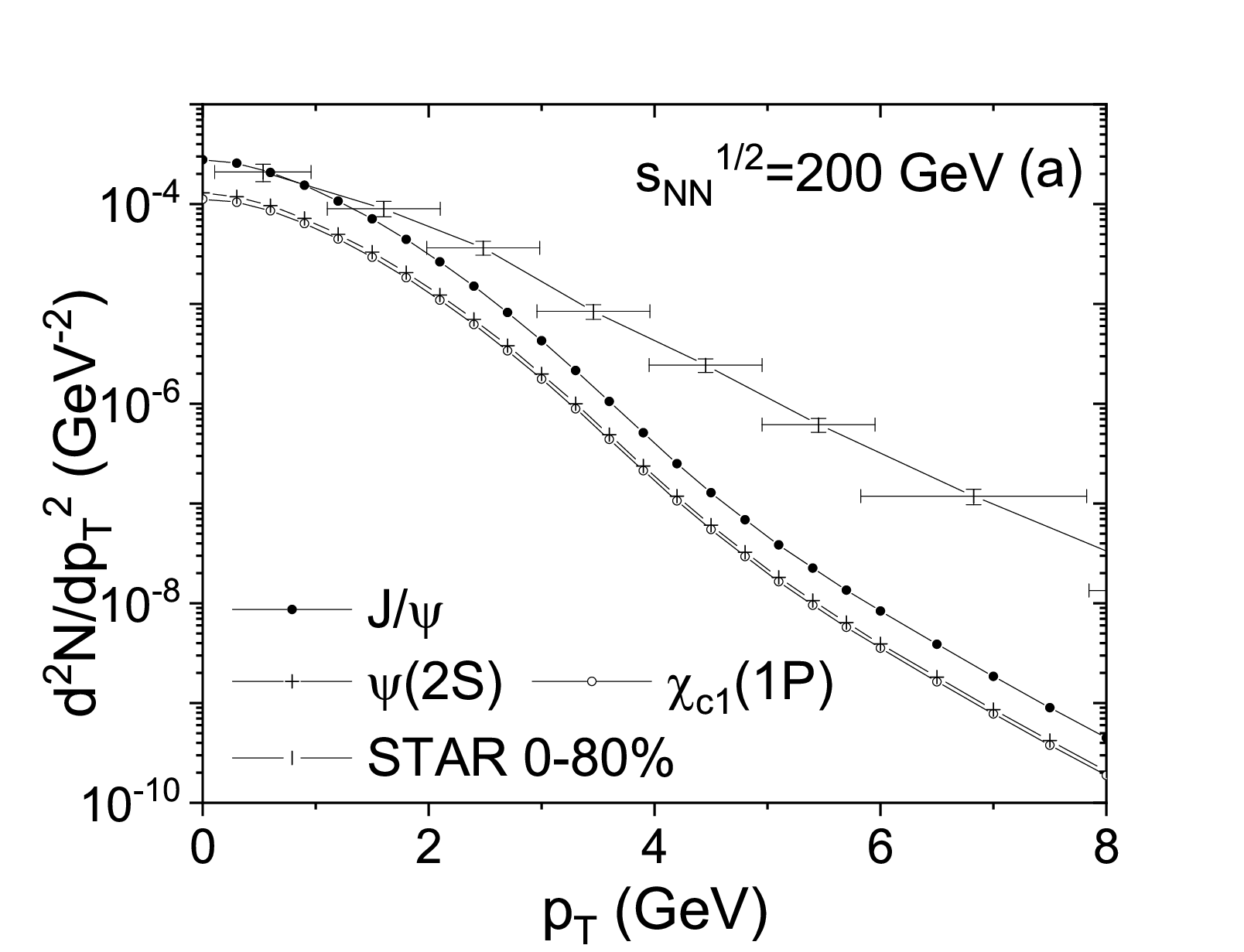}
\includegraphics[width=0.52\textwidth]{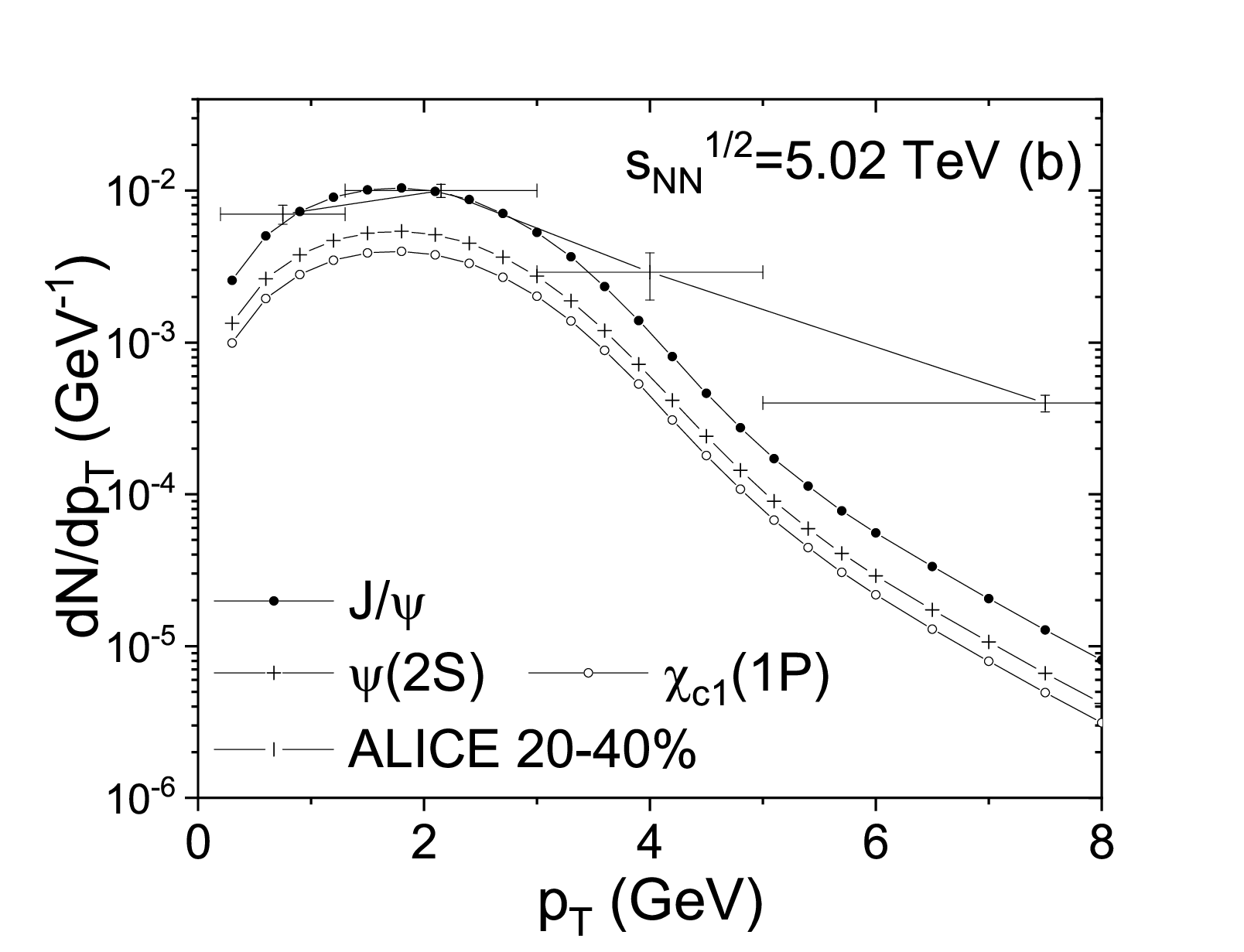}
\end{center}
\caption{Transverse momentum distributions, $d^2N/dp_T^2$ of
charmonium states at midrapidity in 0-80 $\%$ centrality at RHIC,
$\sqrt{s_{NN}}=200$ GeV (a) and those multiplied by $2\pi p_T$,
$dN/dp_T$ at midrapidity in 20-40 $\%$ centrality at LHC,
$\sqrt{s_{NN}}=5.02$ TeV (b) evaluated with new oscillator
frequencies, $\omega$=0.024 and 0.020 for RHIC and LHC,
respectively. We also show experimental measurements of the
$J/\psi$ transverse momentum distribution at RHIC,
$\sqrt{s_{NN}}=200$ GeV \cite{Adam:2019rbk} at midrapidity in 0-80
$\%$ centrality (a) and the transverse momentum distribution of
the $J/\psi$ multiplied by $2\pi p_T$, $dN/dp_T$ measured at LHC,
$\sqrt{s_{NN}}=5.02$ TeV, $|y|<$0.9 in centrality 20-40 $\%$
\cite{Acharya:2019lkh} (b). } \label{pT_charmonia_new}
\end{figure}

Comparing the transverse momentum distribution of the $J/\psi$
evaluated with Eqs. (\ref{CoalTransSim}), (\ref{WigIntdr}), and
(\ref{dNcdpT}) to the experimental measurement of the $J/\psi$
transverse transverse momentum distribution at low transverse
momentum region in Fig. \ref{pT_charmonia} we obtain new
oscillator frequencies, 0.024 GeV in centrality 0-80 $\%$ at RHIC
and 0.020 GeV in 20-40 $\%$ centrality at LHC.

\begin{widetext}

\begin{figure}[!h]
\begin{center}
\includegraphics[width=0.495\textwidth]{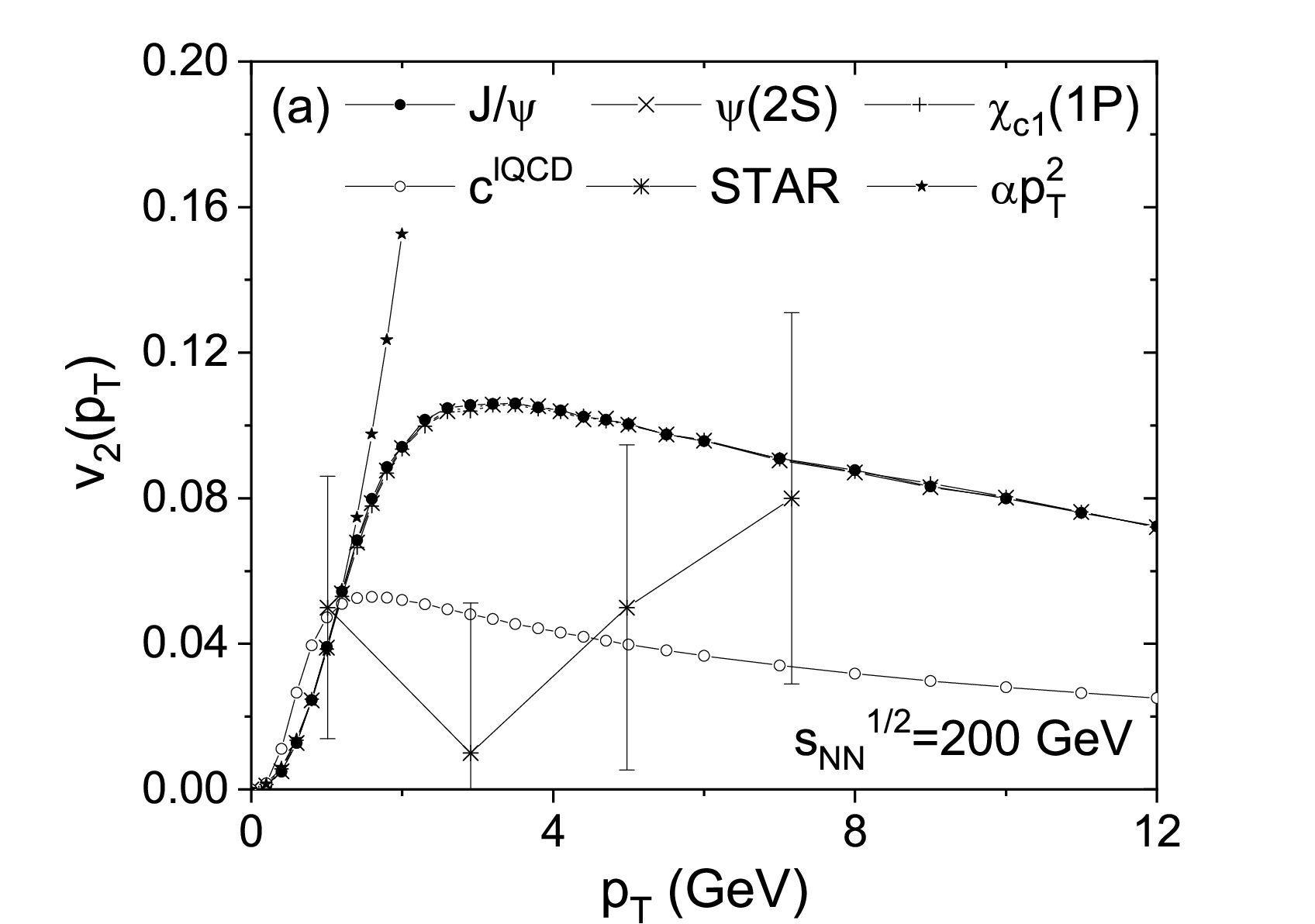}
\includegraphics[width=0.495\textwidth]{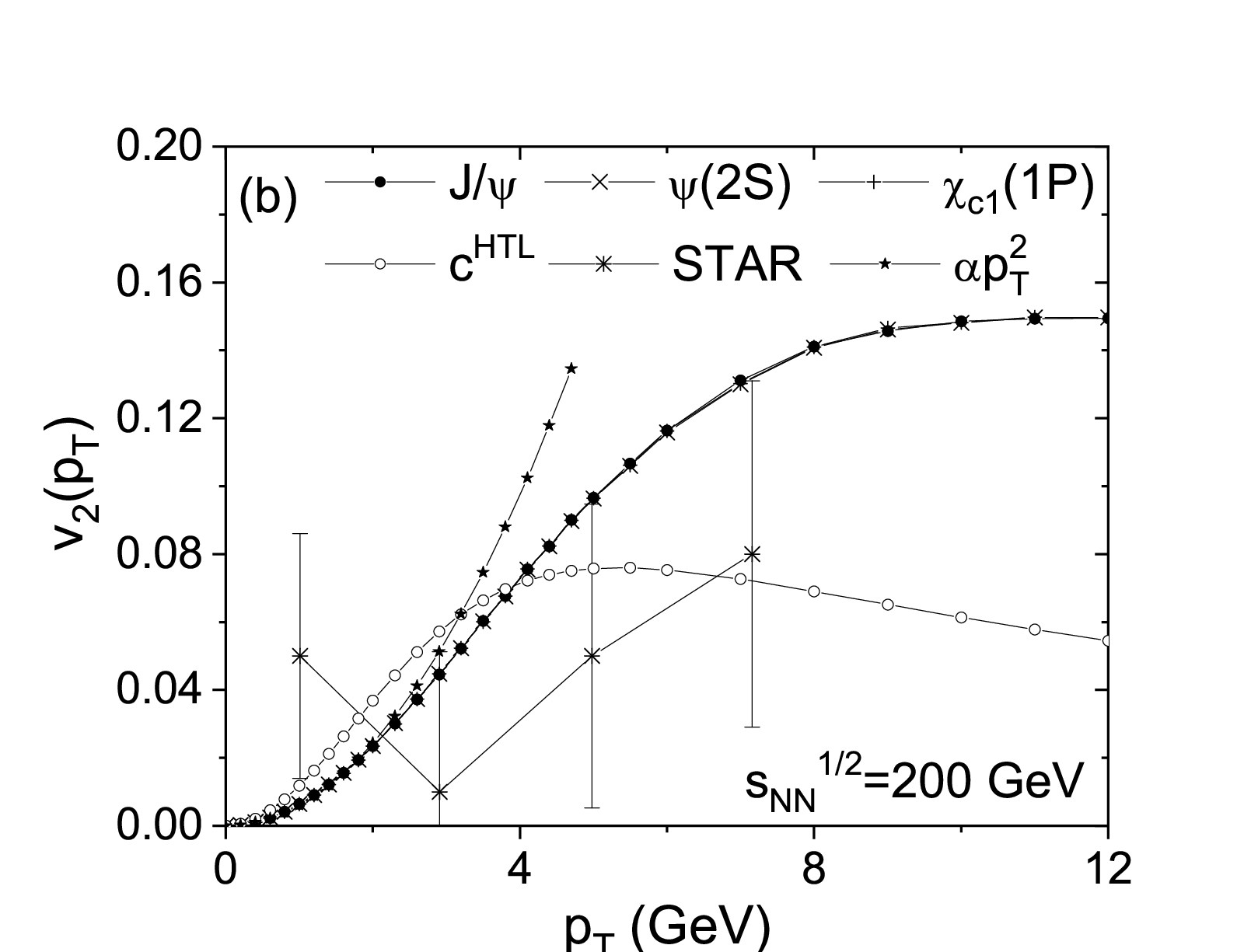}
\includegraphics[width=0.495\textwidth]{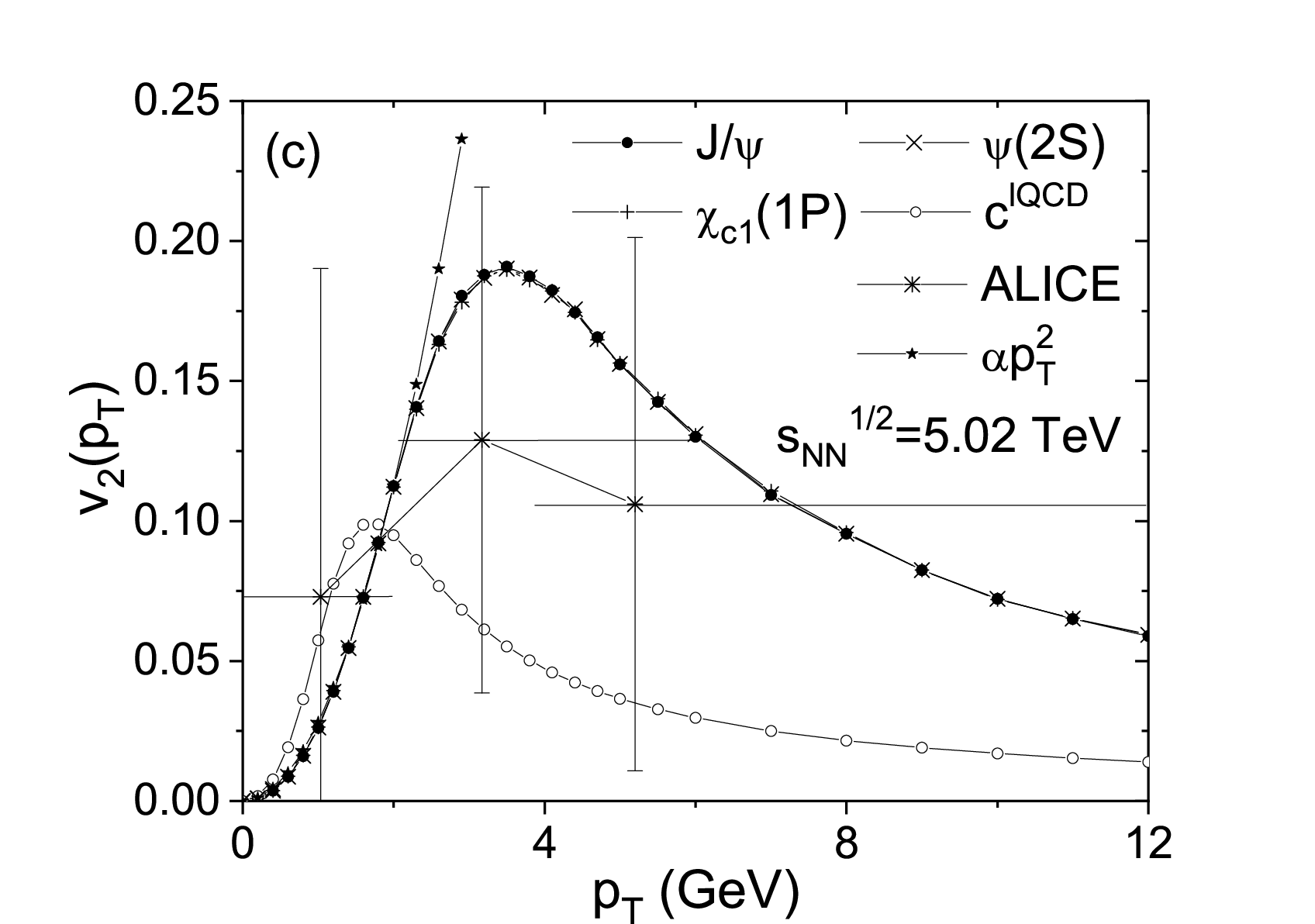}
\includegraphics[width=0.495\textwidth]{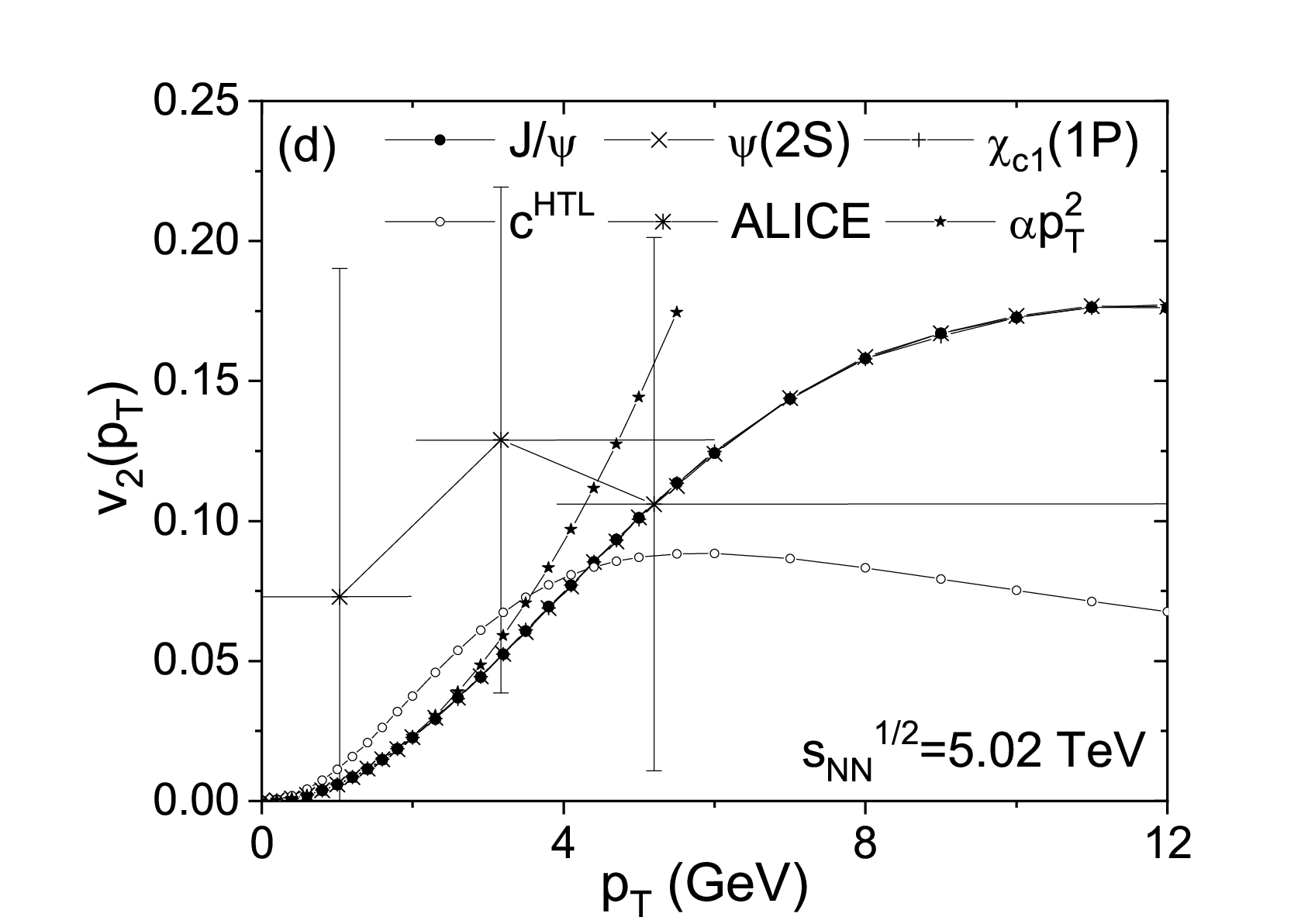}
\includegraphics[width=0.495\textwidth]{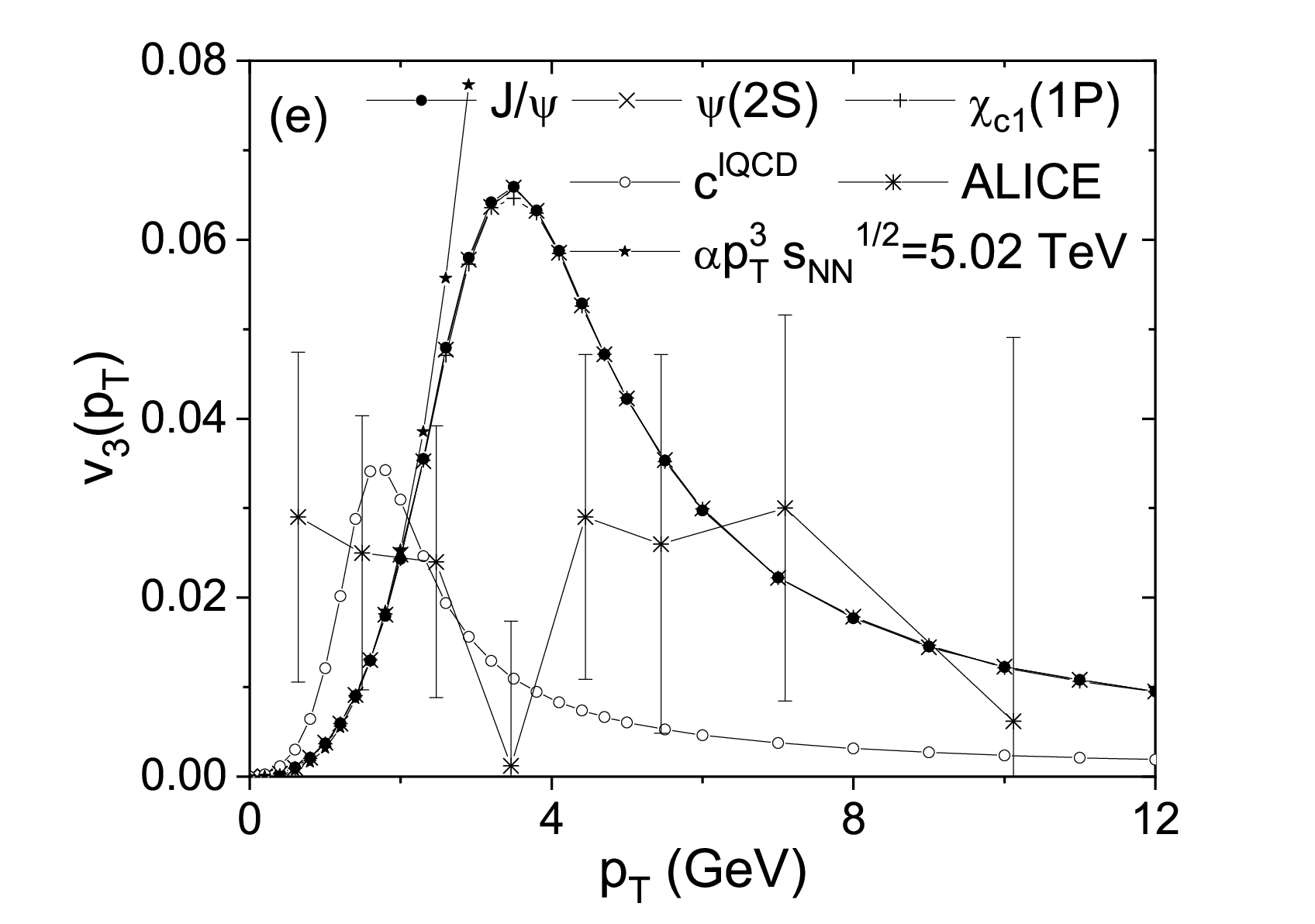}
\includegraphics[width=0.495\textwidth]{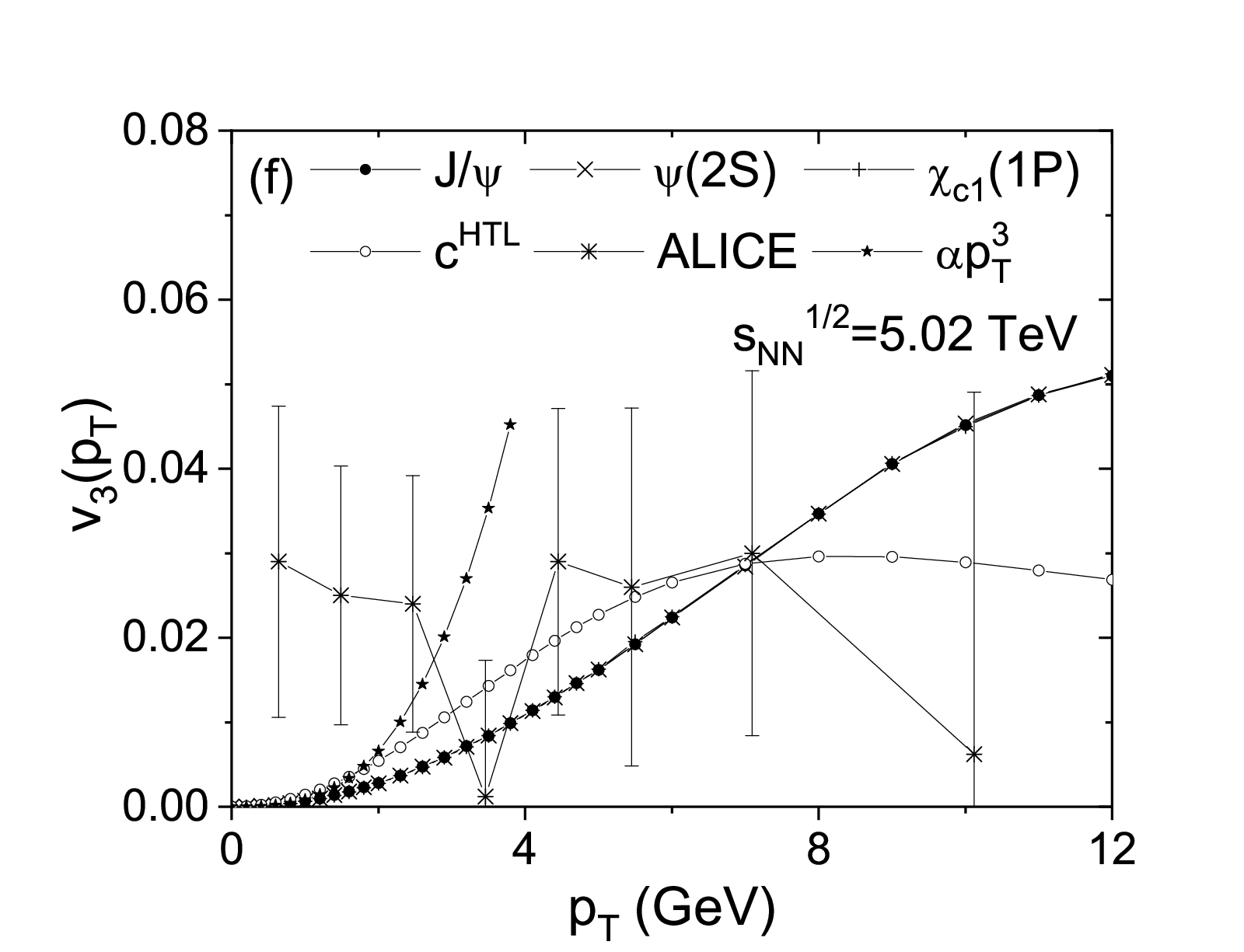}
\end{center}
\caption{Elliptic flows of charmonium states calculated from those
of charm quarks based on the lQCD (a) and HTL (b) transport
coefficients in the POWLANG transport setup \cite{Beraudo:2017gxw}
at RHIC, $\sqrt{s_{NN}}$=200 GeV, together with the measurement of
the $J/\psi$ elliptic flow at the same RHIC top energy in rapidity
$|y|<$1 in 0-80 $\%$ centrality \cite{Adamczyk:2012pw}. Also
elliptic flows of charmonium states calculated from those of charm
quarks based on the lQCD (c) and HTL (d) transport coefficients in
the POWLANG transport at LHC, $\sqrt{s_{NN}}$=5.02 TeV, together
with the measurement of the $J/\psi$ elliptic flow at the same LHC
energy in rapidity $|y|<$0.9 in 20-40 $\%$ centrality
\cite{Acharya:2017tgv}. Finally, triangular flows of charmonium
states calculated from those of charm quarks based on the lQCD (e)
and HTL (f) transport coefficients in the POWLANG transport at
LHC, $\sqrt{s_{NN}}$=5.02 TeV, together with the measurement of
the $J/\psi$ triangular flow in rapidity range of 2.5$<y<$4.0 in
30-50$\%$ centrality at the same LHC energy
\cite{Acharya:2020jil}. We also show corresponding flow harmonics
of bare charm quarks in each figure for comparison. }
\label{v23_charmonia}
\end{figure}

\end{widetext}

We show in Fig. \ref{pT_charmonia_new} transverse momentum
distributions, $d^2N/dp_T^2$ of charmonium states, the $J/\psi$,
$\psi(2S)$ and $\chi_c(1P)$ meson at midrapidity in 0-80 $\%$
centrality at RHIC, $\sqrt{s_{NN}}=200$ GeV and those multiplied
by $2\pi p_T$, $dN/dp_T$ at midrapidity in 20-40 $\%$ centrality
at LHC, $\sqrt{s_{NN}}=5.02$ TeV evaluated with new oscillator
frequencies, $\omega$=0.024 and 0.020 GeV for RHIC and LHC,
respectively. We have considered also the feed-down contribution
from heavier charmonium states, $\psi(2S)$ and $\chi_c(1P)$ mesons
in adjusting oscillator frequencies for the transverse momentum
distribution of the $J/\psi$. We also show in Fig.
\ref{pT_charmonia_new} experimental measurements of the $J/\psi$
transverse momentum distribution at RHIC, $\sqrt{s_{NN}}=200$ GeV
at midrapidity in 0-80 $\%$ centrality \cite{Adam:2019rbk} and the
transverse momentum distribution of the $J/\psi$ multiplied by
$2\pi p_T$, $dN/dp_T$ measured at LHC, $\sqrt{s_{NN}}=5.02$ TeV,
$|y|<$0.9 in 20-40 $\%$ centrality \cite{Acharya:2019lkh}.

As shown in Fig. \ref{pT_charmonia_new} the reasonable agreement
has been made between the experimental measurements and the
evaluation of the $J/\psi$ transverse momentum distribution up to
about $p_T=2$ GeV. We note that at high transverse momentum region
the transverse momentum distribution of the $J/\psi$ does not
agree well with the experimental measurements, owing to the
non-negligible contributions to the $J/\psi$ production at higher
transverse momenta from the decay of heavier bottomed hadrons,
which are not taken into account in our analysis.

The new oscillator frequencies, $\omega$=0.024 and 0.020 GeV are
smaller than those obtained previously, $\omega$=0.078 and 0.076
GeV at RHIC and LHC, respectively, thereby giving rise to more
production of charmonium states; charmonium states are expected to
be more abundant as the centrality range is increased from 0-10 to
0-80 $\%$ at RHIC, and also as the collision energy is increased
from $\sqrt{s_{NN}}=2.76$ to 5.02 TeV at LHC.

% Considering that flow harmonics of charm quarks in centrality
% 30-50 $\%$ are similar to those in centrality 20-40 $\%$ at LHC,
% $\sqrt{s_{NN}}$=5.02 TeV, we combine in Eq. (\ref{vn_averaged})
% the transverse momentum distribution of charm quarks in 20-40 $\%$
% centrality shown in Fig. \ref{pT_charmonia_new} with flow
% harmonics of charm quarks in centrality 30-50 $\%$ shown in Fig.
% \ref{v23_charm}, and calculate the elliptic and triangular flow of
% charmonium states in 20-40 $\%$ centrality at LHC
% $\sqrt{s_{NN}}$=5.02 TeV. As the transverse momentum distribution
% and flow harmonics of charm quarks at RHIC, $\sqrt{s_{NN}}$=200
% GeV are all prepared in the same centrality 0-80 $\%$, we
% calculate consistently the elliptic flow of charmonium states in
% centrality 0-80 $\%$.

In addition to flow harmonics of charm quarks, we also need to
know the information on the event plane, Eq. (\ref{EPangle}) in
order to evaluate the flow harmonics of charmonium states. We
clearly see that flow harmonics of charmonium states, Eq.
(\ref{vn}) are actually sensitive to the change of the event
plane, Eq. (\ref{EPangle}). Experimentally, the event plane is
determined from all hadrons observed in the same event based on
the relation between the impact parameter and transverse plane. On
the other hand, it has been argued that the event plane method may
give rise to discrepancies from the true values, and thus the way
of using the flow vector $\vec Q$ instead of evaluating the event
plane has been suggested \cite{Luzum:2012da}.

Therefore, we consider here event plane blind flow harmonics, or
event-averaged flow harmonics of charmoium states; since the flow
harmonics of charm quarks shown in Fig. \ref{v23_charm} evaluated
in the POWLANG transport study have also been averaged over events
\cite{Beraudo:2017gxw}, it is also natural to consider here
elliptic and triangular flow of charmonium states formed from
those of charm quarks on the same condition that events are
averaged,

\begin{equation}
v_n(p_T)=\frac{\frac{n}{2\pi}\int_{-\frac{\pi}{n}}^{
\frac{\pi}{n}} \int d\psi \cos(n(\psi-\Psi_n))
\frac{d^2N}{dp_T^2}d\Psi_n}{\frac{n}{2\pi}\int_{-\frac{\pi}{n}}^{
\frac{\pi}{n}} \int d\psi \frac{d^2N}{dp_T^2}d\Psi_n}.
\label{vn_averaged}
\end{equation}

Though the event-averaged triangular flow of charm quarks is
adopted in evaluating that of charmonium states, the effects of
event-by-event fluctuation on the triangular flow of charmonium
states are taken into account here; triangular flow of charm
quarks was obtained from collisions between nucleons in numerous
different initial positions reflecting event-by-event fluctuation
in the POWLANG transport study \cite{Beraudo:2017gxw}. In that
sense we adopt the event-by-event fluctuation effects on the
triangular flow of charmonium states entirely from the charm quark
triangular flow bearing the effects caused by different initial
configurations of colliding nucleons.

%% ---------------------------------------------------------
%
% 1. weak dependence of triangular flow on centrality & roughly
% equal to elliptic flow \cite{Petersen:2010cw}

With these in mind we calculate the elliptic and triangular flow
of charmonium states at RHIC and LHC using both the transverse
momentum distribution of charm quarks shown in Fig.
\ref{pT_charmonia_new} and flow harmonics of charm quarks shown in
Fig. \ref{v23_charm}. As the transverse momentum distribution and
flow harmonics of charm quarks at RHIC, $\sqrt{s_{NN}}$=200 GeV
are all prepared in the same centrality 0-80 $\%$, we calculate
consistently the elliptic flow of charmonium states in centrality
0-80 $\%$. However, for the flow harmonics of charmonium states at
LHC, $\sqrt{s_{NN}}$=5.02 TeV, we assume that flow harmonics of
charm quarks in centrality 30-50 $\%$ are similar to those in
centrality 20-40 $\%$ at LHC, $\sqrt{s_{NN}}$=5.02 TeV, and
calculate the elliptic and triangular flow of charmonium states in
20-40 $\%$ centrality at LHC $\sqrt{s_{NN}}$=5.02 TeV by combining
the transverse momentum distribution of charm quarks in 20-40 $\%$
centrality shown in Fig. \ref{pT_charmonia_new} with flow
harmonics of charm quarks in centrality 30-50 $\%$ shown in Fig.
\ref{v23_charm}.

We show in Fig. \ref{v23_charmonia} (a) and (b) the elliptic flows
of charmonium states calculated from those of charm quarks based
on, respectively, the lQCD and HTL transport coefficients in the
POWLANG transport setup \cite{Beraudo:2017gxw} at RHIC,
$\sqrt{s_{NN}}$=200 GeV, together with the measurement of the
$J/\psi$ elliptic flow at the same RHIC top energy in rapidity
$|y|<$1 in 0-80 $\%$ centrality \cite{Adamczyk:2012pw}. Also the
elliptic flows of charmonium states calculated from those of charm
quarks based on, respectively, the lQCD and HTL transport
coefficients in the POWLANG transport at LHC, $\sqrt{s_{NN}}$=5.02
TeV are shown in Fig. \ref{v23_charmonia}(c) and (d), together
with the measurement of the $J/\psi$ elliptic flow at the same LHC
energy in rapidity $|y|<$0.9 in 20-40 $\%$ centrality
\cite{Acharya:2017tgv}. Finally, we show in Fig.
\ref{v23_charmonia}(e) and (f) the triangular flows of charmonium
states calculated from those of charm quarks based on,
respectively, the lQCD and HTL transport coefficients in the
POWLANG transport at LHC, $\sqrt{s_{NN}}$=5.02 TeV, together with
the measurement of the $J/\psi$ triangular flow in the rapidity
range of 2.5$<y<$4.0 in 30-50$\%$ centrality at the same LHC
energy \cite{Acharya:2020jil}. We also show corresponding flow
harmonics of bare charm quarks in each figure for comparison.

\begin{widetext}

\begin{figure}[!h]
\begin{center}
\includegraphics[width=0.495\textwidth]{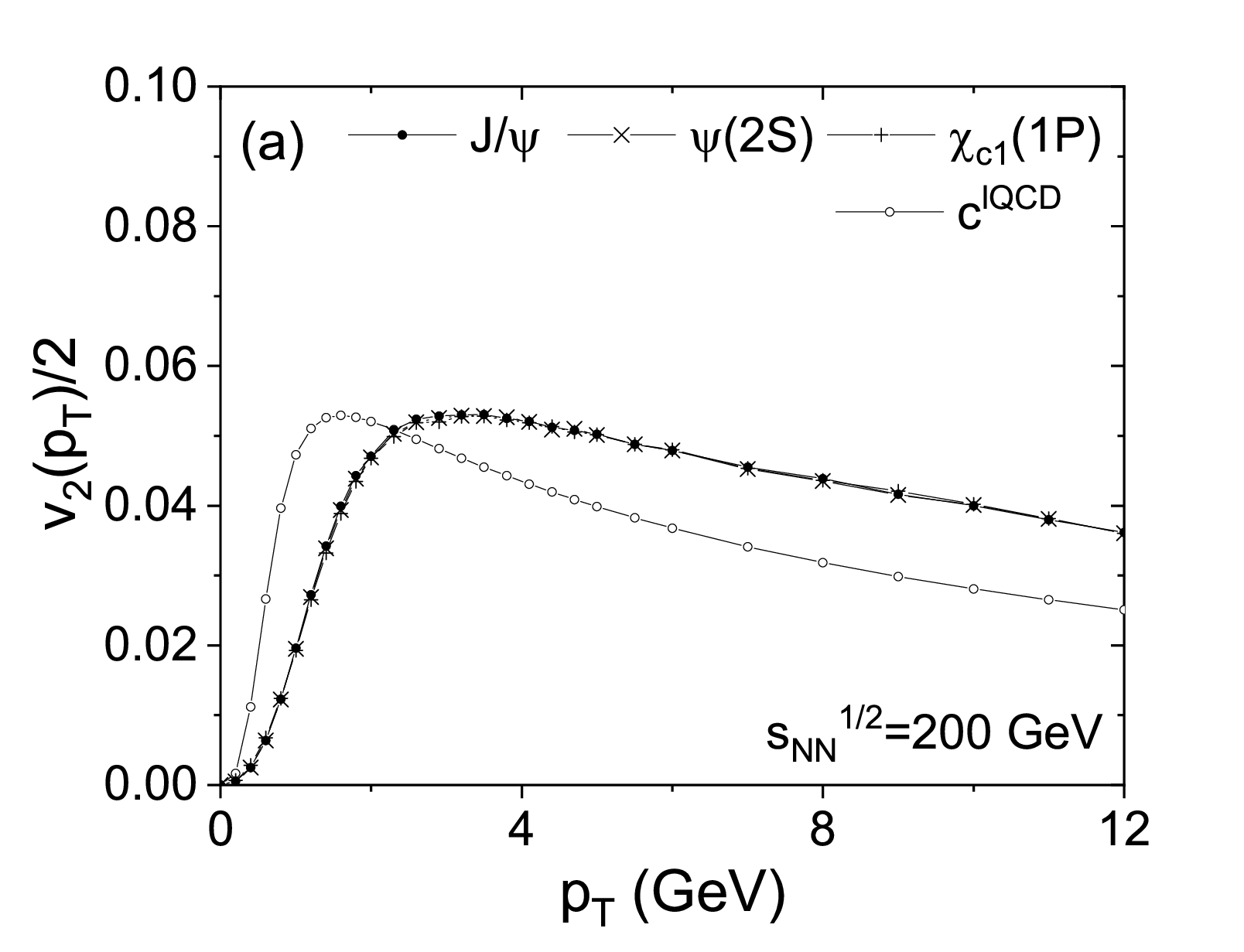}
\includegraphics[width=0.495\textwidth]{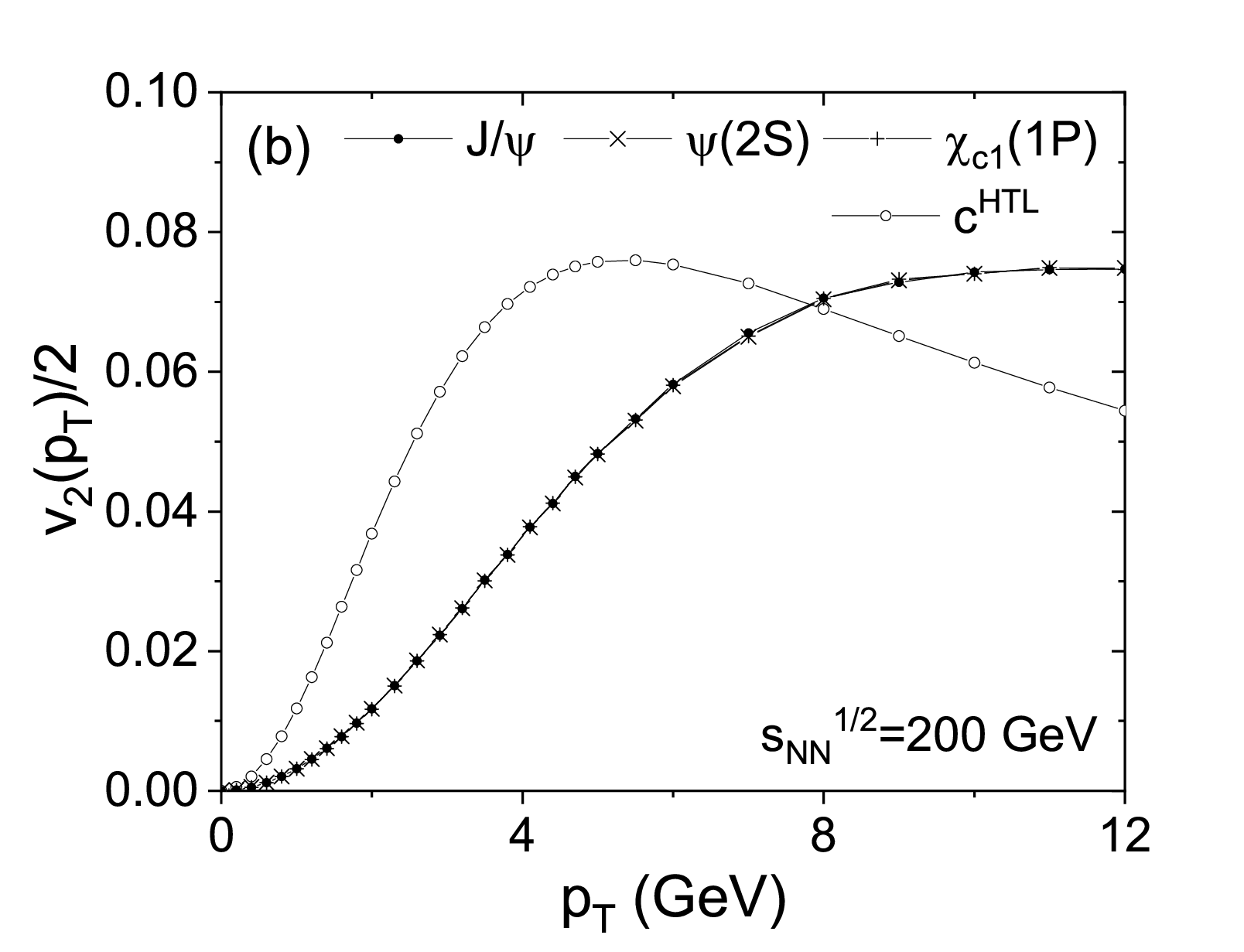}
\includegraphics[width=0.495\textwidth]{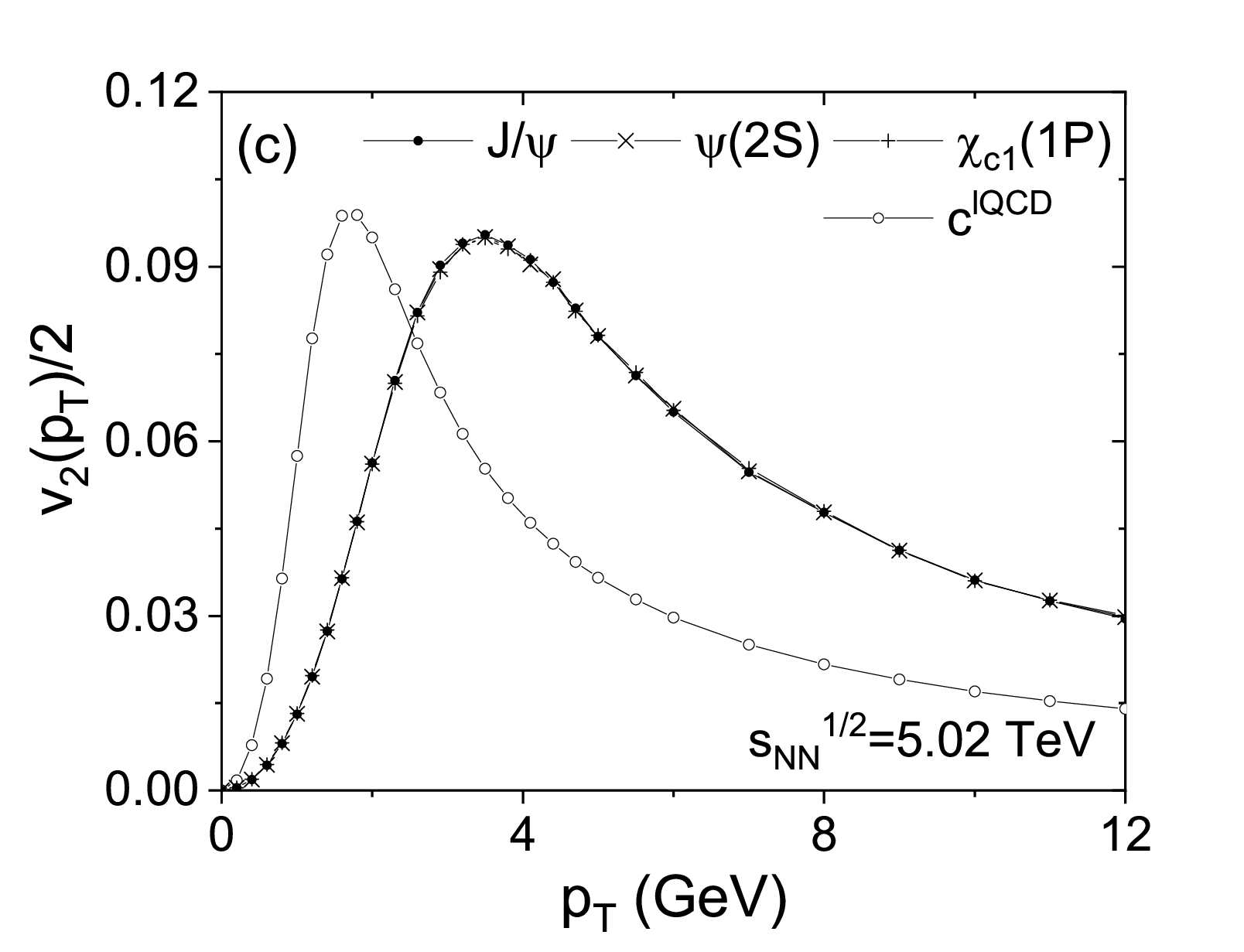}
\includegraphics[width=0.495\textwidth]{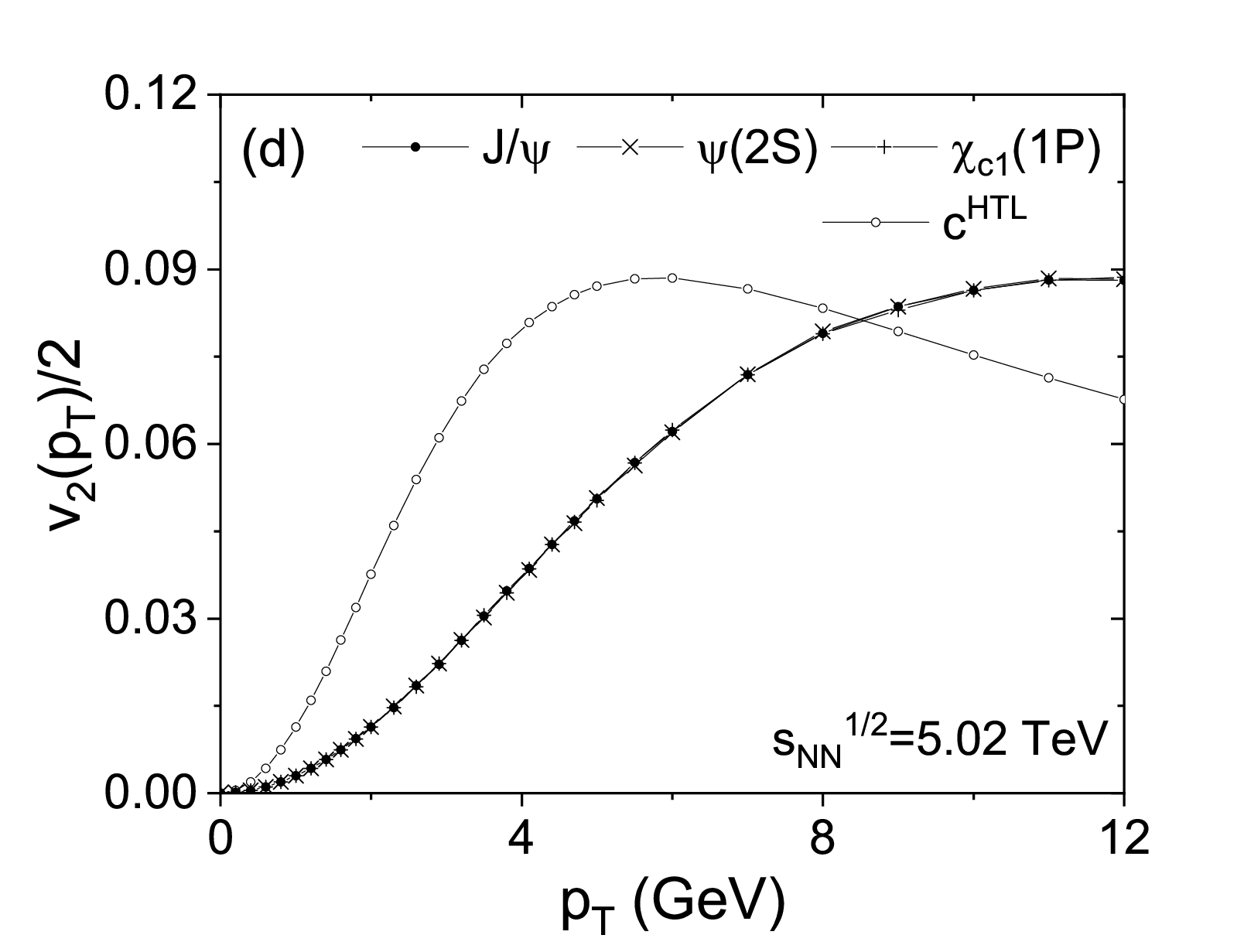}
\includegraphics[width=0.495\textwidth]{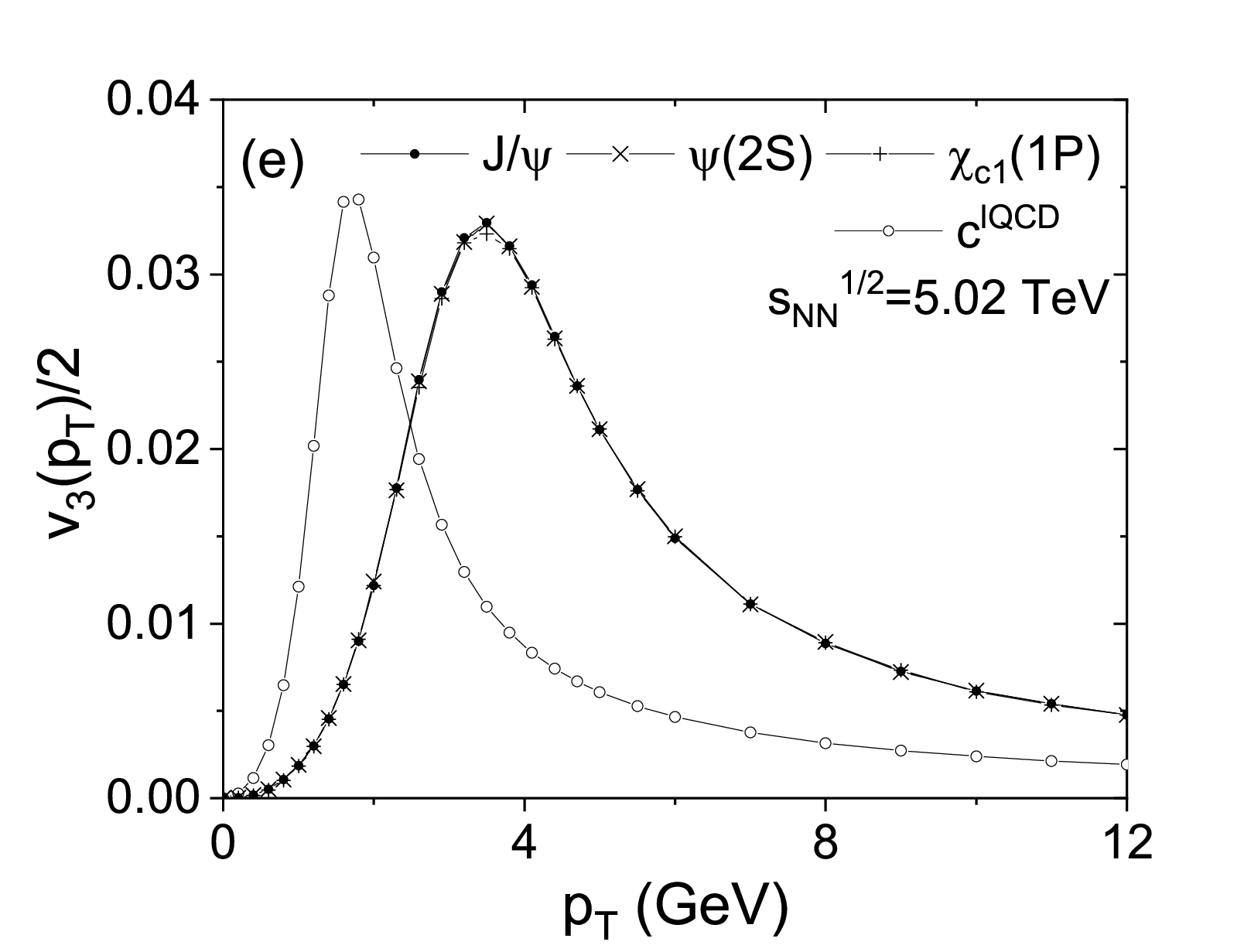}
\includegraphics[width=0.495\textwidth]{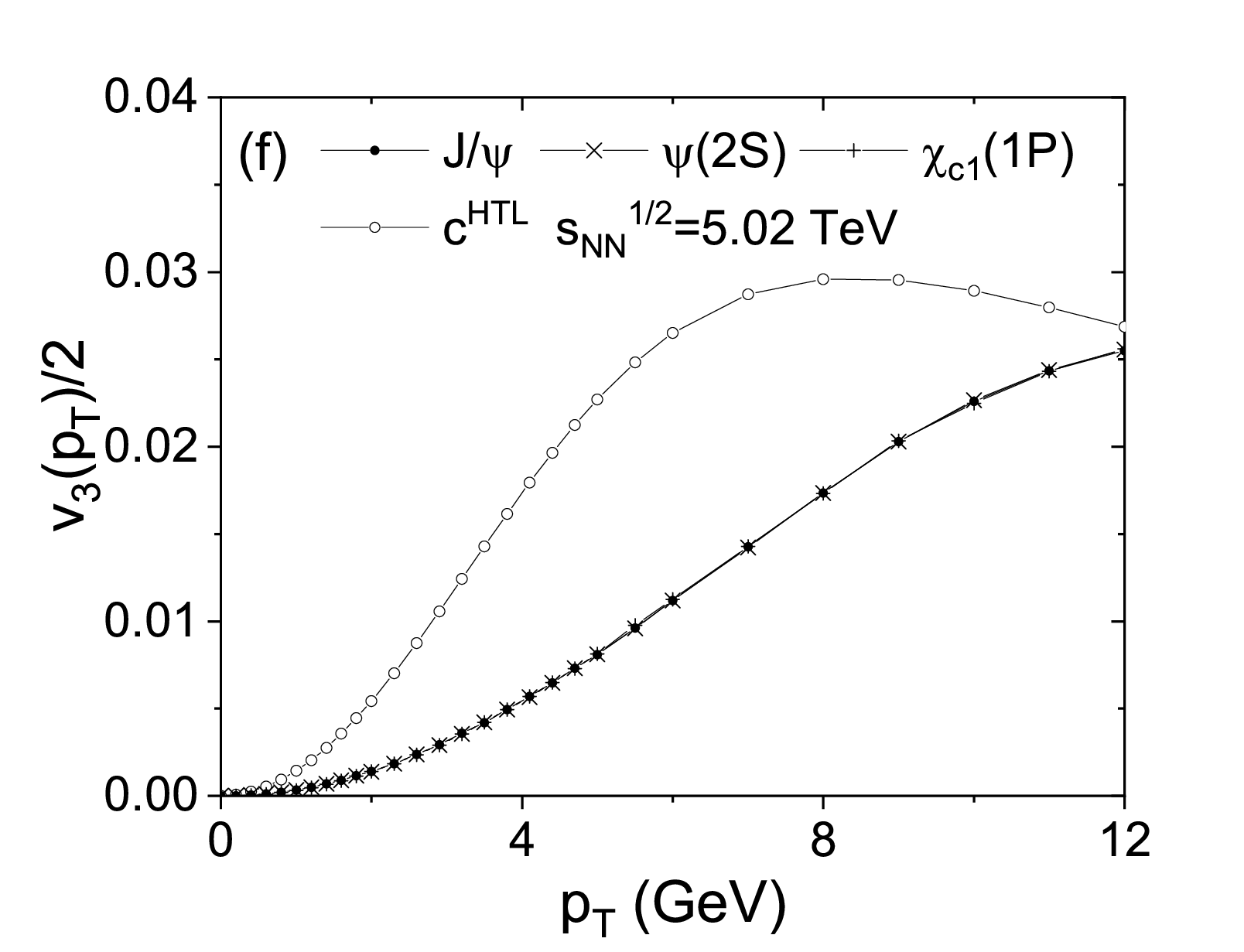}
\end{center}
\caption{We present elliptic flows of charmonium states shown in
Fig. \ref{v23_charmonia} divided by 2, the number of constituents,
calculated from those of charm quarks based on the lQCD (a) and
HTL (b) transport coefficients in POWLANG transport setup
\cite{Beraudo:2017gxw} at RHIC, $\sqrt{s_{NN}}$=200 GeV. Also
elliptic flows of charmonium states divided by 2, calculated from
those of charm quarks based on the lQCD (c) and HTL (d) transport
coefficients in POWLANG transport setup \cite{Beraudo:2017gxw} at
LHC, $\sqrt{s_{NN}}$=5.02 TeV are shown. Finally, triangular flows
of charmonium states divided by 2, calculated from those of charm
quarks based on the lQCD (e) and HTL (f) transport coefficients in
POWLANG transport setup \cite{Beraudo:2017gxw} at LHC,
$\sqrt{s_{NN}}$=5.02 TeV are presented. We also show corresponding
flow harmonics of bare charm quarks in each figure for comparison.
 } \label{v23_charmonia_QNS}
\end{figure}

\end{widetext}

As shown in Fig. \ref{v23_charmonia} the elliptic flow at RHIC is
smaller than that at LHC for the POWLANG analysis with the lQCD
transport coefficients while the elliptic flow at RHIC is similar
to that at LHC for the POWLANG analysis with the HTL transport
coefficients. When the medium is strongly coupled, the larger
collision energy at LHC than at RHIC seems more efficient in
making charmonium states to flow in the medium. On the other hand,
the collision energy does not matter too much for the flow of
chamonium states in a weakly coupled medium.

We also see in Fig. \ref{v23_charmonia} that the flow harmonics of
charmonium states are dominant at low and intermediate transverse
momentum regions when the larger diffusion has been taken into
account at low transverse momenta, or when the flow harmonics of
charm quarks in the non-perturbative lQCD transport coefficient in
the POWLANG transport analysis are adopted, Fig.
\ref{v23_charmonia} (a), (c) and (e), whereas the flow harmonics
of charmonium states based on those of charm quarks prevailing at
higher transverse momentum region due to the weak coupling
transport coefficient based on HTL, are found to be still being
generated at low and intermediate transverse momentum regions,
Fig. \ref{v23_charmonia} (b), (d) and (f).

We notice in Fig. \ref{v23_charmonia} that flow harmonics of
charmonium states, the $J/\psi$, $\psi(2S)$ and $\chi_c(1P)$ meson
are almost same at both RHIC and LHC, irrespective of two
occasions, flow harmonics of constituent charm quarks considered
in two different interaction with the medium, the lQCD and HTL
transport coefficients in the POWLANG transport analysis.

It is reasonable to expect that the elliptic and triangular flow
of different charmonium states are different owing to the
different transverse momentum distribution, Eq. (\ref{vn})
originated from the different internal structures of charmonium
states, or the different wave function distribution in momentum
space, Eq. (\ref{WigIntdr}). However, even though the transverse
momentum distributions are different for each charmonium state as
shown in Fig. \ref{Jpsitopsi2Sratio}, the elliptic and triangular
flow of different charmonium states are found to be almost same in
the entire transverse momentum range.

We find that the reason why we obtain the almost same flow
harmonics for charmonium states, the $J/\psi$, $\psi(2S)$ and
$\chi_c(1P)$ meson is due to the characteristics of the $v_n$
itself. Since the $v_n$ in Eq. (\ref{vn}) is the average of cosine
functions in different orders $n$ over transverse momentum with
the weight of the transverse momentum distribution of charmonium
states, the contribution of transverse momentum distributions
originated from different internal structure of charemonium states
are averaged out for all charmonium states.

We clearly see the dependence of both the numerator and
denominator part of $v_n$ in Eq. (\ref{vn}) on the internal
structure, or the wave function distribution in momentum space of
each charmonium state. Nevertheless, the flow harmonics of all
charmonium states become almost same when those are evaluated by
the definition of $v_n$ in Eq. (\ref{vn}). The similar amounts of
contribution to both the numerator and denominator of the $v_n$
from different transverse momentum distributions of charmonium
states are cancelled out each other, resulting in almost the same
flow harmonics for all charmonium states. We discuss in detail on
this issue in Sec. IV.

Since it has been found that the elliptic flow at low transverse
momentum region is close to the square function of transverse
momentum, $p_T^2$, and similarly the triangular flow also at low
transverse momentum region is close to the cube function of $p_T$,
$p_T^3$ \cite{Dinh:1999mn} we also show in Fig.
\ref{v23_charmonia} the square and cube function fits as functions
of the $p_T$, $\alpha p_T^2$ and $\alpha p_T^3$ with parameters,
$\alpha$ summarized in Table \ref{alphatn}, in order to
investigate the behavior of the elliptic and triangular flow at
low transverse momentum region.

\begin{table}[!t]
\caption{Fitting functions of the $J/\psi$ meson $v_n$ at low
transverse momentum region, $\alpha p_T^n$ shown in Fig.
\ref{v23_charmonia} at RHIC $\sqrt{s_{NN}}=200$ GeV and LHC
$\sqrt{s_{NN}}=5.02$ TeV. } \label{alphatn}
\begin{center}
\begin{tabular}{c|c|c|c|c|c|c}
\hline \hline & \multicolumn{2}{|c}{RHIC} &
\multicolumn{4}{|c}{LHC} \\
\cline{2-7} $\alpha p_T^n$ & \multicolumn{2}{|c}{$n$=2} &
\multicolumn{2}{|c}{$n$=2} &
\multicolumn{2}{|c}{$n$=3} \\
\cline{2-7}
& lQCD & HTL & lQCD & HTL & lQCD & HTL \\
\hline
$\alpha$ & 0.038 & 0.0061 & 0.028 & 0.0058 & 0.0032 & 0.00082 \\
\hline \hline
\end{tabular}
\end{center}
\end{table}

As shown in Fig. \ref{v23_charmonia} the elliptic (triangular)
flow obtained from that of charm quarks in the strongly coupled
medium, or in the lQCD transport coefficients from the POWLANG
analysis behaves like the $p_T^2$ ($p_T^3$) up to about $p_T=2$
GeV while that obtained from charm quarks in the weakly coupled
medium, or in the HTL transport coefficients from the POWLANG
shows the similar behavior of the $p_T^2$ ($p_T^3$) up to about
$p_T=3$ ($p_T=1$) GeV at low transverse momentum region.
Considering that the $v_n$ usually increases with $p_T^n$ up to
about $p_T=M$ for a hadron with its mass $M$ \cite{Dinh:1999mn},
we see that this tendency applies more suitably for charmonium
states in a weakly coupled medium, in the HTL from the POWLANG.

When we remind the relation between the flow coefficients and the
interaction of constituent quarks with other quarks in QGP, we
expect that the elliptic flow of charmonium states increases
rather steeply at $p_T$ smaller than their corresponding masses
$M$ when the larger diffusion has been adopted from the
non-perturbative lQCD transport coefficient in the POWLANG
transport analysis. Similarly, the elliptic flow of charmonium
states based on those of charm quarks prevailing at higher
transverse momentum region due to the weak coupling transport
coefficients is found to increase rather gradually with $p_T^2$ up
to about their corresponding masses $M$. This tendency is opposite
for the triangular flow since the cube function can be fit more
easily when the triangular flow increase more steeply at low
transverse momentum region as shown in Fig. \ref{v23_charmonia}
(e) and (f). We discuss further on this issue in Sec. IV.

We also present in Fig. \ref{v23_charmonia_QNS} elliptic flows of
charmonium states shown in Fig. \ref{v23_charmonia} divided by 2,
the number of constituents, calculated from those of charm quarks
based on the lQCD (a) and HTL (b) transport coefficients in
POWLANG transport setup \cite{Beraudo:2017gxw} at RHIC,
$\sqrt{s_{NN}}$=200 GeV. Also elliptic flows of charmonium states
divided by 2, calculated from those of charm quarks based on the
lQCD (c) and HTL (d) transport coefficients in POWLANG transport
setup \cite{Beraudo:2017gxw} at LHC, $\sqrt{s_{NN}}$=5.02 TeV are
shown. Finally, triangular flows of charmonium states divided by
2, calculated from those of charm quarks based on the lQCD (e) and
HTL (f) transport coefficients in POWLANG transport setup
\cite{Beraudo:2017gxw} at LHC, $\sqrt{s_{NN}}$=5.02 TeV are
presented. We also show in each figure corresponding flow
harmonics of bare charm quarks presented in Fig. \ref{v23_charm}
for comparison.

Reminding the well-known relation between the elliptic flow of
mesons and that of constituent quarks,
$v_{2,M}(p_T)\approx2v_{2,q}(p_T/2)$ \cite{Molnar:2003ff,
Fries:2003kq}, we find that the elliptic flow of charmonium states
satisfy the similar relation to that of charm quarks,
$v_{2,c\bar{c}}(p_T)\approx2v_{2,c}(p_T/2)$; the transverse
momentum at the peak in the elliptic flow of charmonium states is
almost twice that at the peak position of the elliptic flow of
charm quarks as shown in Fig. \ref{v23_charmonia}(a), (b), (c),
and (d). Similarly, it has been found that the above relation also
holds for the triangular flow of charmonium states,
$v_{3,c\bar{c}}(p_T)\approx2v_{3,c}(p_T/2)$ as shown in Fig.
\ref{v23_charmonia}(e) and (f).

\section{Discussion}

\subsection{Dependence of charmonia production on their
 structures; $J/\psi$ vs. $\psi(2S)$}

Here we discuss in more detail on the cause for the larger
production of the $\psi(2S)$, contrary to the statistical
hadronization model expectation, and confirm that the yield of the
$\psi(2S)$ must be large, as comparable as that of the $J/\psi$
when charmonium states are produced from charm quarks by
recombination at the quark-hadron phase transition.

Reminding that the hadron production by recombination is dominant
at low transverse momentum region, we expect that the larger
contributions at lower transverse momentum from the wave function
square in a momentum space, $|\tilde{\psi}(p)|^2$ in Eq.
(\ref{CoalTransSim}) leads to the larger yield, or the larger
transverse momentum distribution. Therefore, the bigger peak, or
the larger contribution in the charmonia wave function square in a
momentum space, especially in the low transverse momentum region,
is very crucial to the yield or transverse momentum distribution
of charmonium states when they are produced from charm quarks by
recombination.

\begin{figure}[!t]
\begin{center}
\includegraphics[width=0.495\textwidth]{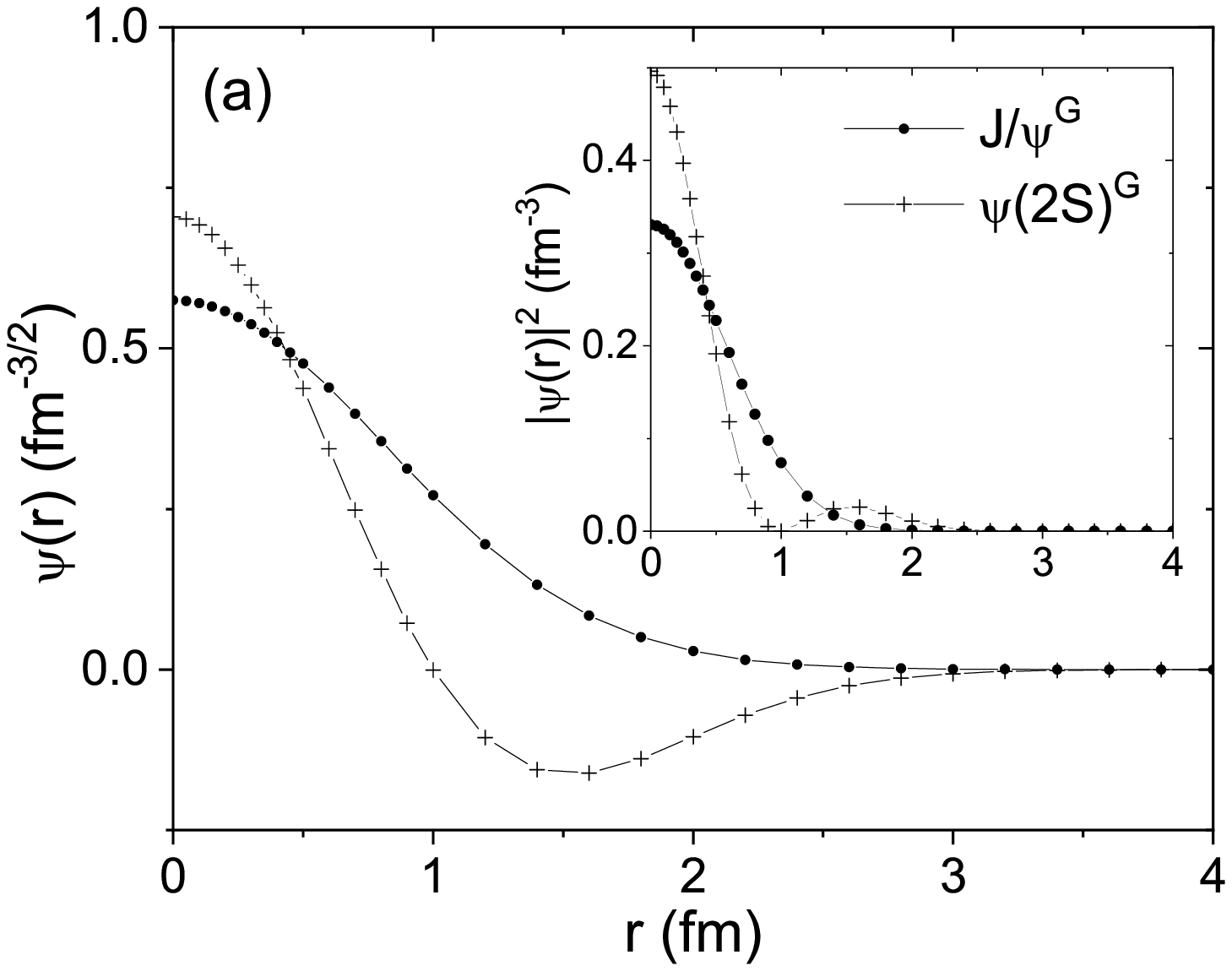}
\includegraphics[width=0.495\textwidth]{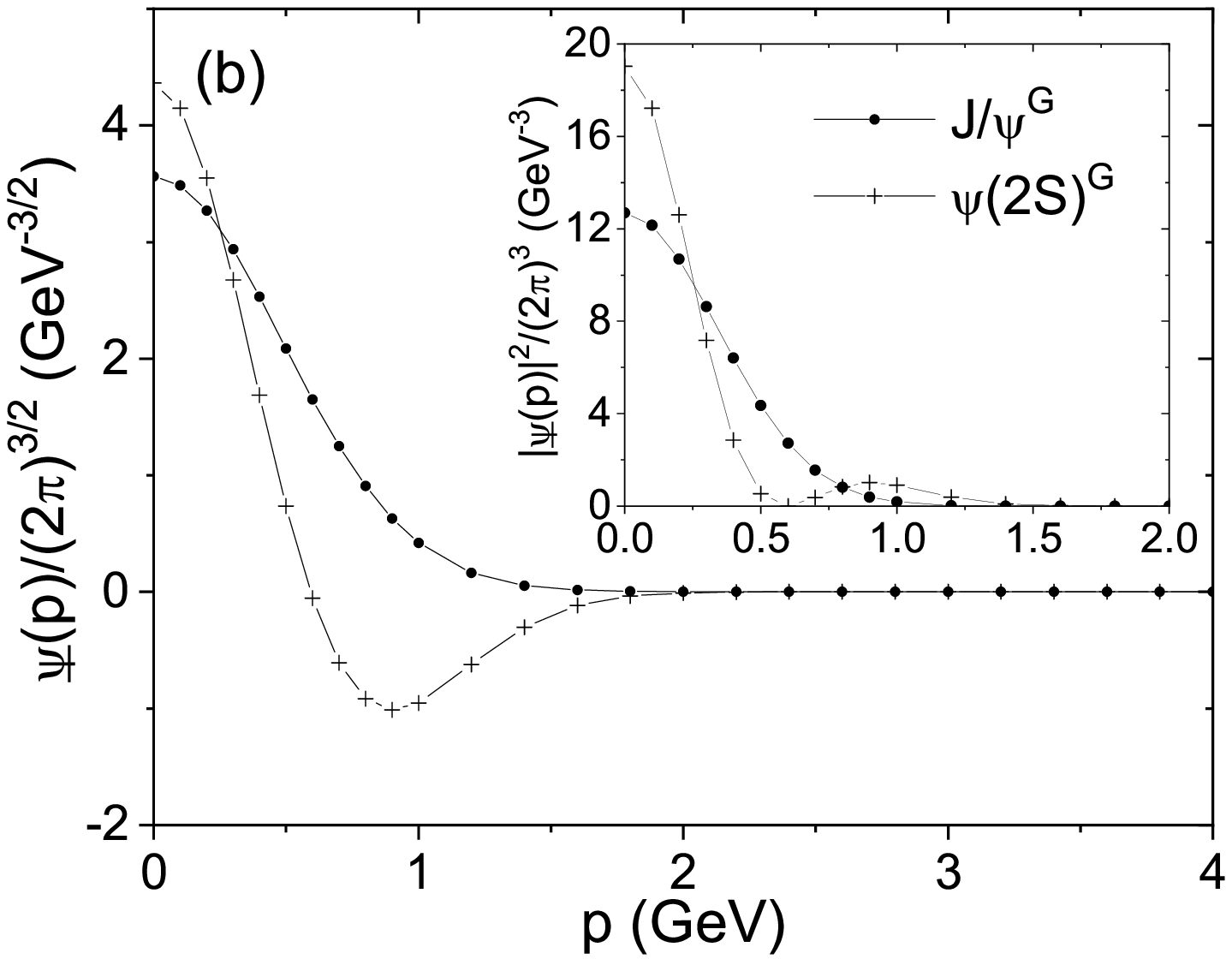}
\end{center}
\caption{(a) Gaussian wave function distributions of the $J/\psi$
and $\psi(2S)$ in a coordinate space, and (b) those of the
$J/\psi$ and $\psi(2S)$ divided by $(2\pi)^{3/2}$ in a momentum
space. In the inset of the figure (a) the wave function squares in
a coordinate space, $|\psi(r)|^2$ for the $J/\psi$ and $\psi(2S)$
are shown. Also in the inset of figure (b) the wave function
squares in a momentum space, $|\tilde{\psi}(p)|^2/(2\pi)^3$ for
the $J/\psi$ and $\psi(2S)$ are shown.} \label{rpwavefucntions}
\end{figure}

Fig. \ref{rpwavefucntions} displays the wave function distribution
of $J/\psi$ and $\psi(2S)$ mesons in both coordinate and momentum
spaces. The wave function squares in a coordinate space,
$|\psi(r)|^2$ for the $J/\psi$ and $\psi(2S)$, and those in a
momentum space, $|\tilde{\psi}(p)|^2/(2\pi)^3$ for the $J/\psi$
and $\psi(2S)$ are also shown in the inset of Fig.
\ref{rpwavefucntions}(a), and in the inset of Fig.
\ref{rpwavefucntions}(b), respectively. The superscript $G$
implies a Gaussian, or a harmonic oscillator wave function adopted
in the construction of the Wigner function, Eq. (\ref{WigIntdr}).

Since the $\psi(2S)$ is a radially excited state of the $J/\psi$,
the wave function of the $\psi(2S)$ has one node as shown in Fig.
\ref{rpwavefucntions}(a). Therefore, the wave function square in a
coordinate space, $|\psi(r)|^2$ of the $\psi(2S)$ has two peaks;
the bigger one located near the center, and the smaller one on the
right hand side of the node as shown in the inset of Fig.
\ref{rpwavefucntions}(a). On the other hand, the wave function of
the $J/\psi$ is localized near the center without a node in a
coordinate space, resulting in one peak in the $|\psi(r)|^2$ for
the $J/\psi$ in a coordinate space as also shown in the inset of
Fig. \ref{rpwavefucntions}(a).

\begin{widetext}

\begin{figure}[!t]

\begin{center}
\includegraphics[width=0.495\textwidth]{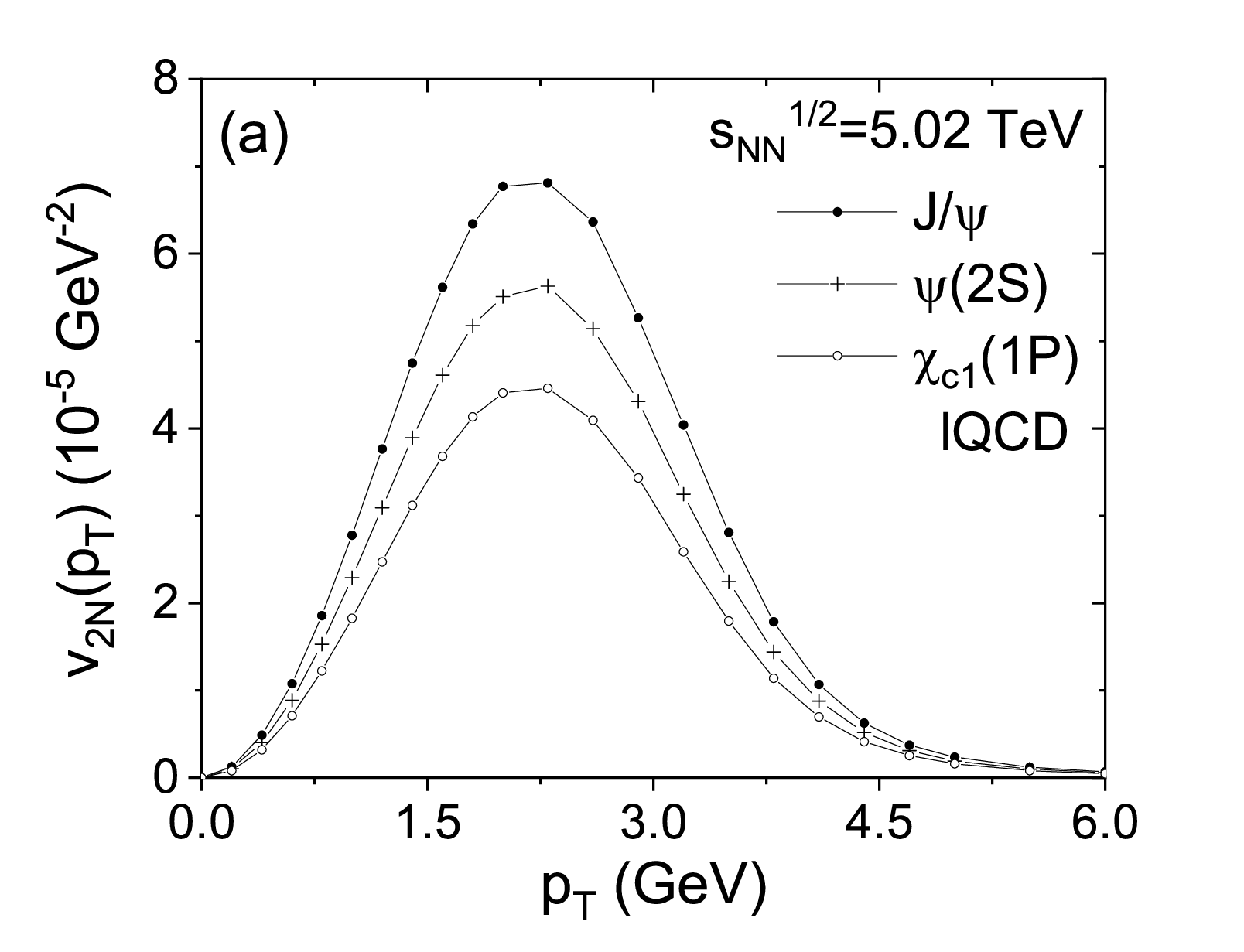}
\includegraphics[width=0.495\textwidth]{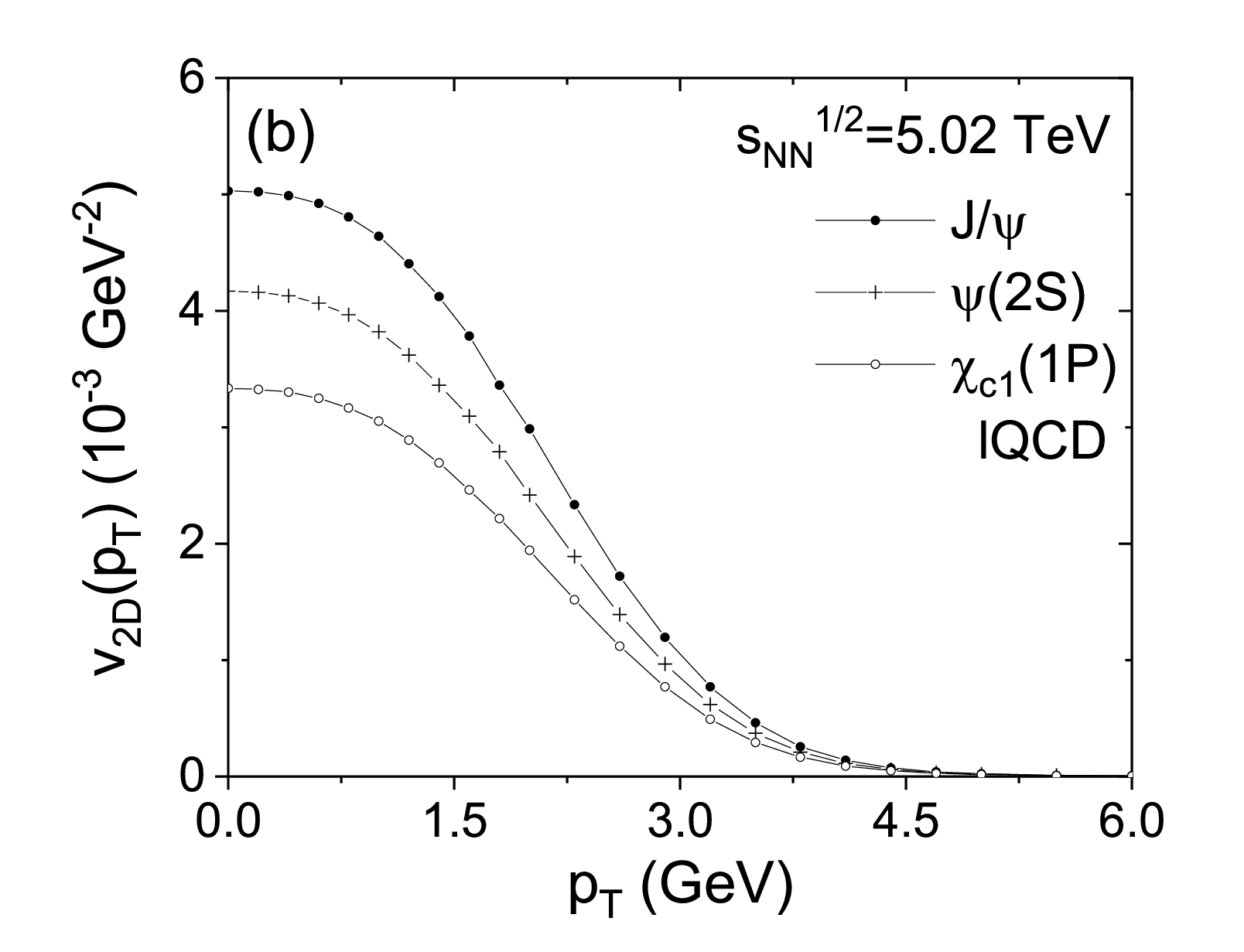}
\includegraphics[width=0.495\textwidth]{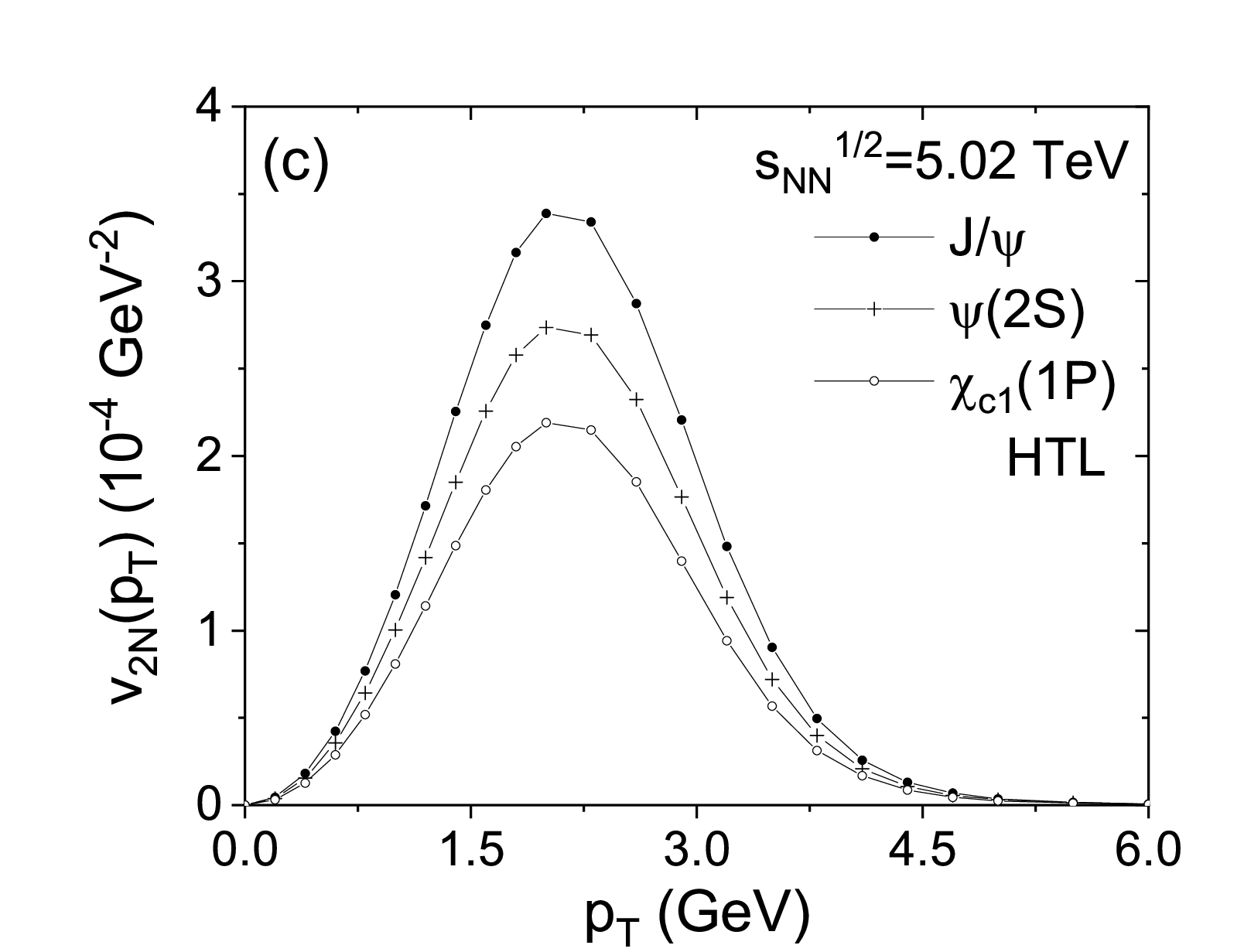}
\includegraphics[width=0.495\textwidth]{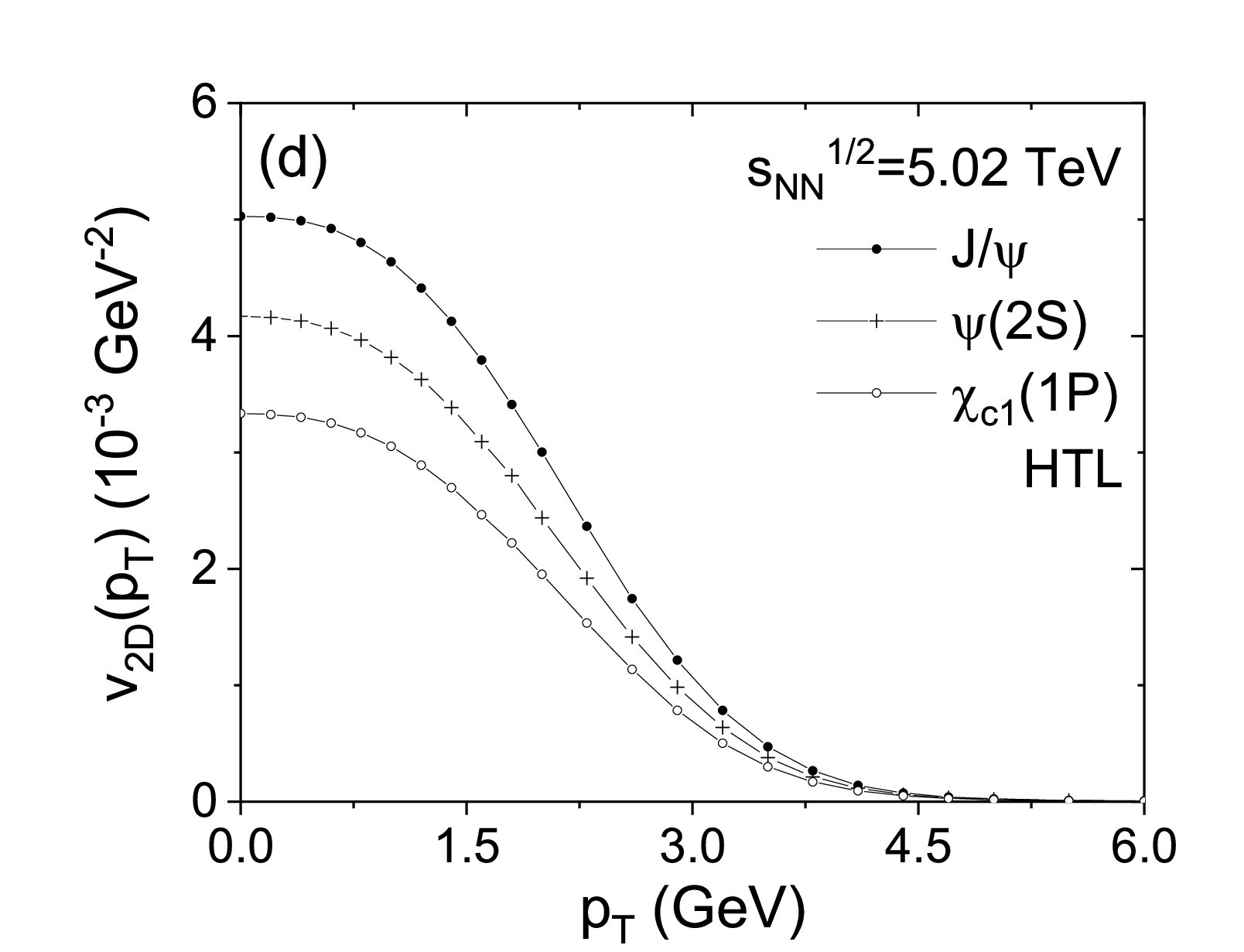}
\end{center}
\caption{Numerator parts of the elliptic flow in Eq.
(\ref{vn_averaged}) denoted by $v_{2N}(p_T)$ for charmonium
states, i.e., the $J/\psi$, $\psi(2S)$, and $\chi_{c1}(1P)$ at
LHC, $\sqrt{s_{NN}}$=5.02 TeV in the lQCD (a) and HTL (c)
transport coefficients in POWLANG transport setup
\cite{Beraudo:2017gxw} are shown. Also shown are denominator parts
of the elliptic flow in Eq. (\ref{vn_averaged}) denoted by
$v_{2D}(p_T)$ of the $J/\psi$, $\psi(2S)$, and $\chi_{c1}(1P)$
meson at LHC, $\sqrt{s_{NN}}$=5.02 TeV in the lQCD (b) and HTL (d)
transport coefficients in POWLANG transport setup. }
\label{v2_numerdinominator}
\end{figure}

\end{widetext}

These differences in the wave function of the $J/\psi$ and
$\psi(2S)$ in a coordinate space play significant roles in making
their production different through the wave function distribution
in a momentum space; the wave function distributions in a
coordinate space can be Fourier transformed into those in a
momentum space as shown in Fig. \ref{rpwavefucntions}(b), and it
is the wave function square in a momentum space,
$|\tilde{\psi}(k)|^2$ that contributes differently to the
production yield, or the transverse momentum distribution for the
$J/\psi$ and $\psi(2S)$ as shown in Eq. (\ref{CoalTransSim}).

Looking into the wave function of the $\psi(2S)$ in a momentum
space in more detail, we still find in Fig.
\ref{rpwavefucntions}(b) one node as in that of the $\psi(2S)$ in
a coordinate space. We also notice that the wave function square
of the $\psi(2S)$ is even bigger than that of the $J/\psi$ near
the zero transverse momentum, causing the bigger peak in the wave
function square of the $\psi(2S)$ compared to the peak in the
$J/\psi$ wave function square in the lower transverse momentum
region as shown in the inset of Fig. \ref{rpwavefucntions}(b).

The transverse momentum of charm quarks in a fire-ball frame,
hidden in the wave function square $|\tilde{\psi}(k)|^2$, Eq.
(\ref{CoalTransSim}) via the relative momentum between charm
quarks, $\vec k=(\vec p_{cT}^{~'}-\vec p_{\bar{c}T}^{~'})/2$ in
the charmonium frame, is integrated out together with the
transverse momentum distribution of charm and anti-charm quarks,
resulting in the transverse momentum distribution of charmonium
states. In this respect, the larger peak in the wave function
square of the $\psi(2S)$ in a momentum space at low transverse
momentum region is attributable to the larger production of the
$\psi(2S)$, thereby the yield or transverse momentum distribution
of the $\psi(2S)$ are as half large as those of the $J/\psi$.

\subsection{Dependence of elliptic and triangular flows of
charmonium states on their structures}

We discuss in this subsection the dependence of elliptic and
triangular flows of charmonium states on their internal
structures. Since the flow harmonics of charmonium states are
dependent on their internal structures via their transverse
momentum distributions, it seems natural to observe different
elliptic and triangular flows originated from different transverse
momentum distributions for different charmonium states. However,
as shown in Fig. \ref{v23_charmonia} almost same flow harmonics
for all charmonium states, the $J/\psi$, $\psi(2S)$ and
$\chi_c(1P)$ have been obtained at both RHIC and LHC, regardless
of two different charm quark interactions with the medium, the
lQCD and HTL transport coefficients in the POWLANG transport
analysis.

Noting that the $v_n$ in Eq. (\ref{vn_averaged}) is the average of
cosine functions in different orders $n$ over the transverse
momentum with the weight of the transverse momentum distribution
of charmonium states, we investigate separately the dependence of
numerator and denominator parts of the elliptic flow $v_2$ on
their internal structures, or their wave function distributions.
We denote the numerator and denominator part of $v_2$ as $v_{2N}$
and $v_{2D}$, respectively from Eq. (\ref{vn_averaged}),
\begin{eqnarray}
&& v_{2N}(p_T)=\frac{2}{2\pi}\int_{-\frac{\pi}{2}}^{
\frac{\pi}{2}} \int d\psi \cos(2(\psi-\Psi_2))
\frac{d^2N}{dp_T^2}d\Psi_2 \nonumber \\
&& v_{2D}(p_T)=\frac{2}{2\pi}\int_{-\frac{\pi}{2}}^{
\frac{\pi}{2}}\int d\psi \frac{d^2N}{dp_T^2}d\Psi_2. \label{v2ND}
\end{eqnarray}
We show in Fig. \ref{v2_numerdinominator} numerator parts of the
elliptic flow $v_{2N}(p_T)$, Eq. (\ref{v2ND}) for charmonium
states, i.e., the $J/\psi$, $\psi(2S)$, and $\chi_{c1}(1P)$ at
LHC, $\sqrt{s_{NN}}$=5.02 TeV in the lQCD (a) and HTL (c)
transport coefficients in POWLANG transport setup
\cite{Beraudo:2017gxw}. Also shown in Fig.
\ref{v2_numerdinominator} are denominator parts of the elliptic
flow $v_{2D}(p_T)$, Eq. (\ref{v2ND}) of the $J/\psi$, $\psi(2S)$,
and $\chi_{c1}(1P)$ meson at LHC, $\sqrt{s_{NN}}$=5.02 TeV in the
lQCD (b) and HTL (d) transport coefficients in POWLANG transport
setup. Here, the feed-down contributions, Eq. (\ref{feeddown})
from more heavier charmonium states are not taken into account in
both $v_{2N}(p_T)$ and $v_{2D}(p_T)$ of the $J/\psi$, $\psi(2S)$,
and $\chi_{c1}(1P)$. We clearly see different $v_{2D}$ as well as
$v_{2N}$ for different charmonium states, showing the explicit
dependence of both $v_{2N}$ and $v_{2D}$ of the $J/\psi$,
$\psi(2S)$ and $\chi_c(1P)$ meson on their transverse momentum
distributions.

The numerator part of $v_2$ is similar to the transverse momentum
distribution multiplied by $2\pi p_T$ as shown in Fig.
\ref{Jpsitopsi2Sratio}; the $v_{2N}$ is larger in the order of the
$J/\psi$, $\psi(2S)$ and $\chi_c(1P)$ due to their transverse
momentum distributions as expected. The shape of both figure is
also similar though the unit in each figure is different. The
behavior of $v_{2N}$ as a function of $p_T$ is entirely affected
by the sum of a charm and an anti-charm quark $v_{2c}$ in the
coalescence, $\sim(v_{2c}+v_{2\bar{c}})$, Eq. (\ref{vn_charm}).

The same characteristics observed in the numerator part of $v_2$
are also found in the denominator part of $v_2$; the $v_{2D}$ is
larger in the order of the $J/\psi$, $\psi(2S)$ and $\chi_c(1P)$
originated from their transverse momentum distributions. In
addition, we note that the $v_{2D}$ of charmonium states is almost
similar to the transverse momentum distribution itself as shown in
Fig. \ref{pT_charmonia}. We can understand this as the $v_{2D}$ is
reflected by the factor $\sim(1+2v_{2c}v_{2\bar{c}})$ in the
coalescence of a charm and an anti-charm quarks, Eq.
(\ref{vn_charm}), and both the $v_{2c}$ and $v_{2\bar{c}}$ are as
small as a few percents as shown in Fig. \ref{v23_charm}.

As discussed so far, the denominator as well as numerator parts of
$v_2$ is explicitly dependent on the transverse momentum
distribution. However, each dependence of both numerator and
denominator on the transverse momentum distribution is cancelled
out when $v_2$ is obtained. By this reason, we find no differences
in the $v_n$ of the $J/\psi$, $\psi(2S)$ and $\chi_c(1P)$ meson in
spite of their different wave function distributions as shown in
Fig. \ref{v23_charmonia}.

\subsection{Elliptic and triangular flows of
charmonium states at low transverse momentum regions}

We discuss here the elliptic and triangular flows of charmonium
states at low transverse momentum region. It has been found that
the elliptic flow at low transverse momentum region behaves like
the square function of transverse momentum, $p_T^2$, and similarly
the triangular flow at low transverse momentum region behaves as
the cube function of $p_T$, $p_T^3$ \cite{Dinh:1999mn}.

If the elliptic flow behaves as $p_T^2$ at low transverse momentum
region, $v_2^{1/2}/p_T$ is expected to be a constant as a function
of transverse momentum at low $p_T$. Therefore, in order to
investigate the elliptic and triangular flows at low transverse
momentum regions, $v_n^{1/n}/p_T$ have been measured for the
$J/\psi$ \cite{Acharya:2020jil} as well as light particles
\cite{ALICE:2018yph} as functions of $p_T$.

\begin{widetext}

\begin{figure}[!h]
\begin{center}
\includegraphics[width=0.495\textwidth]{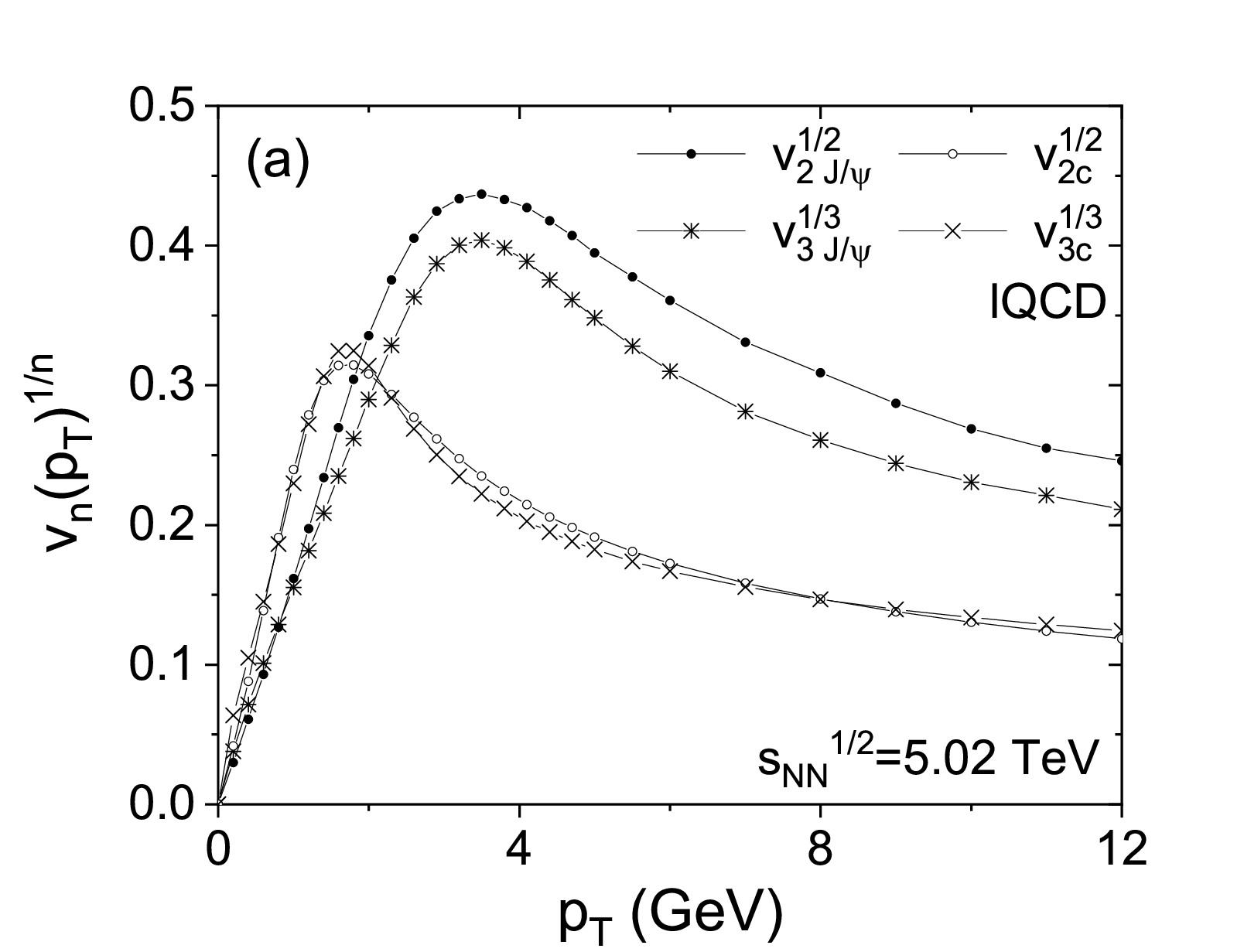}
\includegraphics[width=0.495\textwidth]{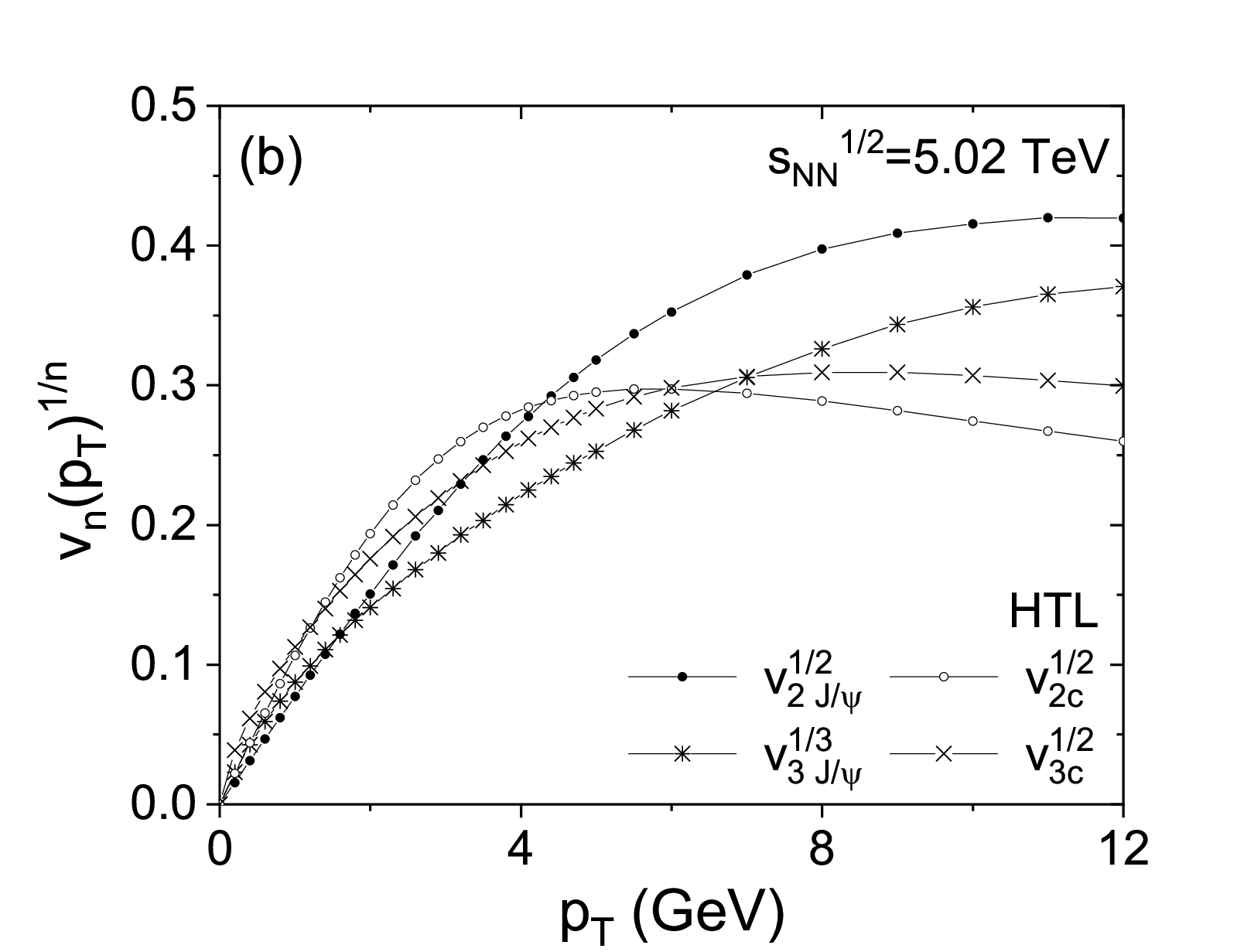}
\includegraphics[width=0.495\textwidth]{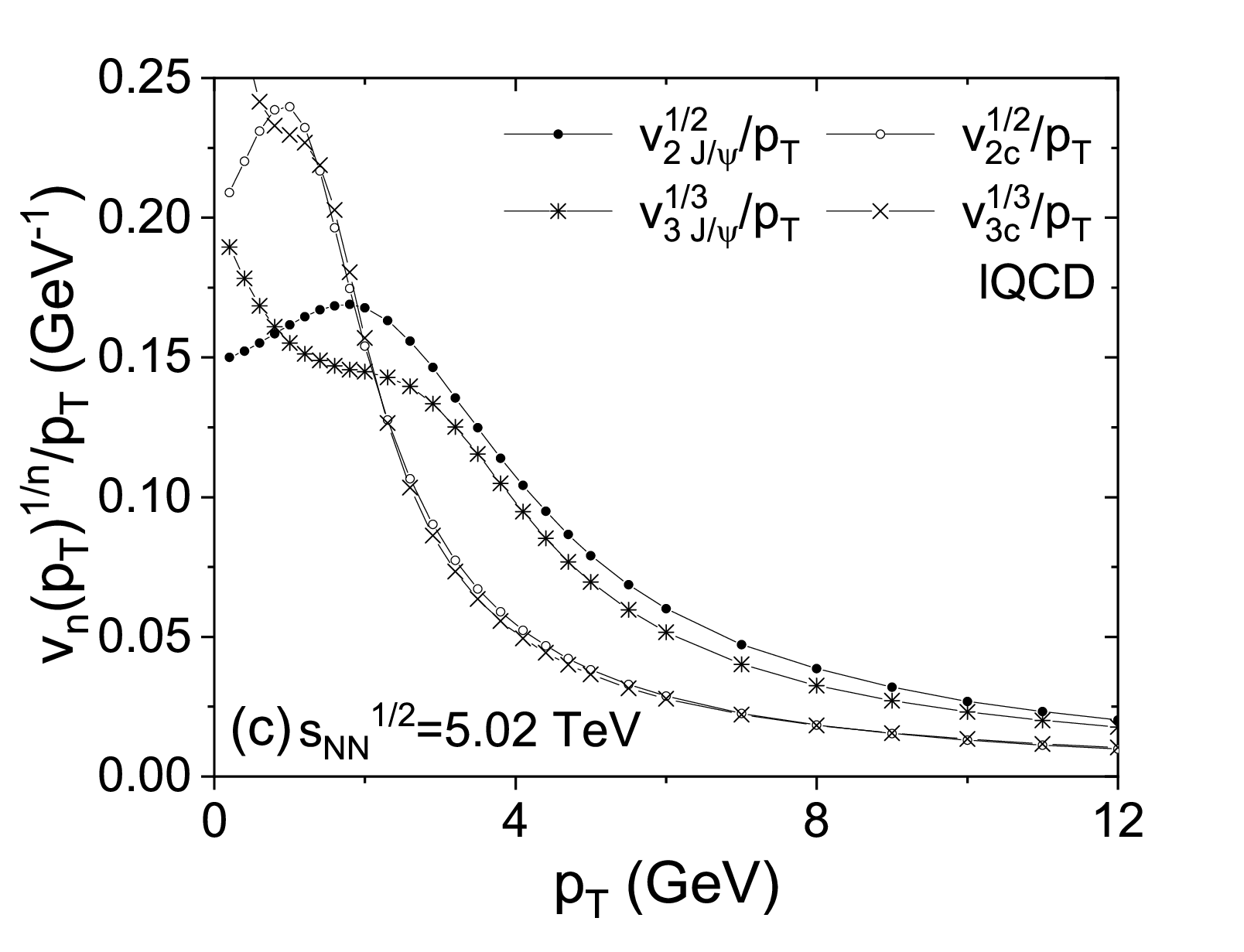}
\includegraphics[width=0.495\textwidth]{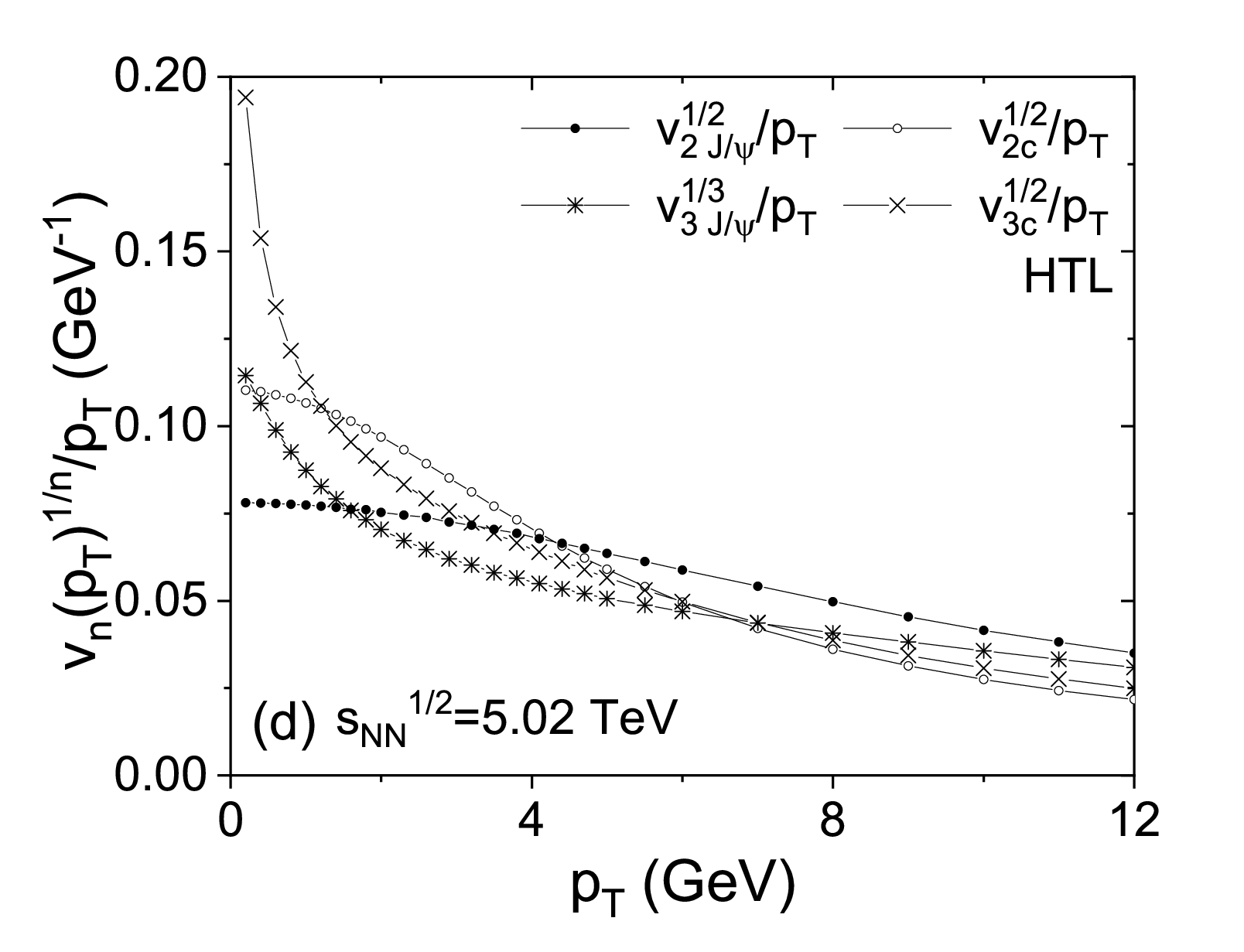}
\end{center}
\caption{Plots of $v_n^{1/n}/p_T$ together with $v_n^{1/n}$ for
the $J/\psi$ meson at LHC, $\sqrt{s_{NN}}$=5.02 TeV evaluated with
the charm quark elliptic and triangular flow from the POWLANG
transport setup in the lQCD (a), (c) and HTL (b), (d) transport
coefficients. Plots of $v_{n}^{1/n}$ as well as $v_n^{1/n}/p_T$ of
bare charm quark elliptic and triangular flow obtained from two
cases in the POWLANG transport setup as a function of transverse
momenta are also shown in each figure for comparison. }
\label{vnto1overn}
\end{figure}

\end{widetext}

We show in Fig. \ref{vnto1overn} plots of $v_n^{1/n}/p_T$ together
with $v_n^{1/n}$ for the $J/\psi$ meson at LHC,
$\sqrt{s_{NN}}$=5.02 TeV evaluated with the charm quark elliptic
and triangular flow from the POWLANG transport setup in the lQCD
(a), (c) and HTL (b), (d) transport coefficients. We also plot
$v_{n}^{1/n}$ as well as $v_n^{1/n}/p_T$ of bare charm quark
elliptic and triangular flow obtained from two cases in the
POWLANG transport setup shown in Fig. \ref{v23_charm} as functions
of transverse momenta for comparison.

As shown in the previous section, the relation between the flow
harmonics of mesons and that of constituent quarks, i.e.,
$v_{n,J/\psi}(p_T)\approx 2v_{n,c}(p_T/2)$ holds for charmonium
states. We see in Fig. \ref{vnto1overn} that the $v_n^{1/n}/p_T$
and $v_n^{1/n}$ of the $J/\psi$ also reflects that of bare charm
quarks via the above relation $v_{n,J/\psi}(p_T)\approx
2v_{n,c}(p_T/2)$. Nevertheless, we observe that the above relation
holds differently at low and intermediate transverse momentum
regions for the elliptic and triangular flow depending on
interaction strength between charm quarks and the medium.

Focusing on the $v_n^{1/n}/p_T$, we find that it is not a constant
at low transverse momentum region in a strongly coupled medium
with the lQCD transport coefficient in POWLANG as shown in Fig.
\ref{vnto1overn}(c). The triangular flow of the $J/\psi$,
$v_3^{1/3}/p_T$ in a weakly coupled medium with the HTL transport
coefficient is not also a constant, whereas the elliptic flow of
the $J/\psi$, $v_2^{1/2}/p_T$ is found to be a constant up to
about 3 GeV as shown in Fig. \ref{vnto1overn}(d).

Even though we obtain at low transverse momentum region best fit
functions of $p_T^2$ or $p_T^3$ for the elliptic and triangular
flow, Fig. \ref{v23_charmonia} and Table \ref{alphatn}, we find
that the $v_n^{1/n}/p_T$ is not a constant except the
$v_2^{1/2}/p_T$ in a weakly coupled medium with the HTL transport
coefficient in the POWLANG. This is possible as the fit function
usually passes through between the data points. For example, the
elliptic flow lies below the $\alpha p_T^2$ fit function,
$\alpha^{1/2}=0.028^{1/2}\approx 0.167$ up to about 1 GeV with the
$\alpha$ from Table \ref{alphatn}, while that lies above the fit
function between about 1 and 2 GeV, resulting in the increasing
$v_2^{1/2}/p_T$ up to about 2 GeV as shown in Fig.
\ref{vnto1overn}(c). In the case of $v_2^{1/2}/p_T$ in the HTL
transport coefficient shown in Fig. \ref{vnto1overn}(d) the
$p_T^2$ fit function describes almost exactly the elliptic flow of
the $J/\psi$, leading to the constant
$\alpha^{1/2}=0.0058^{1/2}\approx 0.076$ up to about 3 GeV.

It should be noted that the relation between the flow harmonics of
the $J/\psi$ and those of charm quarks, i.e.,
$v_{n,J/\psi}(p_T)\approx 2v_{n,c}(p_T/2)$ holds here; when the
$v_n^{1/n}/p_T$ of charm quarks is not a constant at low
transverse momenta, that of the $J/\psi$ cannot be a constant at
two times the low transverse momenta as shown in Fig.
\ref{vnto1overn}.

% Therefore, it is necessary to understand, first of all, why the
% elliptic or triangular flow of quarks is a constant at low
% transverse momentum region in order to see whether hadrons
% composed of those quarks show a constat for their $v_n^{1/n}/p_T$.
% In case of the light hadrons, the $n$th order flow harmonics of
% quarks might be proportional to $p_T^n$ at low transverse momentum
% region due to the same transverse momentum distribution of light
% quarks as they are in thermal equilibrium. On the other hand,
% charm quarks are not in thermal equilibrium, and

\begin{figure}[!t]
\begin{center}
\includegraphics[width=0.495\textwidth]{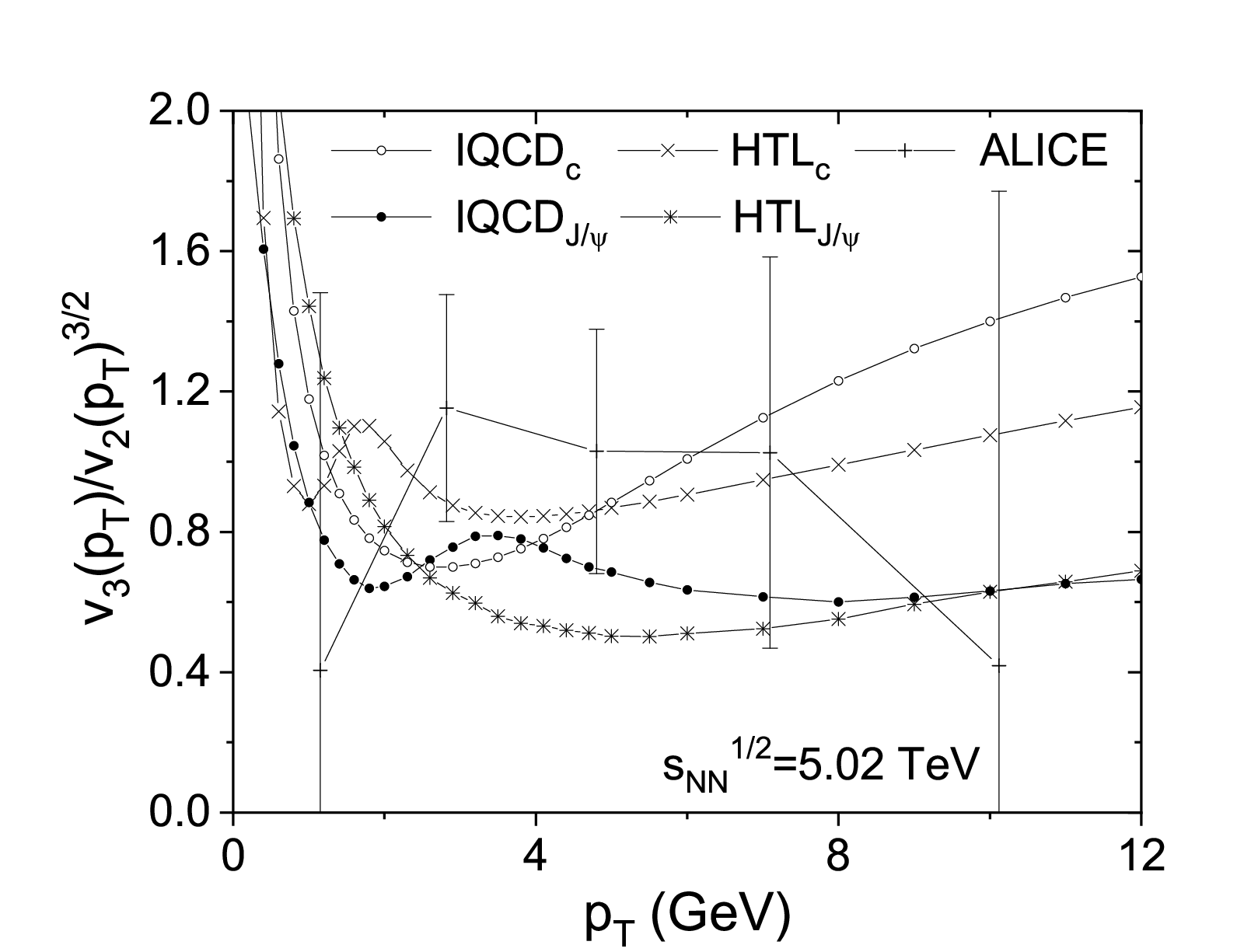}
\end{center}
\caption{The plot of the ratio between the elliptic and triangular
flow of the $J/\psi$ meson, $v_3/v_2^{3/2}$ at LHC,
$\sqrt{s_{NN}}$=5.02 TeV as a function of transverse momenta,
together with the experimental measurement of $v_3/v_2^{3/2}$ by
ALICE Collaboration \cite{Acharya:2020jil} are shown. We also show
the plot of the ratio, $v_3/v_2^{3/2}$ between bare charm quark
elliptic and triangular flow obtained from the POWLANG transport
setup, Fig. \ref{v23_charm} in the HTL and lQCD transport
coefficients \cite{Beraudo:2017gxw} as a function of transverse
momenta for comparison.} \label{vnratio}
\end{figure}

When the above relation holds for the elliptic and triangular
flow, $v_{2,J/\psi}(p_T)\approx 2v_{2,c}(p_T/2)$ and
$v_{3,J/\psi}(p_T)$ $\approx$ $2v_{3,c}(p_T/2)$, the ratio
$v_{2,J/\psi}^{1/2}(p_T)/v_{3,J/\psi}^{1/3}(p_T)$ is estimated to
be about $2^{1/2}/2^{1/3} v_{2,c}^{1/2}(p_T/2)/
v_{3,c}^{1/3}(p_T/2)$. As we see in Fig. \ref{vnto1overn}(a) that
$v_{2,c}^{1/2}(p_T/2)$ is very similar to $v_{3,c}^{1/3}(p_T/2)$,
we find $v_{2,J/\psi}^{1/2}(p_T)/v_{3,J/\psi}^{1/3}(p_T)$ varying
a little around $2^{1/2}/2^{1/3}\approx 1.12$ in the range between
1.10 and 1.19 inside the medium with the lQCD transport
coefficient in the POWLANG. On the other hand, the same ratio
increases from 0.9 to 1.25, and approaches to 1.12 as $p_T$
increases due to different behaviors of $v_{2,c}^{1/2}(p_T/2)$ and
$v_{3,c}^{1/3}(p_T/2)$ in the medium with the HTL transport
coefficient as shown in Fig. \ref{vnto1overn}(b).

The variation of the ratio $v_{2,J/\psi}^{1/2}(p_T)
/v_{3,J/\psi}^{1/3}(p_T)$ as a function of transverse momentum
presented in Fig. \ref{vnto1overn}(a) and (b) implies that the
elliptic and triangular flow of charmonium states in the strongly
coupled medium, i.e., in the lQCD transport coefficient develops
in the same way as functions of transverse momentum, whereas
elliptic and triangular flow of charmonium states develops easily
in a different way in a weakly coupled medium, i.e., in the HTL
transport coefficient in the POWLANG. Therefore, if the medium,
the quark-gluon plasma is strongly coupled, the elliptic and
triangular flow of the $J/\psi$ behave in the similar way, leading
to the constant ratio for $v_{2,J/\psi}^{1/2}(p_T)
/v_{3,J/\psi}^{1/3}(p_T)$.

In order to investigate in more detail the transverse momentum
dependence of the elliptic and triangular flow, we evaluate the
ratio $v_3/v_2^{3/2}$ of the $J/\psi$ which are found to be more
sensitive than $v_3^{1/3}/v_2^{1/2}$ \cite{ALICE:2018yph}. We plot
the above ratio between the elliptic and triangular flow of the
$J/\psi$ meson, $v_3/v_2^{3/2}$ at LHC, $\sqrt{s_{NN}}$=5.02 TeV
as a function of transverse momentum, together with the
experimental measurement of $v_3/v_2^{3/2}$ by ALICE Collaboration
\cite{Acharya:2020jil}. We also show the plot of the ratio,
$v_3/v_2^{3/2}$ between bare charm quark elliptic and triangular
flow obtained from the POWLANG transport setup, Fig.
\ref{v23_charm} in the HTL and lQCD transport coefficients
\cite{Beraudo:2017gxw} as a function of transverse momentum for
comparison.

As the measurement of the ratio, $v_3/v_2^{3/2}$ at LHC is made at
forward rapidity region of 2.5$<$y$<$4 in 0-50$\%$ centralities,
direct comparisons between the measurements at LHC and our results
obtained at midrapidity in 20-40$\%$ centralities cannot be made.
Nevertheless, when the relation between the flow harmonics of the
$J/\psi$ and those of charm quarks, i.e.,
$v_{n,J/\psi}(p_T)\approx 2v_{n,c}(p_T/2)$ holds, the ratio
$v_3/v_2^{3/2}$ is expected to be approximately
$2/2^{3/2}~v_{3,c}(p_T/2)/v_{2,c}^{3/2}(p_T/2)$. As shown in Fig.
\ref{vnto1overn} the ratio does not exhibit a constant at low
transverse momentum region, but it approaches to
$2/2^{3/2}\approx0.71$ as both the $v_{3,c}(p_T/2)$ and
$v_{2,c}^{3/2}(p_T/2)$ converges to equal value due to the same
path-length dependent energy loss with increasing transverse
momentum \cite{ALICE:2012vgf}.

\section{Summary and Conclusion}

We have discussed in this paper elliptic and triangular flow of
$J/\psi$, $\psi(2S)$, and $\chi_c(1P)$ mesons in heavy ion
collisions based on the coalescence model. Starting from the
evaluation of transverse momentum distributions and yields of
those charmonia, we have calculated elliptic and triangular flow
of charmonium states when they are produced by charm quark
recombination. We have investigated the dependence of not only
transverse momentum distributions and yields, but also elliptic
and triangular flow of the $J/\psi$, $\psi(2S)$, and $\chi_c(1P)$
on their internal structures via their wave function distributions
in momentum space. We have also discussed further the quark number
scaling of elliptic and triangular flow for charmonium states, and
have studied in detail the elliptic and triangular flow at low
transverse momentum region.

Following the argument that the internal structure, or the wave
function difference is the main factor affecting yields and
transverse momentum distributions of different charmonium states,
i.e., the $J/\psi$, $\psi(2S)$, and $\chi_c(1P)$ meson when they
are produced from the same number and kind of constituents, a
charm and anti-charm quarks by regeneration at the phase boundary
\cite{Cho:2014xha}, we first investigate in detail the dependence
of transverse momentum distributions and yields of the $J/\psi$,
$\psi(2S)$ and $\chi_{c1}(1P)$ meson on their wave function
distributions.

We consider three different Wigner functions constructed from the
wave function of the $J/\psi$, $\psi(2S)$, and $\chi_c(1P)$ meson,
and calculate the transverse momentum distributions and yields of
those charmonium states. The explicit dependence of transverse
momentum distributions and yields of charmonium states on their
internal structures via their wave function distributions is
observed, and as a result, the origin for the possibly large
production of the $\psi(2S)$ meson, as half large as that of the
$J/\psi$ is explained.

Using different transverse momentum distributions of the $J/\psi$,
$\psi(2S)$ and $\chi_{c1}(1P)$ meson, we then explore charmonia
flow harmonics such as $v_2$ and $v_3$ as an attempt to understand
possible dependence of the elliptic and triangular flow of
charmonium states also on their internal structures via their
different transverse momentum distributions.

We adopt here elliptic and triangular flow of charm quarks
obtained by the POWLANG transport analysis, which describes time
evolutions of heavy quarks in heavy ion collisions through the
relativistic Langevin equation. We consider two kinds of flow
harmonics of charm quarks based on both weak coupling transport
coefficients from Hard Thermal Loop (HTL) re-summation analysis,
and transport coefficients from non-perturbative Lattice Quntum
Chromodynamics (QCD) calculation in POWLANG transport setup
\cite{Beraudo:2017gxw}.

The elliptic and triangular flow of $J/\psi$, $\psi(2S)$ and
$\chi_c(1P)$ mesons are found to be only slightly different at
both RHIC and LHC energies, irrespective of two cases, flow
harmonics of constituent charm quarks considered in two different
interactions in the medium, the lQCD and HTL transport
coefficients in the POWLANG analysis.

Thereby, further examination on the $v_n$ itself is made, and both
the numerator and denominator part of $v_n$ are shown to be
actually dependent on the internal structure, or the wave function
distribution in momentum space of each charmonium state, whereas
flow harmonics of all charmonium states are evaluated to be almost
same, or independent of the internal structure of each charmonium
when those are calculated with the wave function distribution
dependent numerator and denominator. It can be realized that the
similar amounts of contribution to both the numerator and
denominator of the $v_n$ from different transverse momentum
distributions of charmonium states are cancelled out each other,
resulting in almost the same flow harmonics for all charmonium
states.

In addition to the $v_n$, flow harmonics of charmonium states
divided by the number of constituents, $v_n/2$ are also studied.
The relation similar to the well-known relation between the
elliptic flow of mesons and that of constituent quarks can be
found for charmonium states, the quark number scaling of flow
harmonics; the flow harmonics of charmonium states are
approximately twice that of charm quarks,
$v_{n,c\bar{c}}(p_T)\approx 2v_{n,c}(p_T/2)$.

We then investigate whether the behavior of elliptic and
triangular flow at low transverse momentum region, $v_n\sim p_T^n$
also holds for charmonium states by focusing those of the $J/\psi$
at LHC energies. For that purpose, the behavior of the $J/\psi$
$v_n^{1/n}/p_T$ is examined as a function of transverse momentum,
and is also compared to that of bare charm quarks. It can be seen
that the $v_n^{1/n}/p_T$ behaves differently at low and
intermediate transverse momentum regions, depending on interaction
strengths between charm quarks and the medium. It has been found
that the elliptic and triangular flow of charmonium states, the
$v_n^{1/n}/p_T$ in the lQCD transport coefficient vary in the
similar ways as functions of transverse momentum, whereas those of
charmonium states vary easily in different manners in a weakly
coupled medium with the HTL transport coefficients.

% We see that the $v_n^{1/n}/p_T$ of charmonium states in the lQCD
% transport coefficient describes better s an infer whether the
% medium, the quark-gluon plasma is strongly coupled or not by
% investigating the behavior of both the elliptic and triangular
% flow of the charmonium states.

In relation to the $v_n^{1/n}/p_T$, the transverse momentum
dependence of the ratio between the elliptic and triangular flow,
$v_3^{1/3}/v_2^{1/2}$ as well as $v_3/v_2^{2/3}$ is also studied.
It is found that the ratio $v_3/v_2^{3/2}$ is approximately
$2/2^{3/2}~v_{3,c}(p_T/2)/v_{2,c}^{3/2}(p_T/2)$, again simply from
the relation, $v_{n,c\bar{c}}(p_T)\approx 2v_{n,c}(p_T/2)$, and
approaches to $2/2^{3/2}\approx0.71$ as the transverse momentum
increases, and thereby both the $v_{3,c}(p_T/2)$ and
$v_{2,c}^{3/2}(p_T/2)$ experiences the same path-length dependent
energy loss, regardless of two different charm quark interactions
with the medium, the lQCD and HTL transport coefficients in the
POWLANG transport analysis.

% We note that since the relation between the flow harmonics of the
% $J/\psi$ and those of charm quarks, i.e.,
% $v_{n,J/\psi}(p_T)\approx 2v_{n,c}(p_T/2)$ holds here, the
% $v_n^{1/n}/p_T$ of the $J/\psi$ cannot be a constant at two times
% the transverse momentum of the charm quarks unless that of charm
% quarks is a constant at low transverse momentum region. Then, as
% expected, we can infer inversely properties of bare charm quarks
% from elliptic or triangular flow of chrmonium states as a function
% of transverse momentum.

Given the relation between the flow harmonics of the $J/\psi$ and
those of charm quarks, i.e., $v_{n,J/\psi}(p_T)\approx
2v_{n,c}(p_T/2)$ here, it is natural to observe that the
$v_n^{1/n}/p_T$ of the $J/\psi$ cannot be a constant at two times
the transverse momentum of the charm quarks unless that of charm
quarks is a constant at low transverse momentum region. By the
same token, it seems possible to infer various properties of bare
charm quarks inversely from elliptic or triangular flow of
chrmonium states as a function of transverse momentum.

Therefore, the question on why the $v_n^{1/n}/p_T$ of charm quarks
is a constant, or why the $v_n$ of charm quarks is proportional to
$p_T^n$ at low transverse momentum region should be answered
before the investigation on $v_n^{1/n}/p_T$ of the $J/\psi$ as a
function of transverse momentum is made, which needs further
studies in the future. It should be much worth to extract as much
information of charm quarks as possible from charmonium states
composed of a charm and an anti-charm quarks as properties of
charm quarks at the moment of hadronization is not well understood
compared to those of light quarks.

As shown previously, the elliptic and triangular flow of
charmonium states are all found to be almost identical with very
slight differences, and the relation, $v_{n,c\bar{c}}(p_T)\approx
2v_{n,c}(p_T/2)$ holds for all charmonium states. Thus, it does
not seem easy to understand recent measurements by CMS
Collaboration on different elliptic flow for the $J/\psi$ and
$\psi(2S)$ \cite{CMS:2022gvy} entirely from the consideration of
different internal structure effects when different charmonium
states are regenerated from a charm and an anti-charm quarks at
the quark hadron phase boundary.

% Therefore, in order to explain the measurements by CMS
% Collaboration, it seems necessary to consider other production
% mechanism of charmonium states as well, e.g., production from the
% decay of heavier bottom hadrons and production by fragmentation.
% Moreover, it seems also necessary to find other factors reflecting
% different internal structures of charmnoium states on not only
% their flow harmonics but also transverse momentum distributions.
% In addition, we also hope the more precise measurement of the
% anisotropic flow of the $J/\psi$ and $\psi(2S)$ meson at low
% transverse momentum region in the near future to help us to make a
% better understanding on different behaviors of different charmonia
% flow harmonics in heavy ion collision experiments.

Therefore, in order to explain the measurements by CMS
Collaboration, it would be necessary to find other ways to
accommodate the effects from different internal structures as well
as other different features of charmnoium states on not only their
flow harmonics but also transverse momentum distributions. Along
with the development in theory, we also hope the more precise
measurement of the anisotropic flow of the $J/\psi$ and $\psi(2S)$
meson at low transverse momentum region in the near future to help
us to make a clear understanding on the transverse momentum
dependence of different charmonia flow harmonics in heavy ion
collision experiments.

% Considering that the flow harmonics of charmonium states is one
% of important quantities directly connected to that of charm ^
% quarks by the relation, $v_{n,c\bar{c}}(p_T)\approx
% 2v_{n,c}(p_T/2)$ at low and intermediate transverse momentum
% region, we insist that studying the charmonia anisotropic flow as
% well as the charmonia transverse momentum distribution when
% charmonium states are regenerated from a charm and anti-charm
% quarks at the quark hadron phase boundary, provides good chances
% to probe not only flow harmonics but also other properties of
% charm quarks in heavy ion collisions.

Considering that the elliptic or triangular flow of charmonium
states is one of important observables directly connected to that
of charm quarks by the relation, $v_{n,c\bar{c}}(p_T)\approx
2v_{n,c}(p_T/2)$ at low and intermediate transverse momentum
regions, we insist that studying the charmonia anisotropic flow
provides good opportunities to probe not only flow harmonics but
also other properties of charm quarks in heavy ion collisions.

Furthermore, as the charmonia production is one of valuable cases
to observe the production of different hadrons with different
internal structures but produced from both the same number and
kind of constituents, investigating not only the transverse
momentum distributions, yields, and anisotropic flow but also
other closely relevant observables of the $J/\psi$, $\psi(2S)$,
and $\chi_{c1}(1P)$ meson when they are produced from a charm and
an anti-charm quarks by quark coalescence, would be helpful in
understanding in more detail the hadron production mechanism in
heavy ion collisions, finally resulting in the broadening of our
understanding on the properties of the QGP.

\section*{Acknowledgements}

This work was supported by the National Research Foundation of
Korea (NRF) grant funded by the Korea government (MSIT) (No.
2018R1A5A1025563) and (No. 2019R1A2C1087107).

\end{document}